\pdfoutput=1
\documentclass[PAPER, cernpreprint, atlasdraft=false, UKenglish, texmf, orcidlogo]{atlasdoc}
\usepackage{atlaspackage}
\usepackage{float}
\usepackage{atlasbiblatex}

\usepackage{atlasphysics}

\usepackage{multirow}
\usepackage{makecell}

\graphicspath{{logos/}{figures/}}

\usepackage{ANA-TOPQ-2023-31-PAPER-defs}  %
\usepackage{tcolorbox}
\AtlasTitle{Differential cross-section measurements of coherent production of singly and doubly resonant top-quark in \(WWbb\) events with one lepton at \(\sqrt{s} = 13\) TeV with the ATLAS detector}

\hypersetup{pdftitle={Differential cross-section measurements of coherent production of singly and doubly resonant top-quark in WWbb events with one lepton at 13 TeV with the ATLAS detector},pdfauthor={The ATLAS Collaboration}}

\AtlasAbstract{%
This paper presents differential cross-section measurements of events containing a charged lepton, missing transverse momentum, two \(b\)-jets,
and two light jets, consistent with the \(W^{+}W^{-}b\bar{b}\) final state. The analysis is based on \(140~\mathrm{fb}^{-1}\) of proton--proton collision data
at \(\sqrt{s} = 13\)~TeV recorded with the ATLAS detector during Run 2 of the Large Hadron Collider (2015–2018). Differential cross-sections
are unfolded at particle level for a range of kinematic observables and in three kinematic regions. Both regions and observables are optimised
for sensitivity to the interference between \(W^{+}W^{-}b\bar{b}\) produced with singly and doubly resonant top quarks.
Results are compared with various theoretical predictions and deviations are observed in regions sensitive to the interference between singly and doubly resonant top-quark production.
The unfolded results provide new information about the modelling of interference and off-shell effects of the \(W^{+}W^{-}b\bar{b}\) final state, improving the
understanding of top-quark production mechanisms and providing new inputs for future Monte Carlo generator development.
}

\AtlasRefCode{TOPQ-2023-31}  %

\PreprintIdNumber{CERN-EP-2025-229}

\AtlasJournalRef{\JHEP 07 (2026) 028}
\AtlasDOI{10.1007/JHEP07(2026)028}

\begin{document}

\maketitle

%

%
%

%

\section{Introduction}%

The study of the top quark is a cornerstone of the Large Hadron Collider (LHC)~\cite{Evans:2008zzb} physics programme.
At the LHC, top quarks are predominantly produced in top--antitop-quark pairs. This process is  %
referred to as doubly resonant top-quark production, or \ttbar, in the following.  Additionally, top quarks can be produced individually via singly resonant top-quark production,
which occurs through three main channels: the \(t\)-, \(s\)-, and \(tW\)-channels. The \(t\)-channel involves the exchange
of a virtual \Wboson boson and is characterised by the scattering of a light quark, making it the most
abundant single-top production mode. The \(s\)-channel proceeds through the production of a virtual \Wboson boson
that yields a top-quark and a bottom quark. Lastly, the \(tW\)-channel involves the production of a single top quark
in association with a \(W\) boson. The top quark decays into a \Wboson boson and a bottom quark (\(b\)) with a branching
fraction of approximately \qty{100}{\%}. When considering all processes at next-to-leading (NLO) accuracy in quantum chromodynamics (QCD),
the associated production of a single resonant top-quark and a \Wboson boson, or \tWb, has the same final
state as doubly resonant top-quark production.
Traditionally these processes have been treated separately in theory calculations and experimental measurements, due to their large cross-section differences and the non-trivial interplay between leading order (LO) and NLO between the two processes%
\footnote{At NLO, the \( gb \rightarrow WWb \) process receives radiative corrections that include contributions overlapping with the LO \( gg \rightarrow \WbWb \) process (\ttbar).
The contribution of the latter, being \ttbar mediated, is much larger than the former.}.

\begin{figure}
\centering
\subfloat[]{
\includegraphics[width=0.3\textwidth]{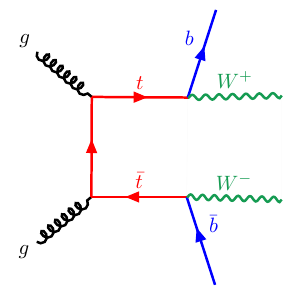}%
\label{fig:intro:feynman:ttbar}
}
\subfloat[]{
\includegraphics[width=0.3\textwidth]{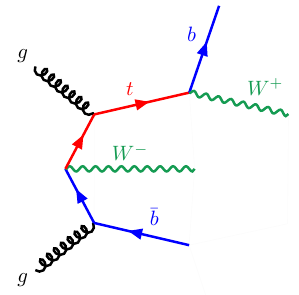}%
\label{fig:intro:feynman:singletop}
}
\subfloat[]{
\includegraphics[width=0.33\textwidth]{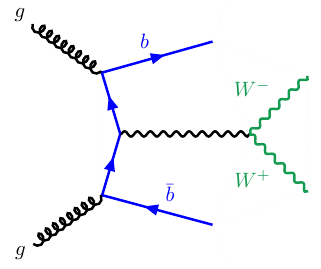}%
\label{fig:intro:feynman:nonresonant}
}
\caption{Feynman diagram of the \WbWb\ production. The \WbWb topology can arise from
\protect\subref{fig:intro:feynman:ttbar} doubly resonant top-quark processes,
\protect\subref{fig:intro:feynman:singletop} singly resonant top-quark processes,
\protect\subref{fig:intro:feynman:nonresonant} diagrams without intermediate top-quarks, and their interference.}%
\label{fig:intro:feynman}
\end{figure}

Differential cross-sections for \ttbar production have been measured by the ATLAS~\cite{TOPQ-2016-01,TOPQ-2018-15,TOPQ-2019-22,TOPQ-2020-09,TOPQ-2024-12} and CMS~\cite{CMS-TOP-16-007,CMS-TOP-16-014,CMS-TOP-17-002,CMS-TOP-18-004} collaborations and accurate theory calculations have been performed~\cite{Czakon:2011xx,Czakon:2015owf,Czakon:2016dgf}.
Differential cross-sections for \tW production have also been measured by the two LHC experiments~\cite{TOPQ-2015-16,TOPQ-2016-12,TOPQ-2018-07,CMS-TOP-17-018,CMS-TOP-20-002,CMS-TOP-21-010,CMS-TOP-17-023}.
A dedicated procedure~\cite{White:2009yt,Frixione:2008yi,Hollik:2012rc,Demartin:2016axk,ATL-PHYS-PUB-2016-020} is used to define single-top-quark
cross-sections and Monte Carlo predictions not including the doubly resonant top-quark production. The separation of single and double top-quark
production is  subject to a certain degree of arbitrariness, related to the treatment
of interferences and off-shell top-quark (finite-width) effects. This theory approximation provides a significant source of uncertainty for many
measurements~\cite{TOPQ-2018-26,TOPQ-2020-06,TOPQ-2018-15,STDM-2018-41} and searches for physics beyond the Standard Model (BSM)~\cite{SUSY-2016-17,SUSY-2014-07,SUSY-2015-01,SUSY-2015-08,SUSY-2016-16,SUSY-2018-12,SUSY-2019-18,EXOT-2019-12,EXOT-2019-08,EXOT-2020-18,SUSY-2018-30}.
For this reason, a coherent treatment of singly and doubly top resonant production is desirable. This can be achieved considering the complete
process \( pp \rightarrow \WbWb \).  The \WbWb topology can arise from doubly resonant top-quark processes (\cref{fig:intro:feynman:ttbar}),
singly resonant top-quark processes (\cref{fig:intro:feynman:singletop}), non-resonant diagrams without intermediate top quarks (\cref{fig:intro:feynman:nonresonant})
and their interference.
The consistent calculation of the \WbWb process was performed up
to NLO order in QCD~\cite{Denner:2010jp} and supplemented with
parton showers~\cite{Jezo:2016ujg,Jezo:2023rht}.
Calculations of the process \(pp\to\ell\nu b b j j\) were performed up to NLO QCD~\cite{Denner:2017kzu} and NLO QCD including NLO electroweak corrections in Ref.~\cite{Mans:2025czh}.
The only existing measurements  to assess these predictions in
a region sensitive to the interplay between singly and doubly resonant top-quark processes have been
performed in final states with two light charged leptons~\cite{TOPQ-2017-05, TOPQ-2020-14, TOPQ-2024-12}.

This paper considers events with at least one leptonically decaying \Wboson boson.
The final state is characterised by a light charged lepton (\(e\) or \( \mu \)), missing transverse momentum, and
three or four jets, of which two identified as originating from \(b\)-quarks (\bjets).
The focus of this measurement is singly and doubly resonant top quark
production: \ttbar and \tWb. Differential cross-sections on key top-quark physics observables are extracted
using an unfolding procedure at particle level. The analysis selections, fiducial volumes
and observables are optimised to probe a kinematic phase space where the interplay (and interference)
between the two processes is enhanced, using dedicated requirements to actively suppress
the pure doubly resonant top-quark production.
The non-resonant \WbWb production is not explicitly enhanced in these selections and has therefore a negligible
contribution to the analysis selections.
In comparison with measurements requiring two light charged leptons in the final state,
the single-lepton channel provides the opportunity to measure the \Wboson boson kinematic properties directly,
since the hadronically decaying \Wboson boson can be fully reconstructed.
In addition, the larger branching ratio gives access to kinematic regions of the phase space characterised by more
extreme momenta.
The analysis therefore focusses on three kinematic regions.
The inclusive region (\IR), designed to study both single and double top-quark
processes and their interference in the broadest possible phase space;
the \WRlong region (\WR), which directly probes the \WbWb process by studying the kinematic properties of
the \Wboson boson and \bjet kinematic observables and thus provides constraints on higher-order corrections
or the modelling of real emissions;
and the search-like region (\SLR), which takes inspiration from a number of BSM searches~\cite{SUSY-2015-02,SUSY-2016-17,SUSY-2014-07,SUSY-2015-01,SUSY-2016-15,SUSY-2015-08} that probe
less inclusive corners of the phase space where the interference effects
between \ttbar and \tWb processes are found to be enhanced.

The results of this paper are compared with various state-of-the-art simulations and can be used in the development of new predictions.
The full data sample collected with the ATLAS detector between 2015 and 2018
at centre-of-mass energy \( \sqrt{s} = \qty{13}{\TeV} \)  with an integrated luminosity of
\( \mathcal{L} = \qty{140}{\ifb} \) is used for the measurement.

%
%
%
%
%
%


\section{Experimental and simulation setups}%
\label{sec:setup}

\subsection{The ATLAS experiment}
The ATLAS experiment~\cite{PERF-2007-01} at the LHC is a multi-purpose particle
detector with a forward--backward symmetric cylindrical geometry and a near  %
\( 4\pi \) coverage in solid angle.\footnote{\AtlasCoordFootnote}
It consists of an inner tracking detector surrounded by a thin superconducting
solenoid providing a \qty{2}{\tesla} axial magnetic field, electromagnetic
and hadronic calorimeters, and a muon spectrometer.
The inner tracking detector (ID) covers the pseudorapidity range \(|\eta| < 2.5\).
It consists of silicon pixel, silicon microstrip, and transition radiation
tracking detectors.
Lead/liquid-argon (LAr) sampling calorimeters provide electromagnetic
(EM) energy measurements with high granularity within the region \(|\eta|< 3.2\).
A steel/scintillator-tile hadronic calorimeter covers the central pseudorapidity
range (\(|\eta| < 1.7\)).
The endcap and forward regions are instrumented with LAr calorimeters
for EM and hadronic energy measurements up to \(|\eta| = 4.9\).
The muon spectrometer surrounds the calorimeters and is based on
three large superconducting air-core toroidal magnets with eight coils each.
The field integral of the toroids ranges between \num{2.0} and
\qty{6.0}{\tesla\metre} across most of the detector.
The muon spectrometer includes a system of precision tracking chambers
up to \(|\eta| = 2.7\) and fast detectors for triggering up to \(|\eta| = 2.4\).
The luminosity is measured mainly by the LUCID--2~\cite{LUCID2} detector which
is located close to the beampipe.
A two-level trigger system is used to select events~\cite{TRIG-2016-01}.
The first-level trigger is implemented in hardware and uses a subset of
the detector information to accept events at a rate close to \qty{100}{\kHz}.
This is followed by a software-based trigger that reduces the accepted
rate of complete events to \qty{1.25}{\kHz} on average depending on the data-taking conditions.
A software suite~\cite{SOFT-2022-02} is used in data simulation,
in the reconstruction and analysis of real and simulated data,
in detector operations, and in the trigger and data acquisition
systems of the experiment.

\subsection{Object reconstruction}%
\label{reconstruction}

Each physics object included in the event selection is defined by a collection of detector information and is reconstructed using dedicated
algorithms. These reconstructed particles are referred to as detector-level
objects in the following text.

Electron candidates are reconstructed from energy deposits (clusters)
in the electromagnetic calorimeter associated with reconstructed tracks
in the inner detector~\cite{EGAM-2018-01}. Candidates that have a transverse momentum
\( \pt > \qty{10}{\GeV} \) and \( |\eta| < 2.47 \) are selected, excluding
the calorimeter transition region \( 1.37 < |\eta| < 1.52 \).
Electrons in this measurement must satisfy the \textit{Tight} likelihood-based
identification criterion~\cite{EGAM-2018-01}.
Muon candidates are reconstructed from track segments in the various layers
of the muon spectrometer, and matched with tracks from the inner
detector~\cite{MUON-2018-03}. The final muon candidates are refitted using
the complete track information from both detector systems, and required
to satisfy \( \pt > \qty{10}{\GeV} \) and \( |\eta| < 2.5 \). Muons are required
to satisfy the \textit{Medium} identification requirements.
Both electrons and muons are required to meet the tight working point of the prompt-lepton isolation discriminant~\cite{HIGG-2017-02},
trained to separate prompt and non-prompt leptons.
Electrons (muons) are further required to have
\( |\zzsth| < \qty{0.5}{\mm} \) and \( |\dzero/\sigma(\dzero)| < 5~(3) \),
where the longitudinal and transverse impact parameters \dzero and \zzero
are computed relative to the primary vertex, and \( \sigma(\dzero) \) is the
resolution of the track transverse impact parameter
\dzero~\cite{Magliocca:2791559}.
Calibrations of the electron and muon energy scale and resolution, as well as corrections to
the reconstruction, identification and isolation efficiencies are applied following Refs.~\cite{EGAM-2018-01,EGAM-2021-02,MUON-2022-01,MUON-2018-03}.
A secondary set of (fake) electrons and muons candidates is used in the estimate of
background arising from misidentified and non-prompt lepton sources. Electron candidates in the fakes category are required to
fulfil all kinematic requirements of standard electrons. They are additionally required to satisfy the \textit{Loose} likelihood-based
identification criterion but fail to meet either the tight identification requirement or the isolation requirements.
Muon candidates in the fakes category are required to fulfil all kinematic and identification requirements of standard muons, but fail to satisfy the isolation requirements.

Jets are reconstructed using the \antikt algorithm~\cite{Cacciari:2008gp}
as implemented in FastJet~\cite{Fastjet} with a jet radius parameter
\( R = 0.4 \). The inputs to this algorithm are particle flow (PFlow)
objects~\cite{PERF-2015-09}, which combine measurements from the ATLAS inner
detector and calorimeters~\cite{PERF-2014-07} to improve the jet energy
resolution and increase the jet reconstruction efficiency, especially at low
jet \pt. Jets are required to have \( \pt > \qty{25}{\GeV} \) and
\( |\eta| < 2.5 \). Jets originating from \pileup interactions  are rejected using the \textit{jet vertex tagger}
(JVT) tool~\cite{PERF-2014-03}, which uses a combination of track-based
variables providing discrimination against pile-up jets. The \texttt{Tight}
working point is used for jets in the momentum range \( 20 < \pt < \qty{60}{\GeV} \)
and \( |\eta| < 2.4 \).
Calibrations (and their associated uncertainties) are
applied in the measurement to the jet energy scale (JES) and resolution
(JER)~\cite{JETM-2018-05}, which include components
derived both from simulation and in situ measurements.
The identification of \bjets is made using the \texttt{DL1r} \(b\)-tagging algorithm~\cite{FTAG-2019-07},
a high-level tagger based on a deep neural network and it is applied to jets satisfying the selections stated above.
In this measurement, a working point with a \(b\)-tagging efficiency of \qty{77}{\%} is used, as evaluated in simulated \ttbar events.
The corresponding rejection factors are 6 for charm-jets and 134 for light-jets, respectively.
Correction factors are applied to the simulated event samples to compensate
for differences between data and simulation in the \(b\)-tagging efficiency for
\(b-\), \(c-\) and light-jets.

The missing transverse momentum, with magnitude \met, is defined as the negative
vector sum of the transverse momenta of all selected and calibrated electrons,
muons and jets in the event, with an extra track-based term added to account for energy
in the event that is not associated with any of these objects~\cite{JETM-2020-03}.

To avoid double-counting in the analysis due to reconstruction ambiguities, objects are removed based
on the distance \( \Delta R_{y} = \sqrt{{(\Delta y)}^2 + {(\Delta \phi)}^2} \),
where \(y\) is rapidity.
If an electron and a muon overlap within \( \Delta R_{y} = 0.1 \), the muon is removed if it is
reconstructed only from an ID track
and calorimeter energy deposits consistent with a minimum-ionising particle, otherwise the electron is removed.
Then, jets within \( \Delta R_{y} = 0.2 \) of an electron are removed.
If a jet within \( \Delta R_{y} = 0.4 \) survives, the electron is discarded.
Muons are removed if \( \Delta R_{y} < 0.4 \) from a jet, but if the jet has
fewer than three tracks, the jet is removed instead, ensuring efficient
selection of high-energy muons affected by calorimeter energy loss.

\subsection{Modelling and simulation of Standard Model processes}%
\label{subsec:simulation}

Monte Carlo (MC) simulated event samples of all major Standard Model (SM) processes are used
to aid in the estimation of the differential cross-sections and the unfolding
procedure. They are additionally used to estimate systematic uncertainties
arising from the theory modelling of such processes.
The ATLAS detector response is simulated using a detailed description
of the detector geometry and response~\cite{SOFT-2010-01} based on
\texttt{Geant4}~\cite{Agostinelli:2002hh} or through a fast simulation using
a parameterisation of the calorimeter response and \texttt{Geant4} for
the other parts of the detector.
Detector-level objects are reconstructed in simulated events using the same
algorithms used to process the data, with appropriate performance corrections and calibrations applied.
The matrix element (ME) generator, parton shower (PS), cross-section
accuracy (CS) and specific tools used for their calculation, parton distribution function (PDF) set and the set of tuned
parameters (tune) for these samples
are given in \cref{tab:MC}, and more details of the generator configurations
can be found in Refs.~\cite{ATL-PHYS-PUB-2017-001,ATL-PHYS-PUB-2017-005,ATL-PHYS-PUB-2018-009,ATL-PHYS-PUB-2016-020}
and are briefly summarised in the following text.
All simulation samples, except for \SHERPA, use \EVTGEN~\cite{Lange:2001uf} to model the decays of bottom and charm
hadrons.
The effect of multiple interactions in the same and neighbouring bunch
crossings (\pileup) is modelled by overlaying the simulated hard-scattering event with
inelastic proton--proton (\(pp\)) events generated with \PYTHIA[8.186]~\cite{Sjostrand:2007gs}  %
using the \NNPDF[2.3lo] set of PDFs~\cite{Ball:2012cx} and the
A3 set of tuned parameters~\cite{ATL-PHYS-PUB-2016-017}.
Events are weighted to reproduce the distribution of the average number
of interactions per bunch crossing (\(\left< \mu \right>\)) observed in the data.
The \(\left< \mu \right>\) value in data is rescaled by a factor
of \(1.03 \pm 0.04\) to improve agreement between data and simulation
in the visible inelastic \pp cross-section~\cite{STDM-2015-05}.

\paragraph{Modelling of the signal processes}
This measurement targets singly and doubly resonant top quark
production:  \tWbparenthesis and \ttbar. Simulation samples of these processes are used
to estimate the unfolding parameters needed to extract fiducial differential
cross-sections at particle-level and to estimate the systematic uncertainties
associated with the modelling of these processes. These simulation samples are
also used to construct the pseudo-data sample used to estimate the closure
and robustness of the unfolding procedure.

The production of both final states is modelled using the
\POWHEGBOX[v2]~\cite{Frixione:2007nw,Nason:2004rx,Frixione:2007vw,Alioli:2010xd} generator,
which provides MEs at NLO in the strong
coupling constant \alphas, and is interfaced with \PYTHIA[8.2]~\cite{Sjostrand:2014zea}
for the parton shower and hadronisation according to the settings described
in \cref{tab:MC}. The generation of both processes uses the five-flavour scheme.
The top quark mass, \mtop, is set to \qty{172.5}{\GeV}.
The \hdamp parameter, which controls the matching in \POWHEG
and effectively regulates the high-\pt radiation against which the
\ttbar or \tW system recoils, is set to \qty{1.5}{\mtop}~\cite{ATL-PHYS-PUB-2016-020}.
The functional form of the renormalisation and factorisation scales is
set to the default dynamic scale \(\sqrt{\mtop^{2} + \pTX[2]}\).
The definition of the \tW sample is chosen to enable combination with
the corresponding \ttbar calculation. This is non-trivial at NLO, where care
must be taken to avoid double-counting \tWb events with \( \mbW \sim \mtop \).
Interference effects between the \tW and \ttbar processes are included in the predictions.
The default scheme for combining the \ttbar and \tW processes at NLO adopted
here is diagram removal (DR)~\cite{Frixione:2008yi}.
Other choices exist (see Ref.~\cite{Demartin:2016axk} for a review), for example
the case where doubly resonant contributions are canceled out by gauge-invariant
subtraction terms (diagram subtraction, DS)~\cite{Frixione:2008yi,ATL-PHYS-PUB-2016-020}.
The \ttbar sample is normalised to a cross-section of \(831.8^{+29.3}_{-36.5}\)~pb.
The cross-section prediction is calculated at next-to-next-to-leading-order (NNLO) in QCD
including the resummation of next-to-next-to-leading logarithmic (NNLL) soft-gluon terms calculated using
\TOPpp[2.0]~\cite{Beneke:2011mq,Cacciari:2011hy,Baernreuther:2012ws,Czakon:2012zr,Czakon:2012pz,Czakon:2013goa,Czakon:2011xx}.
The \tW sample is normalised to a cross-section of \(79.3^{+2.9}_{-2.8}\)~pb.
This cross-section prediction is calculated at NLO in QCD with
NNLL soft-gluon corrections~\cite{Kidonakis:2021vob,Kidonakis:2010ux,Kidonakis:2013zqa}.

Additional uncertainties associated with the modelling of \ttbar and
\tWbparenthesis processes have been considered using alternative simulation samples.

The impact of using a different parton shower and hadronisation model
is evaluated with a simulation sample produced with an alternative shower
generator (\HERWIG~\cite{Bahr:2008pv,Bellm:2015jjp,Bellm:2017jjp,Bellm:2019zci}), see \cref{tab:MC}.
Furthermore, effects associated with the parameters that regulate the matching
between ME and PS generators are estimated by varying the \pthard parameter
according to the studies in Ref.~\cite{Hoche:2021mkv}. The top-quark recoil
model is also varied in such a way that the parton shower of the nominal \ttbar sample
employs the recoiling against the top-quark instead of \bquarks.
Uncertainties from missing higher orders are
evaluated~\cite{Bothmann:2016nao} using two-point variations of the QCD
factorisation and renormalisation scales in the matrix elements by
factors of 0.5 and 2.

To assess the uncertainties related to the modelling of the initial-state
radiation (ISR), the sample is reweighted using the Var3c up variation
from the A14 tune, which changes the scales in the ISR shower in \PYTHIA.
The impact of final-state radiation (FSR) is evaluated by varying
the renormalisation scale for emissions from the parton shower up and down by
a factor of two.
The uncertainties in the cross-section due to the PDF and \alphas are calculated
using the \PDFforLHC[15] prescription~\cite{Butterworth:2015oua} and are added
in quadrature to the effect of the scale uncertainty.
Alternative simulation samples with varied \hdamp parameter (\qty{3}{\mtop})
value and varied top-quark mass values (equivalent to a variation of \( \pm 0.5 \mtop \))
were also used to estimate the impact of these parameters on the modelling of
the \ttbar and \tW processes.
An alternative \tW sample generated with the DS scheme is used to estimate the
impact of the double-counting removal procedure used to generate \tWbparenthesis events.
It is also used as an alternative prediction when comparing to the final result.

Finally, a sample aims to model \ttbar processes at NNLO QCD and
incorporates all-order radiative corrections through parton-shower simulations
as detailed in Ref.~\cite{Mazzitelli:2020jio}.
This sample is generated using a procedure that consists of three steps. In the first step \ttbar pairs plus one light parton
are generated using the \POWHEGBOX[v2] program, which performs an NLO calculation and includes a second light parton.
The second step involves computing the limit in which the emitted light partons become collinear with the \ttbar\ system.
This step incorporates Sudakov form factors and relevant higher-order corrections to ensure the result remains well-behaved
and accurate at NNLO for inclusive \ttbar\ production.
The third step explicitly generates the kinematic properties of the second radiated parton using \POWHEGBOX[v2], since this was only
treated inclusively in the initial stage. This ensures that the NLO accuracy of the \ttbar+jet cross-section is maintained.
Following this step, additional radiation is simulated using the PS method of Pythia.
This sample is referred to as the ``MiNNLOPS'' sample in the paper and is used only in the final comparison with the unfolded results.
The sample is normalised to the same cross-section at NNLO+NNLL in pQCD calculated with the Top++ 2.0 program, as previously mentioned.

\paragraph{Modelling of the background processes}

All relevant SM background processes, with the exception of the QCD multi-jet
background, are estimated with the aid of MC simulation samples.
The most important processes are the production of vector bosons in association
with jets (\Vjets), diboson production (\(VV\)), \ttH, \ttV and single top
(\(s\)- and \(t\)-channel) production.
The production of single-boson and diboson processes is simulated using the
\SHERPA[2.2.1]~\cite{Bothmann:2019yzt}.
The \Vjets (\(VV\)) processes are simulated at NLO MEs for up to two (one)
partons, and LO matrix elements for up to five (three) partons
calculated with the \AtlasMC{Comix}~\cite{Gleisberg:2008fv} and \OPENLOOPS~\cite{Buccioni:2019sur,Cascioli:2011va,Denner:2016kdg}
libraries.%
The matrix element calculations are matched and merged with the \SHERPA parton
shower based on Catani--Seymour dipole factorisation~\cite{Gleisberg:2008fv,Schumann:2007mg}  %
using the \MEPSatNLO prescription~\cite{Hoeche:2011fd,Hoeche:2012yf,Catani:2001cc,Hoeche:2009rj}.
The diboson simulation samples include off-shell effects and Higgs boson
contributions, where appropriate.
The \Vjets simulation samples include electroweak virtual corrections~\cite{Kallweit:2014xda,Kallweit:2015dum}.
An additional data-driven normalisation correction is applied to the \Wjets\ processes as described in \cref{sec:unfolding}.
Single-top-quark (\(s\)- and \(t\)-channel) processes are simulated using
the \POWHEGBOX[v2] generator interfaced with \PYTHIA.
The \ttH and \ttV processes are simulated using the \POWHEGBOX[v2] and
\MGNLO~\cite{Alwall:2014hca} generators, respectively.

\begin{table}[htbp]
\caption{Summary of the generator settings used for the Monte Carlo simulation
samples. The columns indicate the physics process,
the matrix element (ME) generator, the parton shower (PS) generator,
the cross-section accuracy (CS), the parton distribution function (PDF) set,
the tune used for the underlying event, and the program used for
the \(b\)/\(c\) hadron decays. \MGNLO is here referred to as MG.\@
See text for further details.}%
\label{tab:MC}
\begin{center}
\resizebox{\textwidth}{!}{\small
\begin{tabular}{m{21mm}|m{23mm}|m{19mm}|m{38mm}|m{23mm}|m{36mm}|m{19mm}}  %
\toprule
Process                                                     & ME                                   & PS             & CS                                                  & PDF             & Tune                     & \(b\)/\(c\)  decays \\

\midrule
\ttbar                                                      & \POWHEGBOX[v2]                       & \PYTHIA[8.230] & NNLO+NNLL (\TOPpp[2.0])                             & \NNPDF[3.0nlo]  & A14 tune + \NNPDF[2.3lo] & \EVTGEN[1.6.0]      \\
\tW                                                        & \POWHEGBOX[v2]                       & \PYTHIA[8.230] & NLO+NNLL                                            & \NNPDF[3.0nlo]  & A14 tune + \NNPDF[2.3lo] & \EVTGEN[1.6.0]      \\
\midrule
\ttbar (alt.)                                               & \POWHEGBOX[v2]                       & \HERWIG[7.1.3] & NNLO+NNLL                                           & \NNPDF[3.0nlo]  & H7 def. + \MMHT[lo]      & \EVTGEN[1.6.0]      \\
\tW (alt.)                                                 & \POWHEGBOX[v2]                       & \HERWIG[7.2.3] & NLO+NNLL                                            & \NNPDF[3.0nlo]  & H7 def. + \MMHT[lo]         & \EVTGEN[1.6.0]      \\
\ttbar MiNNLOPS                                             & MiNNLOPS + \POWHEGBOX[v2]            & \PYTHIA[8.230] & NNLO+NNLL                                           & \NNPDF[3.0nlo]  & A14 tune + \NNPDF[2.3lo] & \EVTGEN[1.6.0]      \\
\midrule
\(V+\)jets                                                  & \SHERPA[2.2.11]                      & \SHERPA        & NLO                                                 & \NNPDF[3.0nnlo] & \SHERPA tune                                   \\

\(VV\)                                                      & \SHERPA[2.2.11]                      & \SHERPA        & NLO                                                 & \NNPDF[3.0nnlo] & \SHERPA tune                                   \\

\midrule
\(s\) and \(t\)- channel single-\(t\)                                                & \POWHEGBOX[v2]                       & \PYTHIA[8.230] & \makecell[l]{NLO\\(\HATHOR[2.1]~\cite{Aliev:2010zk,Kant:2014oha})} & \NNPDF[3.0nlo]  & A14 tune + \NNPDF[2.3lo]                 & \EVTGEN[1.6.0]      \\
\ttH                                                        & \POWHEGBOX[v2]                       & \PYTHIA[8.230] & NLO QCD \& EW~\cite{deFlorian:2016spz}              & \NNPDF[3.0nlo]  & A14 tune + \NNPDF[2.3lo]                 & \EVTGEN[1.6.0]      \\
\ttV                                                        & MG 2.3.3 & \PYTHIA[8.210] & NLO QCD \& EW                                       & \NNPDF[3.0nlo]  & A14 tune + \NNPDF[2.3lo]                 & \EVTGEN[1.2.0]      \\

\bottomrule
\end{tabular}
}
\end{center}
\end{table}

The background arising from QCD multi-jets events containing misidentified or non-prompt leptons
is estimated using a data-driven \emph{fake-factor} method as described in \cref{sec:unfolding}.
Simulation samples as described in this section are used to subtract the prompt-lepton contamination
of the non-prompt estimation region and to estimate the systematic uncertainties,
including the ones associated with the extrapolation to the phase space of this measurement.


\section{Event selection}%
\label{sec:selection}
The results presented in this paper are based on data collected with the ATLAS
experiment between 2015 and 2018 during Run 2 of the LHC.\@
The data sample corresponds to an integrated luminosity of \qty{140}{\ifb}~\cite{DAPR-2021-01,LUCID2}
of \pp collisions at a center-of-mass energy of \qty{13}{\TeV}. The measurement of the integrated
luminosity is obtained using the LUCID-2 detector for the primary measurement and complemented
by measurements in the inner detector and calorimeters.
Every detector subsystem must have been in operation during data collection.
Moreover, stringent data quality criteria are applied to ensure the reliability
of the data~\cite{DAPR-2018-01}.
Events are required to be triggered by a collection of single-electron or
single-muon triggers~\cite{TRIG-2016-01,TRIG-2018-05,TRIG-2018-01}
operating at different thresholds of transverse momentum and identification
criteria through the data-taking periods.
Events are required to contain at least one collision vertex with two or more associated
tracks with \(\pT > \qty{500}{\MeV} \), and the vertex with the highest \( \pt^2 \) sum of the associated
tracks is taken as the primary vertex (PV)~\cite{ATL-PHYS-PUB-2015-026} of the event.

\begin{table}[htbp]
\caption{Summary of the selections used in the three measurement regions:
inclusive region (\IR), \WRlong region (\WR) and search-like region (\SLR).
Details on the definition of the hadronic \Wboson candidate, \Whad, and the invariant mass quantities \mt and \mblmin
are given in the text. }%
\label{tab:selection}
\centering
\begin{tabular}{c|cc}  %
\toprule
\multicolumn{3}{c}{Requirements}                                                                                                                       \\ \midrule
\multirow{4}{*}[-0.6em]{\rotatebox[origin=B]{90}{\emph{preselection}}}  & \( N_{\bjets} \)                                                        & \( \geq 2 \)                   \\[0.9ex]
& \( N_\ell (\pt > \qty{10}{GeV})\)                                                          & \(=1 \)            \\ [0.5ex]
& \( \met + \mt \)                                                      & \( \geq \qty{60}{\GeV} \)      \\ [0.5ex]
& \( \pt(\ell) \)                                                       & \( \geq \qty{30}{\GeV} \)      \\[0.9ex]
\midrule[0.08em]
\multirow{1}{*}{\rotatebox[origin=B]{90}{\emph{\IR}}}  & \( N_\mathrm{jets} \)                                                 & \( \geq 4 \)                   \\[0.9ex]
\midrule
\multirow{3}{*}[-0.4em]{\rotatebox[origin=B]{90}{\emph{\WR}}}  & \( N_\mathrm{jets} \)                                                 & \( \geq 3 \)                   \\  [0.5ex]
& \( \pt(\Whad) \)                                                      & \( \geq \qty{80}{\GeV} \)      \\[0.5ex]
& \multicolumn{2}{c}{\( 60 < m(\Whad) < \qty{100}{\GeV} \)}                                              \\ [0.9ex]
\midrule
\multirow{6}{*}[-0.5em]{\rotatebox[origin=B]{90}{\emph{\SLR}}} & \( N_\mathrm{jets} \)                                                 & \( \geq 4 \)                   \\  [0.5ex]
& \mblmin                                                               & \( \geq \qty{150}{\GeV} \)     \\ [0.5ex]
& \( \pt(j_1), \pt(j_2) \)                                              & \( \geq \qty{50}{\GeV} \)      \\  [0.5ex]
& \( \pt(b_1), \pt(b_2) \)                                              & \( \geq 160, \qty{50}{\GeV} \) \\ [0.5ex]
& \multicolumn{2}{c}{\( 1.4 < \Delta R (b_1, b_2) \)}                                              \\ [0.5ex]
& \multicolumn{2}{c}{\( \met \geq -0.625\pt(\ell) + \qty{270}{\GeV} \)}                                  \\

\bottomrule
\end{tabular}
\end{table}

The target topology of this measurement are events containing two \Wboson bosons,
one decaying leptonically and one hadronically, and two \bjets, arising from \ttbar or
\tWb production.
To select this topology, the events are required to contain exactly one light
charged lepton (\( \ell \), either an electron or a muon) with
\( \pt > \qty{30}{\GeV} \), which ensures the full efficiency of the triggers.
Any event containing an additional lepton with \( \pt > \qty{10}{\GeV} \)
is rejected. Events are also required to have at least four jets, of which
at least two are \bjets. An additional requirement on the sum of the \met and the \mt,
the transverse mass between the \met and the lepton%
\footnote{This transverse mass is defined as \(\mt = \sqrt{{(\met + E_{\mathrm T,\ell})}^2 - {(\ptmiss+\vec{p}_{\mathrm T,\ell})}^2}\).},
of greater than \qty{60}{\GeV} is applied to suppress contaminations from the misidentified
and non-prompt lepton background.

As introduced before, three kinematic regions are used in this measurement:
the inclusive region \IR, the \WRlong region \WR and the search-like region \SLR.
The selection criteria described so far define the \IR.

The \WR is defined by explicitly requiring a hadronically decaying \Wboson boson
in the final state. A \Wboson boson reconstruction algorithm is optimised
to maximize the overall \Wboson identification purity with a procedure
that is equally easy to apply on detector-level objects and particle-level
objects. This procedure identifies the \Wboson candidate as either the leading
jet, the subleading jet, or their vectorial sum, with \bjets excluded from this procedure.
This approach aims to increase the purity of \Wboson identification in the high-\pt region, where
the boosted \Wboson boson decay products are collimated and the \Wboson boson
is reconstructed as a single jet. The candidate with a mass closest to
the \Wboson mass is selected. The mass of the candidate is required to be
between \( 60 < \mW < \qty{100}{\GeV} \) and its \pt is required to be at least
\qty{80}{\GeV}. These requirements ensure a signal purity of at least \qty{75}{\%}, which becomes larger than \qty{90}{\%}
for events with a \Wboson boson transverse momentum larger than \qty{180}{\GeV}.

The \SLR is optimised for an extreme phase space and it takes inspiration from
BSM physics searches that probed phase spaces where the interference
modelling between \ttbar and \tWb processes resulted in large systematic
uncertainties.
The unfolded results in this peculiar phase space are very sensitive to the modelling of interference effects.
To optimise the selection, the \tWb\ purity and the yield difference between the \tW\ DR and DS simulation samples are used as key metrics. The optimisation aims to minimise this yield difference while maintaining high \tWb\ purity.
The minimum invariant mass obtained by pairing the lepton
with each of the \bjets in the event,  \mblmin, is an important observable to enhance
topologies where at least one top quark is produced off-shell, where interference effects between
\ttbar and \tWb can be probed best.
A requirement of \( \mblmin\geq\qty{150}{\GeV} \)
exploits the characteristic kinematic endpoint of \ttbar events
and suppresses events with an on-shell semileptonically decaying top quark.
Further requirements are embedded into the selection to suppress on-shell
\ttbar events: the transverse momentum of the leading (subleading) \bjet is
required to be greater than \(160\;(50)\;\GeV \).
The leading and subleading \bjets are required to have a relatively large spatial separation
(\( \Delta R (b_1, b_2) > 1.4 \)).
Finally, \met and \(\pt(\ell)\) are required to satisfy the following condition in the \SLR:
\begin{equation*}
\met \geq -0.625\pt(\ell) + \qty{270}{\GeV}.
\end{equation*}
This requirement is chosen to maximise the overall \sampleB acceptance, while enhancing the interference effects in this phase space
using easily-interpretable observables.
The selection criteria for the three measurement regions are summarised in \cref{tab:selection}.

In the \WR selection, observables are chosen study the kinematic
properties of the hadronically decaying \Wboson boson. These are:
\begin{itemize}
\item \ptW: the transverse momentum of the hadronically decaying \Wboson boson candidate.
\item \yW: the rapidity of the hadronically decaying \Wboson boson candidate.
\item \mWbbl: the invariant mass of the \( Wbb\ell \) system.
\item \mbWminimax: variation of the min-max variable from Ref.~\cite{TOPQ-2017-05}, which takes into account the different final state considered in this
paper. The variable is defined using the hadronic \Wboson candidate (\Whad) and the two leading \bjets as:
\[
\mbWminimax = \min\left( \max(m_{W_{\textrm{had.}}b_1}, m_{\ell b_2}), \max(m_{\ell b_1}, m_{W_{\textrm{had.}}b_2}) \right).
\]
This observable considers all possible pairings of the two \bjets with either \Whad or the lepton. It exhibits a kinematic endpoint at
\[
\mbWminimax < \sqrt{\mtop^2 - m_W^2}
\]
for doubly resonant top-quark topologies, where \( \mtop \) and \( m_W \) denote the top-quark and \Wboson masses, respectively. This makes it sensitive to regions of phase space enriched in singly resonant processes, where this endpoint is exceeded.

\item \mblselected: a variation of the \mblmin observable which uses the \bjets assignment from the \mbWminimax observable to calculate the invariant mass between the lepton and the selected \bjet.
\end{itemize}

A variation of the \mblmin observable, \mtblmin, is studied in the \IR\ and \SLR selections. It considers also the
\met information to constrain the mass of the leptonically decaying top.
This is defined as the minimum transverse invariant mass between the \met, the lepton and the \bjet.

The total yields observed in data and predicted from the simulation in each
region are shown in \cref{tab:yields}. The breakdown by SM process is also shown.

\begin{table}[htbp]
\caption{Summary of the contributions of all SM processes and observed data
in the three measurement regions.
All statistical and systematic uncertainties are included, with the exception of the signal modelling,
in the total uncertainty in the yields and are summed in quadrature.
Both diagram removal (DR) and diagram subtraction (DS) yields
are shown for the \tW process for comparison.}%
\label{tab:yields}
\centering
\bgroup
\def\arraystretch{1.3}%
\begin{tabular}{
l @{\quad\quad}
r
@{\(\,\,\)}  %
r
@{\quad\quad}
r
@{\(\,\,\)}  %
r
@{\quad\quad}
r
@{\(\,\,\)}  %
r
}
\toprule
Physics process           &
\multicolumn{2}{c}{\IR}   &
\multicolumn{2}{c}{\WR}   &
\multicolumn{2}{c}{\SLR}                                                                                                                                   \\
\midrule
\tW DR                    & \num{139000}  & \numpmerr{+8000}{-8000}     & \num{34000}  & \numpmerr{+1900}{-1900}     & \num{1680} & \numpmerr{+100}{-80} \\
\tW DS (alt.)             & \num{120000}  & \numpmerr{+8000}{-8000}     & \num{28400}  & \numpmerr{+1900}{-1900}     & \num{234}  & \numpmerr{+15}{-15}      \\
\ttbar                    & \num{4000000} & \numpmerr{+200000}{-200000} & \num{850000} & \numpmerr{+40000}{-40000}   & \num{1170} & \numpmerr{+80}{-80}  \\
\Wjets                    & \num{120000}  & \numpmerr{+22000}{-18000}   & \num{10300}  & \numpmerr{+2000}{-1600}     & \num{344}  & \numpmerr{+90}{-56}    \\
\OtherT                   & \num{91000}   & \numpmerr{+24000}{-24000}   & \num{9300}   & \numpmerr{+2400}{-2400}     & \num{104}  & \numpmerr{+27}{-27}    \\
\OtherV                   & \num{25000}   & \numpmerr{+4000}{-16000}    & \num{2210}   & \numpmerr{+230}{-1600}      & \num{61}   & \numpmerr{+4}{-61}     \\
\fakes                    & \num{20000}   & \numpmerr{+8000}{-14000}    & \num{1900}   & \numpmerr{+1300}{-700}      & \num{45}   & \numpmerr{+33}{-34}    \\
\midrule
Total MC w/ DR            & \num{4400000} & \numpmerr{+230000}{-220000} & \num{910000} & \numpmerr{+50000}{-50000} & \num{3400} & \numpmerr{+210}{-190} \\
Data                      & \num{4303650} &      & \num{849115} &       & \num{2403} &   \\
\bottomrule
\end{tabular}
\egroup
\end{table}

In addition to the detector-level objects, corresponding particle-level objects are defined using generator-level
information from stable particles, identified as those with a proper lifetime \( c\tau > 10 \) nm.
These objects are identified without simulating their interactions with the detector and are used to define the
fiducial phase-space regions and observables of interest in the measurement.

Particle-level electrons and muons are required not to originate from hadrons, either directly or through intermediate
\( \tau \)-lepton decays. A lepton is retained if it arises from a \( \tau \)-lepton that itself originates directly from a
\Wboson boson. The lepton four-momenta are corrected for final-state radiation from photons
not originating from hadron decays and located within \(\Delta R = 0.1\) of the lepton direction.
The same transverse momentum and pseudorapidity requirements as for detector-level leptons are applied.

Particle-level jets are reconstructed by clustering all stable particles, excluding those identified as leptons
or as neutrinos from electroweak boson decays, using the anti-\(k_t\) algorithm with a radius parameter of \(R = 0.4\).
The identification of \(b\)-jets is performed via the ghost-association technique: all \(b\)-hadrons are included in the
jet clustering with their momentum set to a negligible value, and jets containing a \(b\)-hadron with \( \pT > \qty{5}{\GeV} \) are classified as
\(b\)-tagged.

The particle-level missing transverse momentum is computed as the negative vector sum of the transverse momenta of all neutrinos,
excluding those originating from hadron decays, either directly or via \( \tau \)-leptons.

An overlap removal is applied among particle-level objects where electrons and muons located within
\(\Delta R = 0.4\) of any jet are removed.

The particle-level selections for the three measurement regions used for the differential measurements are constructed to match the
detector-level regions by applying the same selection criteria.

\FloatBarrier
\section{Unfolding procedure}%
\label{sec:unfolding}
The differential cross-section measurement is performed in each of the three
measurement regions (\IR, \WR and \SLR), using the fiducial volumes definitions
described in the previous section. An unfolding
procedure is applied to extract the results at particle-level, following the methodology outlined in Refs.~\cite{Schmitt:2016orm,Blobel:2011fih}.
The unfolding is performed using the \texttt{TUnfold} package~\cite{Schmitt:2012kp}.

The differential cross-section for a variable \(X\) at particle-level is defined
as follows:
\begin{equation*}%
\dfrac{d \sigma_\mathrm{fid}}{d X_{i}}
= \dfrac{1}{\Lint \cdot \Delta X_{i}} \cdot \frac{1}{\epsilon_{i}} \cdot \sum_{j} M^{-1}_{i,j} f_\mathrm{acc}^{j} (N_{d,j} - N_{b,j}).
\end{equation*}

The index \(i\) runs over all the bins of the particle-level definition
of the variable \(X\) while the index \(j\) runs over all the bins of the
reconstructed-level definition of the variable \(X\).\@ \Lint is the integrated
luminosity. The differential cross-section is obtained from the data
distribution \(N_{d,j}\) after subtracting the background contribution
\(N_{b,j}\). The unfolding efficiency \(\epsilon_{i}\) and the unfolding
acceptance \(f_\mathrm{acc}^{j}\) are obtained from the simulation of
the target processes of the analysis (\ttbar + \tWb). The efficiency is defined
as the fraction of particle-level events that satisfy the detector-level
selections and corrects for detector inefficiencies in the fiducial phase space.
The acceptance is defined as the fraction of detector-level signal
simulated events
that satisfy the particle-level selection in each bin and corrects for events
that are generated outside the fiducial phase-space but satisfy
the detector-level selection.
The migration matrix \(M_{i,j}\), or normalised response matrix, is derived
from the simulation and it describes the fraction of events that migrate
from particle-level bin \(i\) to reconstructed bin \(j\).
The bin width is indicated by \(\Delta X_{i}\).
The bin boundaries are chosen such that, in most cases, the diagonal of the migration matrix
contains at least \qty{60}{\%} of particle-level events of a row.
The number of reconstructed distribution bins is chosen to be twice the number
of particle-level bins, according to the \texttt{TUnfold} default configuration~\cite{Schmitt:2012kp}.
The migration matrix inversion, \(M^{-1}\), is estimated
using a Tikhonov-Phillips regularisation scheme~\cite{Tikhonov:1990,Phillips:1962},
which reduces the statistical uncertainty associated with the inversion
at the price of a possible deviation from the true value (bias). The bias is quantified with a closure test and found
to be negligible.

Typical efficiency values across the three measurement regions range between 20 and \qty{30}{\%} for most of the phase space
and observables considered in the unfolding results.
The acceptance values exhibit stronger variability across regions and phase space.
In the \IR and \WR selections, typical acceptance values range between 60 and \qty{80}{\%},
while in the \SLR selections, the acceptance values are lower, with typical values around \qty{40}{\%}.
The migration matrix is almost diagonal for observables with small reconstruction uncertainties,
such as the lepton \pt, while it is more off-diagonal for observables with larger reconstruction uncertainties,
such as the (\(b\))-jet momenta or the \mblmin observable.

The unfolding is performed in each of the three measurement regions and
the results are compared with the predictions of the simulation on key
observables that allow for a thorough assessment of the data and its comparison
with the MC models across all relevant kinematic phase space.

All SM processes other than \ttbar and \tWb are subtracted from data as
the first step of the unfolding procedure. The dominant processes are \Wjets,
\Zjets and diboson (referred to as ``\OtherV''), \ttV, \ttH and
sub-dominant single-top channels (referred to as ``\OtherT'') and QCD multi-jets
processes with fake and non-prompt leptons (referred to as ``Mis ID.\ lep.'').  %
\Wjets is an important background for
the \IR and \SLR selections. Its normalisation is constrained in a dedicated
control region, \WCR, enriched in this process with the purpose of better
controlling the impact of systematic uncertainties related to this background.
Events in the \WCR are required to fulfil all  requirements of
the \IR selection, except for the jet multiplicity requirement that is reduced
to exactly three jets to ensure orthogonality. The overlap with the \WR
is removed by vetoing events that fulfil the \WR selection,
which is a small fraction of the total events in \WCR. The \Wjets contribution
is further enhanced by requiring \( 40~\GeV < \mt < 100~\GeV \) and
\( \mblmin > \qty{180}{\GeV} \). A purity of \qty{50}{\%} is achieved, with a contamination of \qty{39}{\%} from \sampleB events.
The normalisation of the \Wjets background is estimated by means of a likelihood
fit in the \WCR, where detector and modelling uncertainties of all processes are
treated as nuisance parameters in the fit. The resulting normalisation factor is
\(1.10 \pm 0.10\). The data is within the uncertainty in the expected yields after the fit.
The normalisation factor is used to correct the \Wjets process in all measurement regions
(and is applied also in \cref{tab:yields}).

Misidentified and non-prompt leptons are estimated by using a data-driven method
called \emph{fake-factor method}
that estimates the non-prompt contamination from a control region rich
in this background and with high statistical power as described
in detail in Ref.~\cite{EGAM-2019-01}. This process is found to be the smallest source of background in
every region of the analysis.
The electron and muon requirements used to select this fakes control region are described in \cref{reconstruction}. The prediction is validated in a
dedicated and statistically independent validation region selecting events failing the requirement on \( \met + \mt \) and applying looser requirements on the jet multiplicity.
All other backgrounds are estimated from simulation.

The unfolding procedure is tested using pseudo-experiments and the self-consistency of the procedure (closure)
is quantified. The unfolding is found to be stable and the bias is found
to be negligible. In addition, stress tests are performed to verify
that the unfolding is able to recover an alternative truth-level distribution.
The stress tests include the reweighing of the unfolding observable with
a linear function, the reweighting of the unfolding observable to match the data distribution,
and the reweighing of the unfolding observable with
a histogram of the ratio of observed-to-expected signal in the unfolded variable
at detector level. The unfolding is able to recover an alternative distribution
in all cases, with discrepancies always consistent with the statistical
precision of the test.

\section{Systematic uncertainties}%
\label{subsec:systematics}

The sources of systematic uncertainties that affect the measurements of
the differential cross-sections include the detector-related uncertainties
that affect the calibration and the reconstruction efficiency of detector-level
particles, the background and signal modelling,
and the size of the data and MC samples.
All systematic uncertainties related to the detector performance and calibration
and the modelling of the SM processes (as described in \cref{sec:setup})
are propagated through the unfolding procedure.

The integrated luminosity has an uncertainty of \qty{0.83}{\%}~\cite{DAPR-2021-01},
which is propagated as a systematic uncertainty in the measurement to all physics processes
other than the QCD multi-jet background.
The pile-up uncertainty is assessed through the uncertainty on the factor used for the \( \mu \) reweighting.

Charged leptons have uncertainties in the efficiency of the reconstruction, identification, isolation requirements, and trigger,
which are derived from studies using \( Z\rightarrow ee \) and \( Z\rightarrow\mu\mu \) events~\cite{EGAM-2018-01,MUON-2018-03,TRIG-2018-05,TRIG-2018-01}.
The impact of the lepton energy scale and resolution uncertainty is evaluated using resonance decays~\cite{EGAM-2021-02,MUON-2022-01}.
The JER and JES uncertainties are derived using a combination of data and MC samples, using measurements of (for example)
the jet \pT\ balance in dijet, \(Z\)+jet and \( \gamma \)+jet events~\cite{JETM-2018-05}. Further jet-related uncertainties originate
from efficiency corrections for tagging pile-up jets~\cite{PERF-2014-03} and the identification of \bjets~\cite{FTAG-2018-01,FTAG-2020-08,FTAG-2019-02}.
Uncertainties in the \met are estimated by propagating the uncertainties in the energy and momentum scale of each of the objects entering the calculation of \met,
with additional terms taking into account the uncertainty in the resolution and scale of the soft term~\cite{JETM-2020-03}.

Modelling uncertainties are considered for both signal and background processes.
Uncertainties associated with the modelling of \ttbar and
\tWbparenthesis processes have been considered using alternative simulation samples as described in \cref{subsec:simulation}.
The most important uncertainty is found to be the one related to different schemes used as double-counting removal procedure
in the generation of \tWbparenthesis events (DS versus DR removal schemes), followed by those related to
the use of a different parton shower and hadronisation model and to variations of ISR and FSR\@.
Additional uncertainties include factorisation and renormalisation scale variations,
effects associated with the parameters that regulate the matching
between ME and PS generators (\pthard parameter), top-quark recoil model, and PDF choice.
Since the \ttbar and \tWbparenthesis cross-sections are known at different perturbative orders in QCD and the two processes
are treated here coherently, the modelling uncertainties in the two processes are treated consistently at NLO in QCD\@.
The impact of the assumed value of the top quark mass on the results is evaluated using additional simulation samples with
varied top-quark mass values of \qty{172}{\GeV} and \qty{173}{\GeV}. The resulting uncertainty is usually found to be negligible,
but can grow up to \qty{5}{\%} in the most extreme regions of phase space.
Furthermore, uncertainties related to the flavour composition of extra jets radiated in association
with top pairs are estimated and found to be negligible.

All signal modelling uncertainties are treated as correlated between the two processes.
A dedicated uncertainty is included to account for the approximations used to model the shape of the top-quark mass with a proper width and high off-shell behaviours.
This uncertainty is estimated by comparing two existing implementations from \POWHEG and \MGNLO~\cite{Artoisenet:2012st}.
Uncertainties related to the modelling of additional \bbbar or \ccbar pairs in the parton shower are estimated by
varying their respective contributions in the nominal \ttbar sample and found to be negligible.
The cross-section uncertainties are treated as anti-correlated for the \ttbar
and \tW samples, which is the most conservative approach for a result that is
sensitive to the relative normalisation of the two processes.
The results are also found to be insensitive to whether the \ttbar samples are normalised using NLO+NNLL or NNLO+NNLL cross-sections.

The full uncertainty in the \Wjets normalisation, as obtained from the profile likelihood fit, is propagated
to the unfolding. In addition, the shape component of all systematic
uncertainties associated with the \Wjets background are propagated individually to the unfolding using the corresponding variations.

For all background samples, cross-section uncertainties are estimated by taking into account scales and PDF variations.
The uncertainties due to missing higher-order corrections are evaluated by varying the renormalisation and factorisation scales by factors
of 2.0 and 0.5.
These uncertainties are estimated by keeping one scale at its central value while the other is varied up and down, and vice versa.
The PDF uncertainties are estimated by using the \PDFforLHC[15] prescription. Electroweak correction uncertainties are
estimated for single and multi boson processes~\cite{Kallweit:2014xda,Kallweit:2015dum}.

The uncertainties related to the modelling of the QCD multi-jet background are estimated
by varying the control region selection criteria used in the fake-factor method and by varying the
Monte Carlo predictions by \qty{10}{\%} around their nominal values. These procedures result in a conservative
uncertainty of \qty{50}{\%} to \qty{70}{\%} in the total yield of this background, which does not affect the final results
because of the small contribution of this background in all regions.

Detector and modelling uncertainties in backgrounds are propagated from detector to particle
level by calculating varied unfolding corrections
(\(M_{i,j}\), \(\epsilon_{i}\) and \(f_\mathrm{acc}^{j}\)). The data is then unfolded with these varied corrections and
the difference to the nominal unfolded data defines the uncertainty.
Modelling uncertainties in signal are treated differently. Pseudo-data built
using alternative predictions are unfolded using the nominal setup.
These results are compared with the alternative simulation sample at particle-level
and the difference defines the uncertainty.

Overall, small uncertainties are observed for the variables that are
reconstructed only from leptons or dominated by \ttbar only, thanks to the large number of data events
and small migrations between the bins. The more statistically limited \SLR results in unfolded observables
with relatively large statistical uncertainties.

One of the dominant uncertainties is the uncertainty in the modelling of the
interference between \ttbar and \tWb arising from using
the alternative \tWb DS MC sample in the unfolding procedure.
This comparison estimates whether the unfolding is able to recover
an alternative truth-level distribution and it is particularly important when
no simulation is available that models the data well, as in this case.
This uncertainty particularly strongly affects the SLR selection, where the large
difference both in acceptance and normalisation of the two simulation approaches
changes the relative fractions of \ttbar and \tWb processes in the unfolding
and affects the unfolded result in a significant way. This uncertainty
is presented separately in all unfolded cross-sections in \cref{sec:results} as it is the one with the largest
potential for reduction in the future.

The uncertainties of the unfolded
distributions range approximately from \qty{5}{\%} to \qty{20}{\%} in the higher statistics regions
and from \qty{20}{\%} to \qty{80}{\%} in extreme phase spaces. The more extreme phase spaces,
especially in the \SLR selection, are strongly affected by the DR-DS uncertainty.
The results for these selections achieve a precision at the level of \qty{20}{\%} or better overall
when the DS-DR uncertainty is not taken into account.

\begin{figure}
\centering
\subfloat[]{
\includegraphics[width=0.37\textwidth]{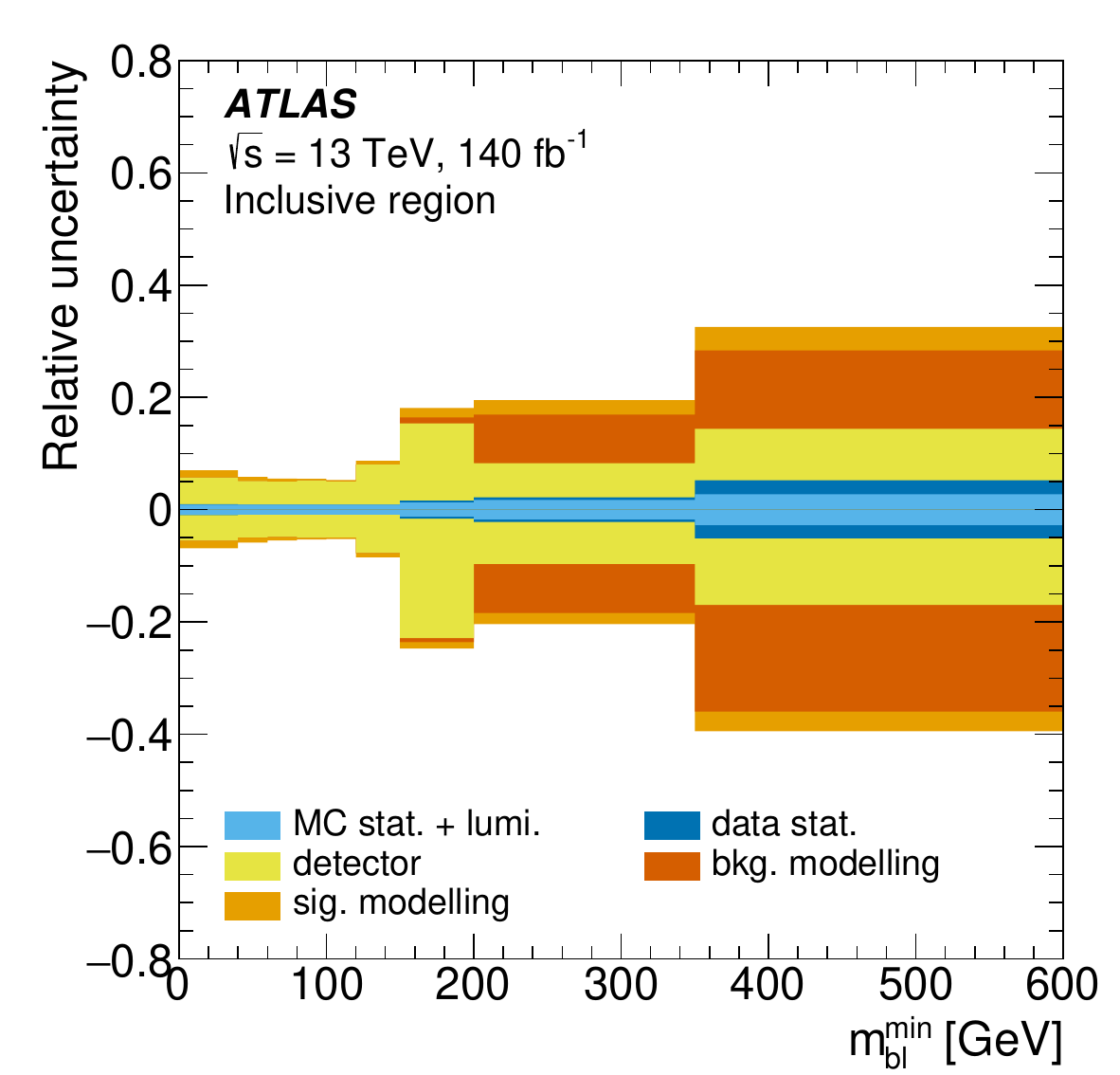}%
\label{fig:IR:mblmin:uncertainties}
}
\subfloat[]{
\includegraphics[width=0.37\textwidth]{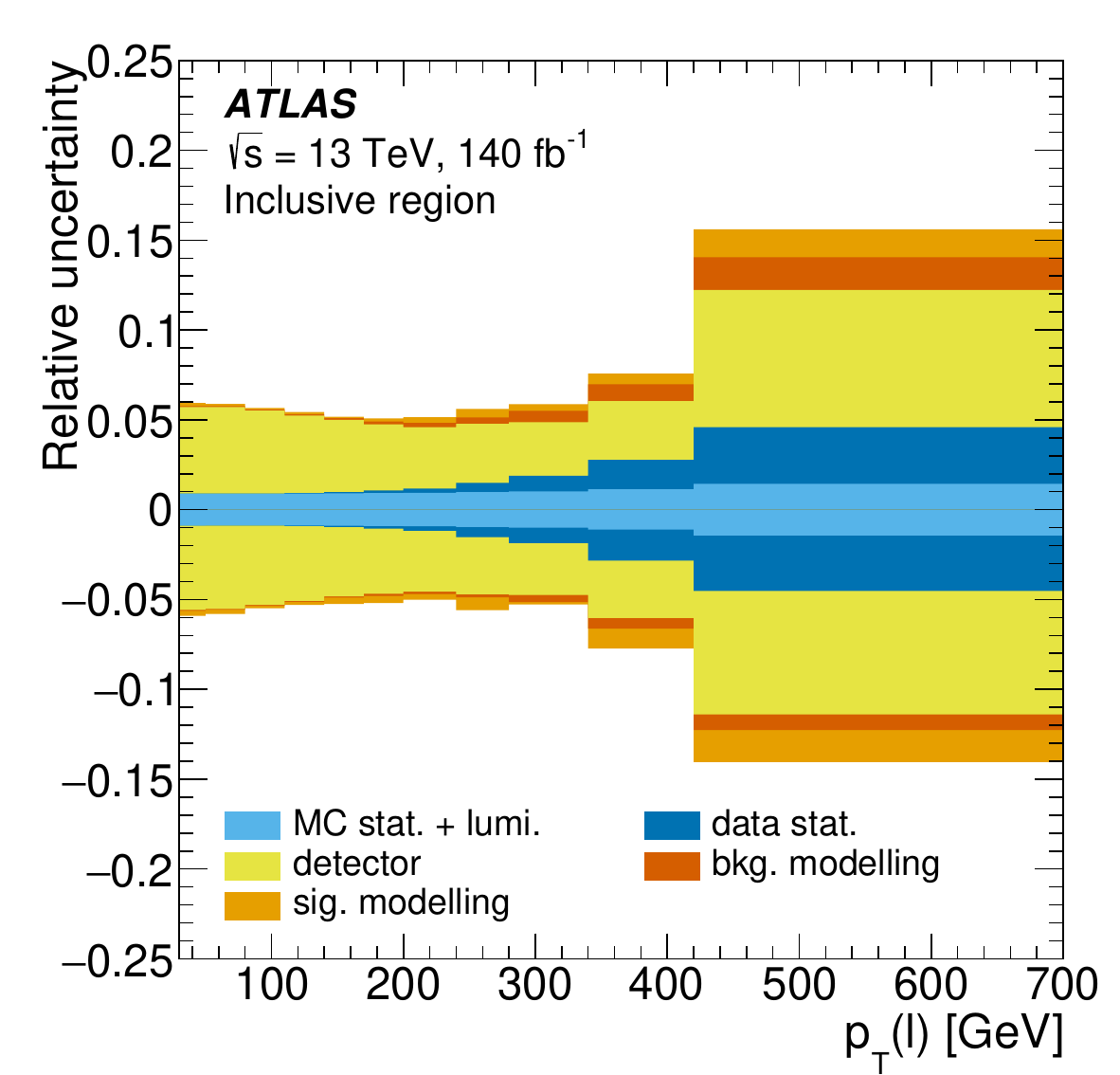}%
\label{fig:IR:ptl1:uncertainties}
}\\
\subfloat[]{
\includegraphics[width=0.37\textwidth]{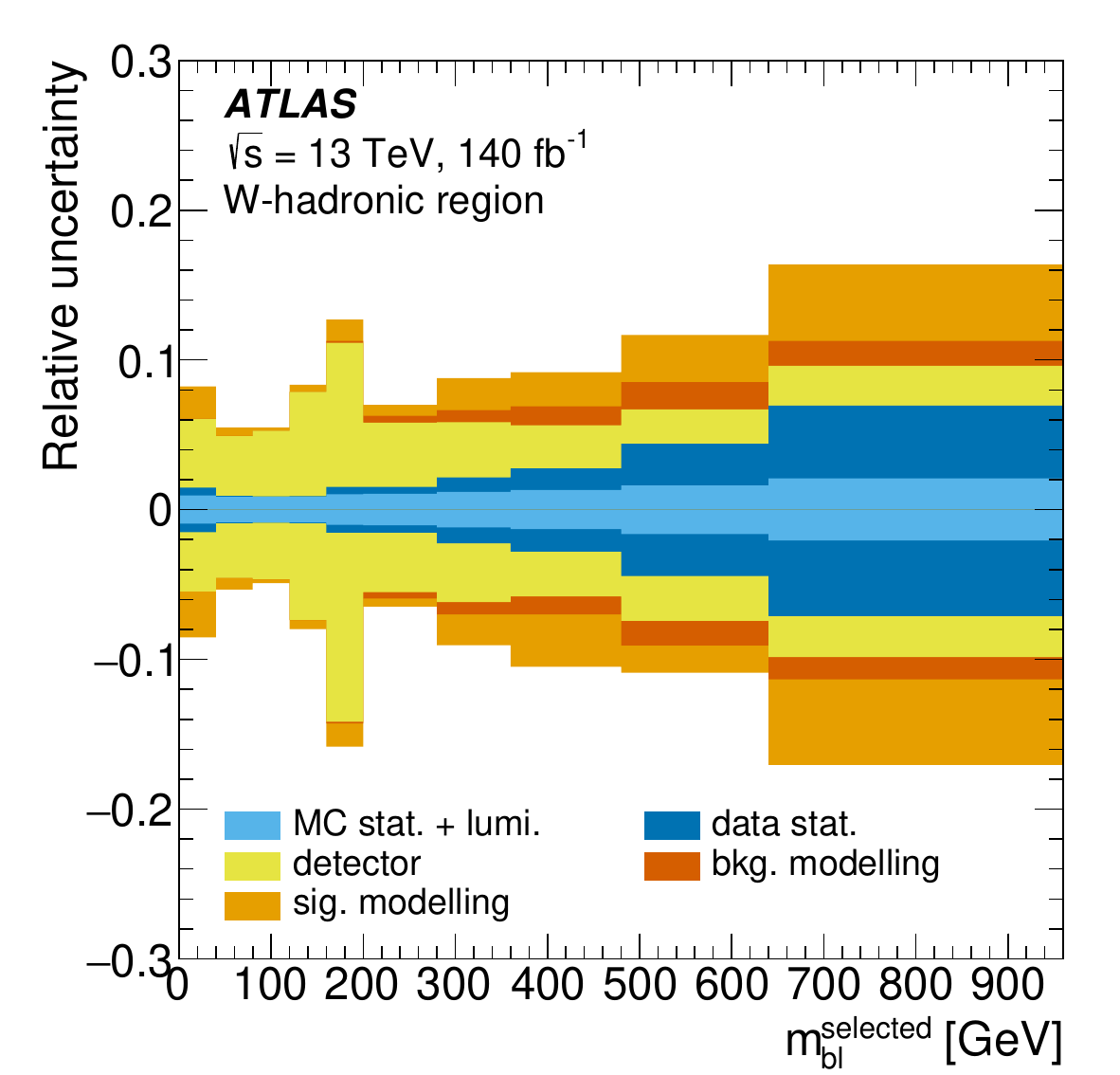}%
\label{fig:WR:mblselected:uncertainties}
}
\subfloat[]{
\includegraphics[width=0.37\textwidth]{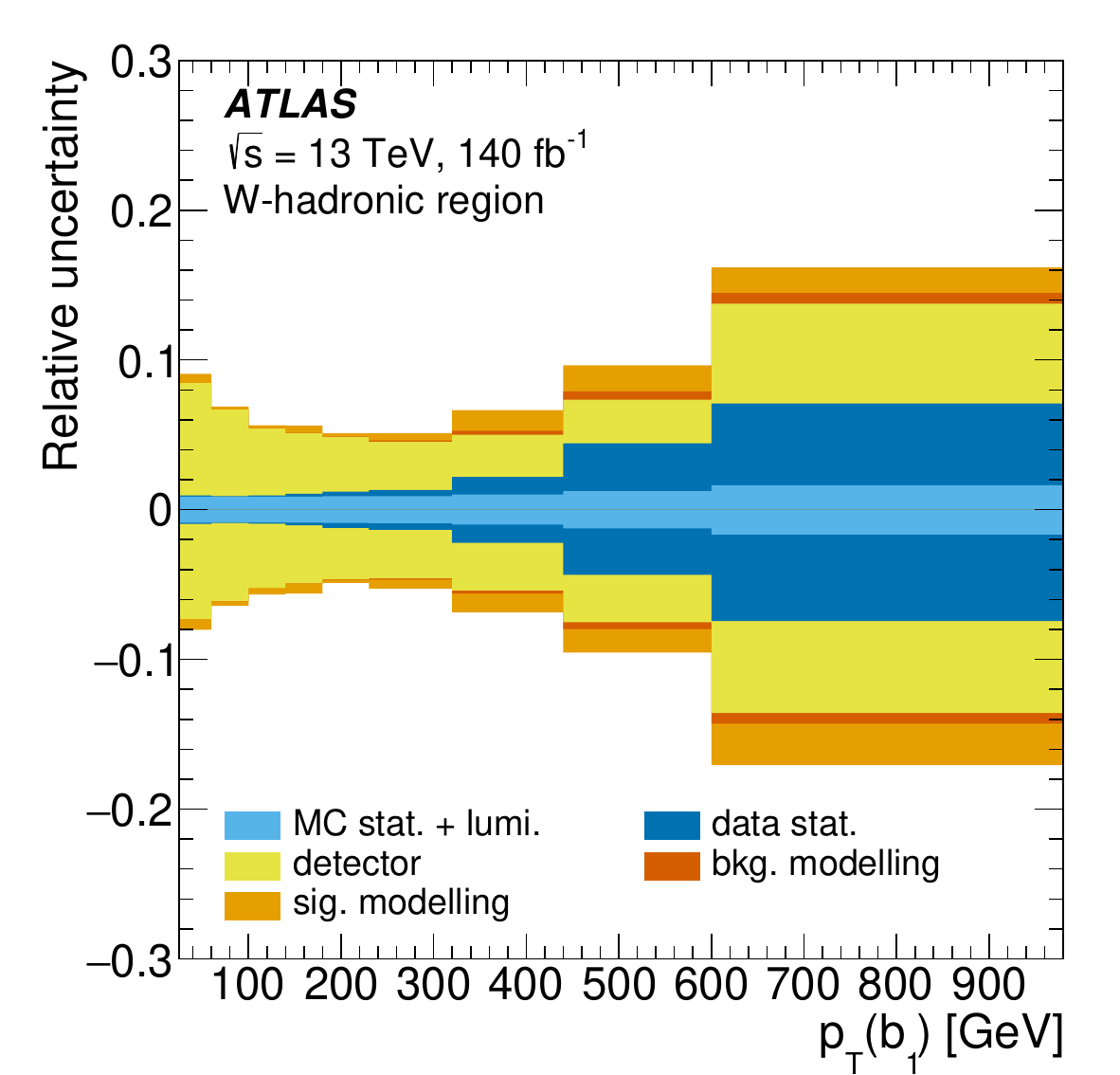}%
\label{fig:WR:ptb1:uncertainties}
}\\
\subfloat[]{
\includegraphics[width=0.37\textwidth]{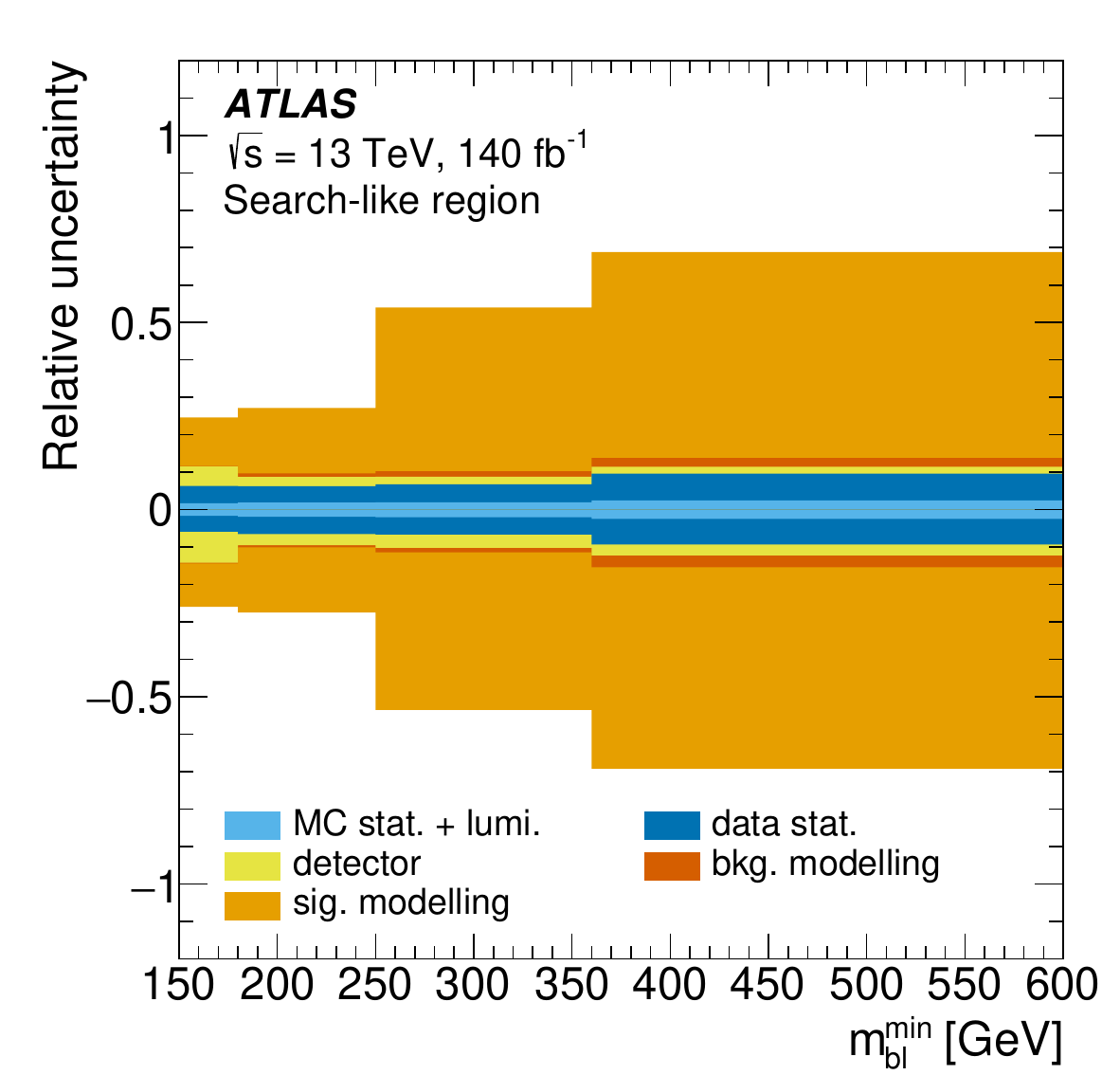}%
\label{fig:SLR:mblmin:uncertainties}
}
\subfloat[]{
\includegraphics[width=0.37\textwidth]{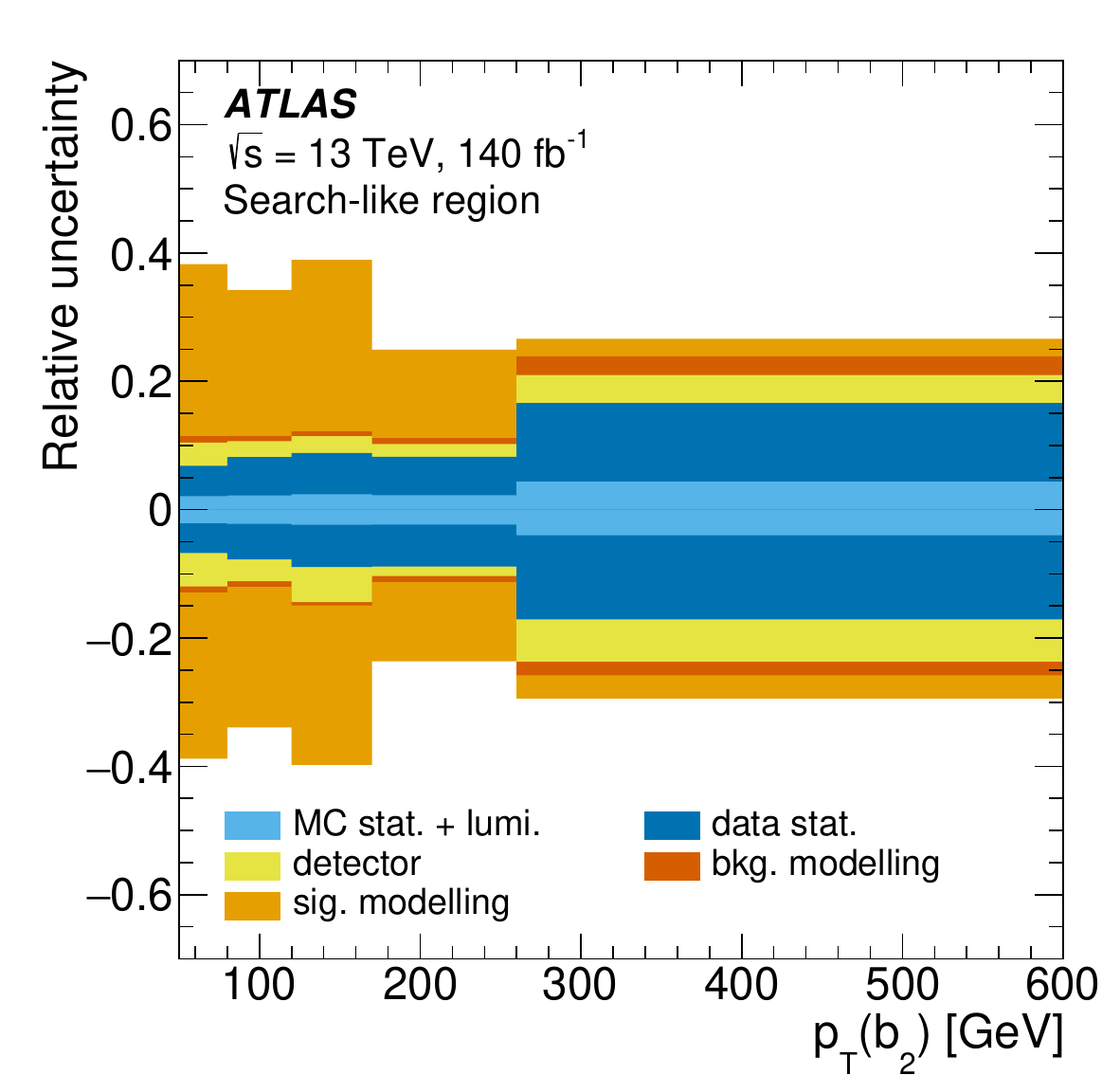}%
\label{fig:SLR:ptb2:uncertainties}
}
\caption{Breakdown of the systematic uncertainties in the unfolded differential
cross-section measurement for
\protect\subref{fig:IR:mblmin:uncertainties} \mblmin for the inclusive region,
\protect\subref{fig:IR:ptl1:uncertainties} the lepton \pt for the inclusive region,
\protect\subref{fig:WR:mblselected:uncertainties} \mblselected for the \WRlong region,
\protect\subref{fig:WR:ptb1:uncertainties} leading \bjet \pt for the \WRlong region,
\protect\subref{fig:SLR:mblmin:uncertainties} \mblmin for the search-like region, and
\protect\subref{fig:SLR:ptb2:uncertainties} the subleading \bjet \pt for the search-like region.
\captionUncertainties}%
\label{fig:uncertainties:block1}
\end{figure}

\begin{figure}
\centering
\subfloat[]{
\includegraphics[width=0.37\textwidth]{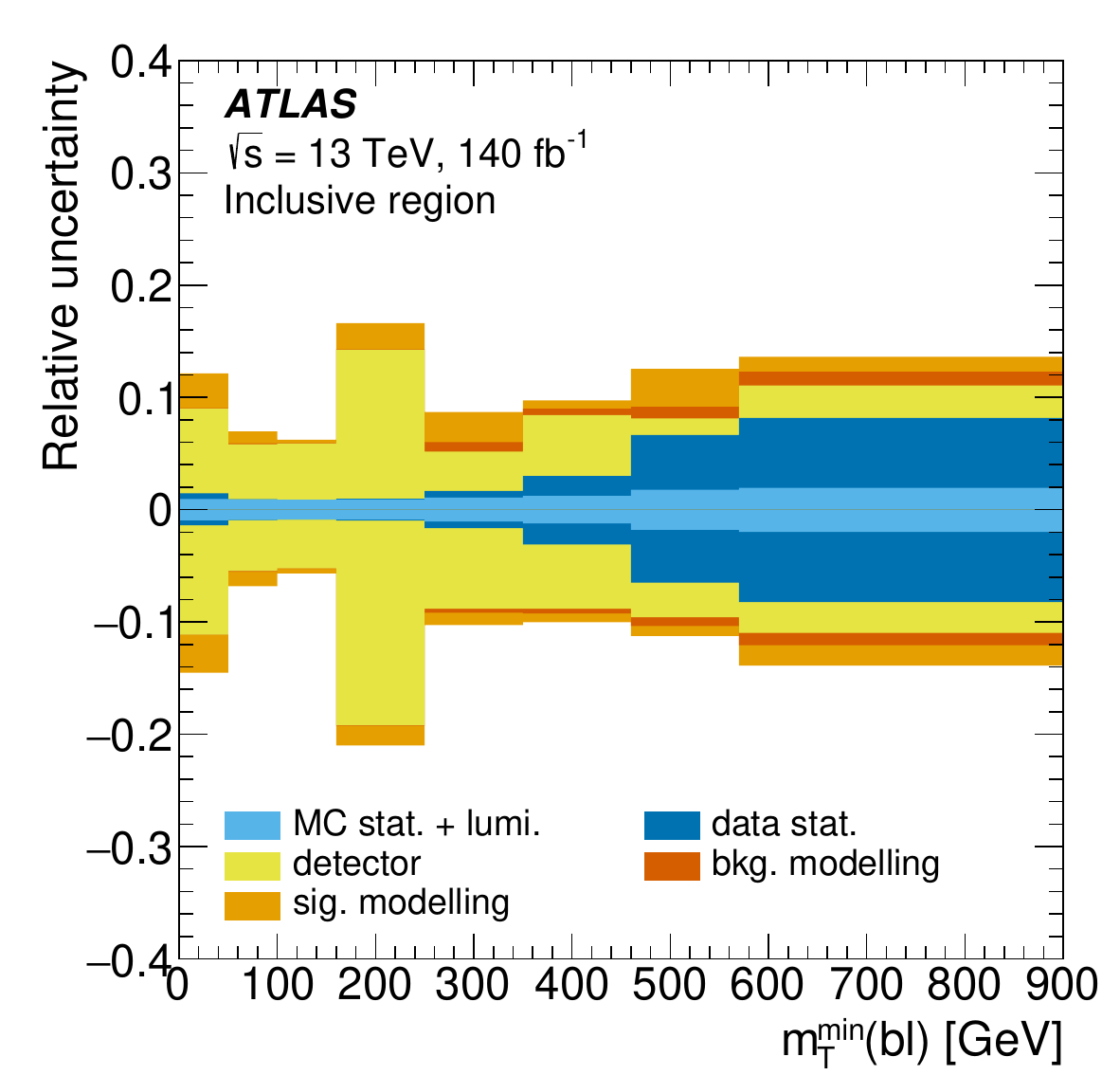}%
\label{fig:IR:mtblmin:uncertainties}
}
\subfloat[]{
\includegraphics[width=0.37\textwidth]{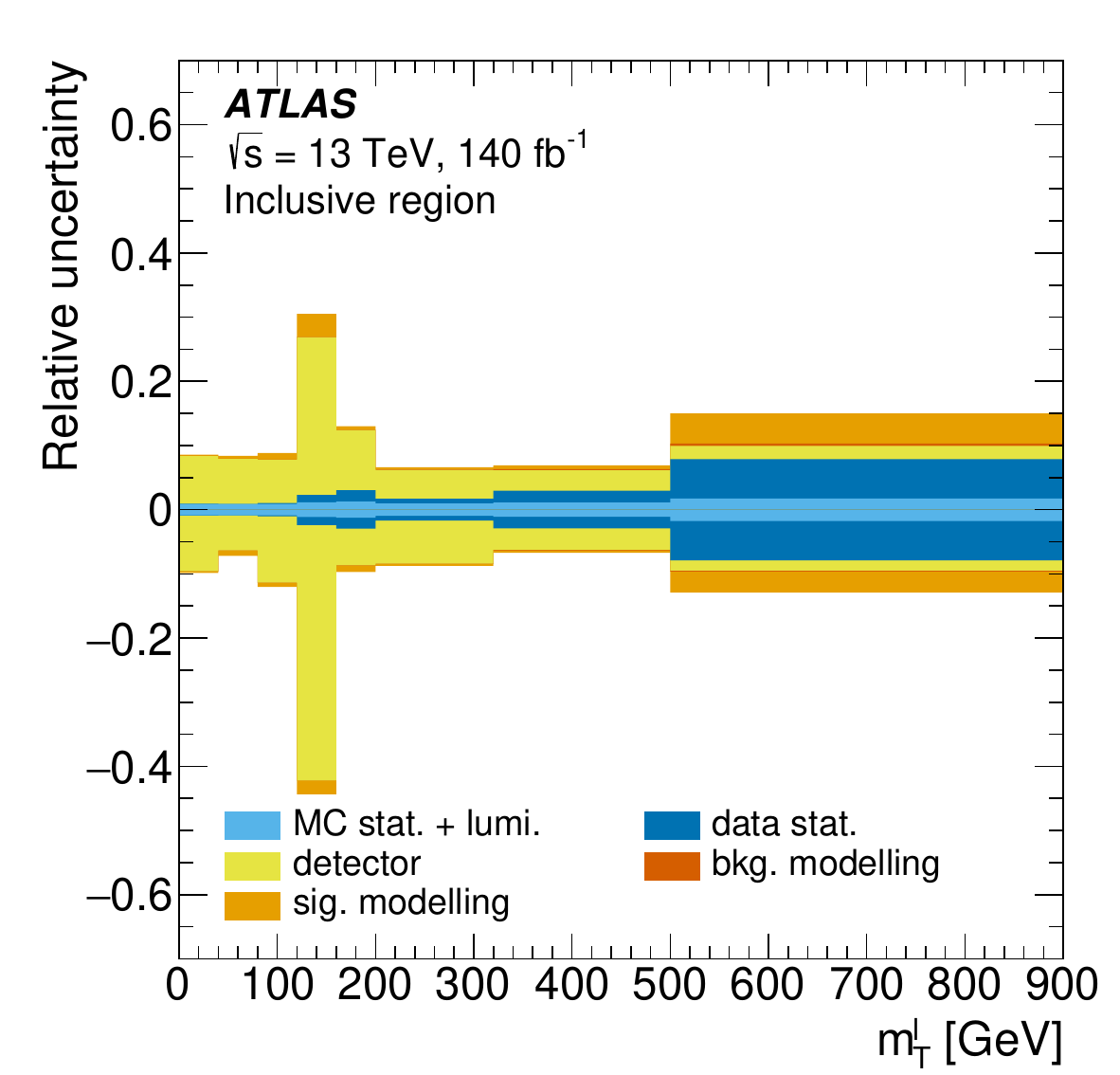}%
\label{fig:IR:mtlep:uncertainties}
}
\\
\subfloat[]{
\includegraphics[width=0.37\textwidth]{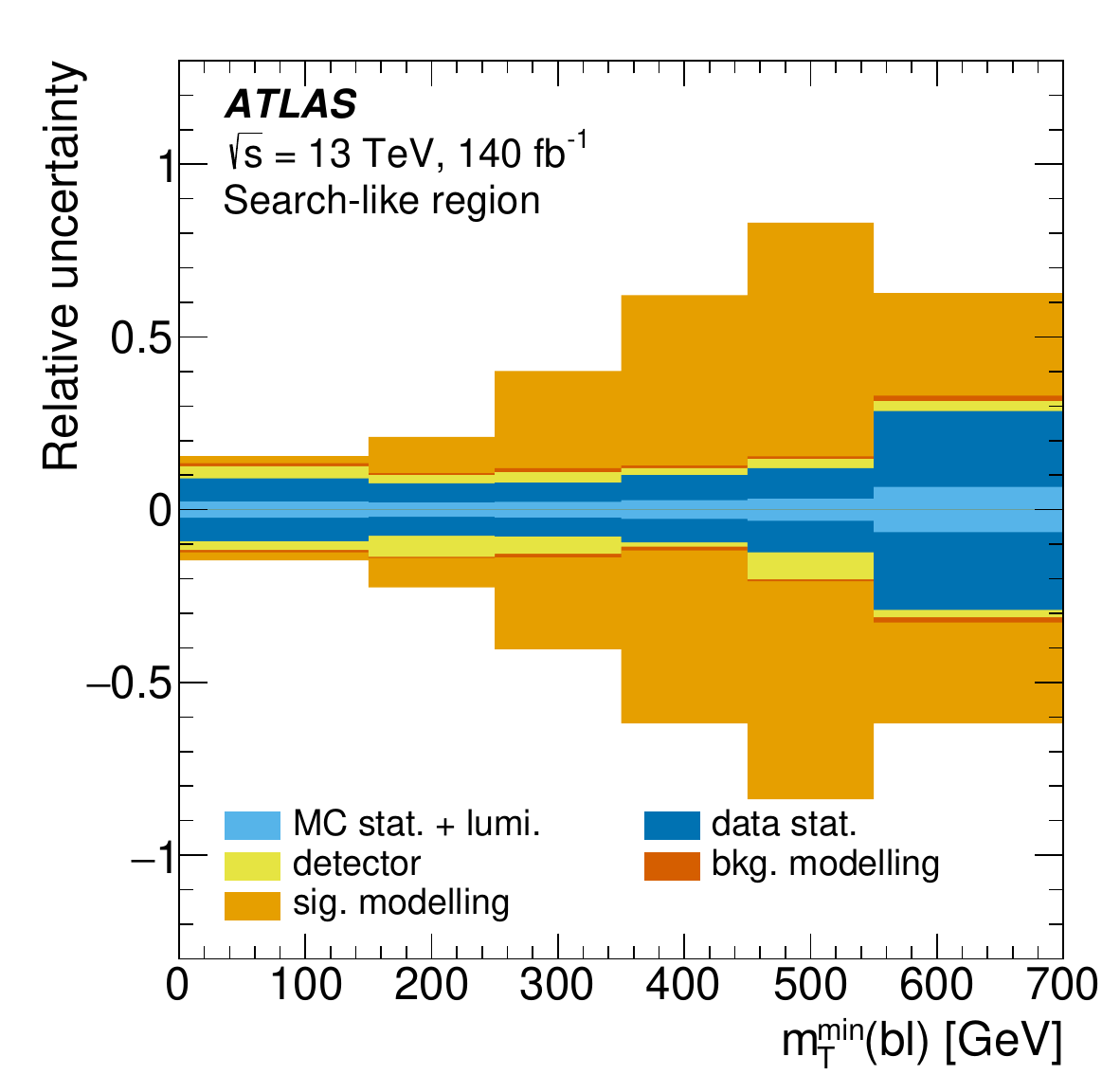}%
\label{fig:SLR:mtblmin:uncertainties}
}
\subfloat[]{
\includegraphics[width=0.37\textwidth]{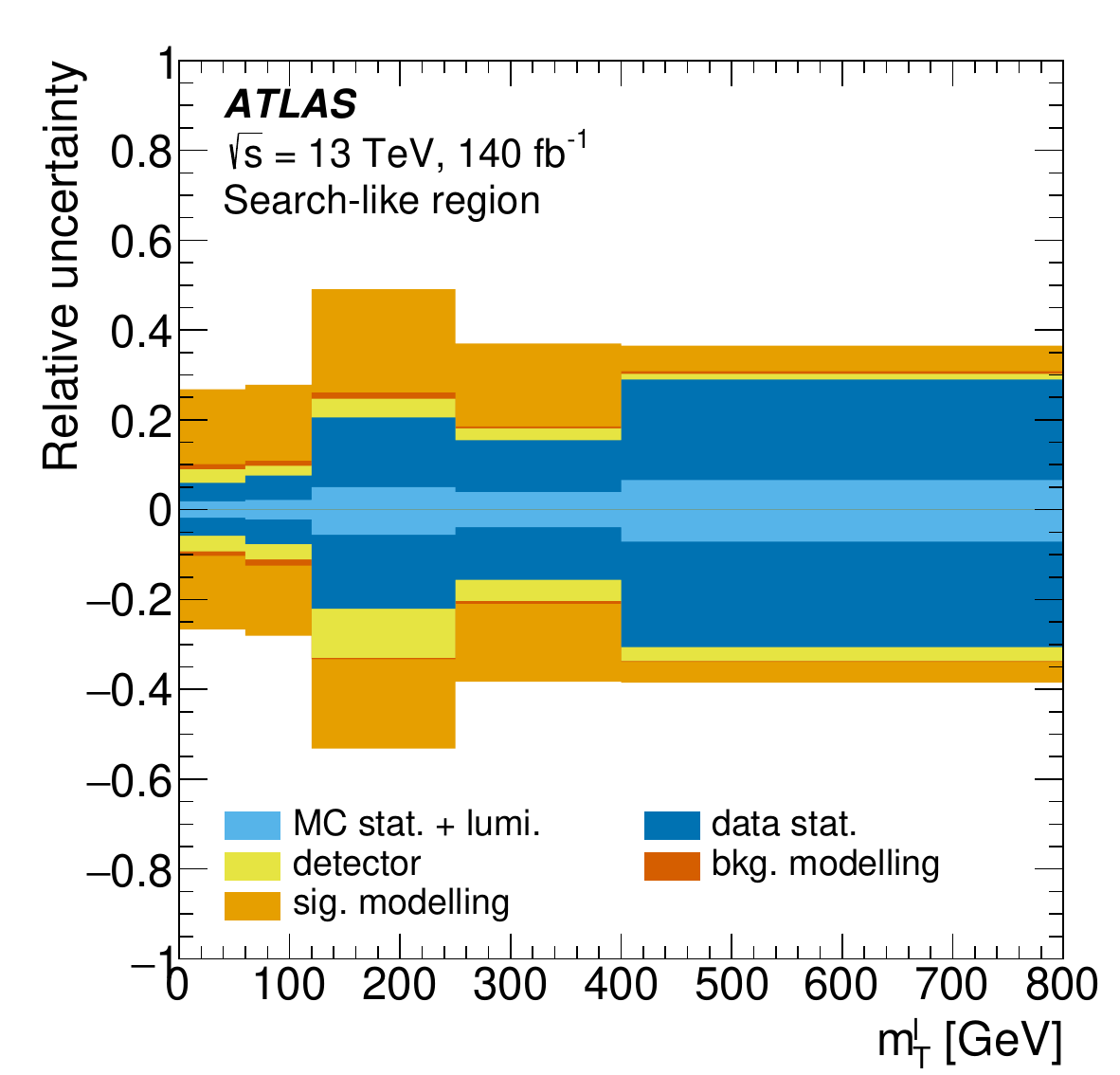}%
\label{fig:SLR:mtlep:uncertainties}
}\\
\subfloat[]{
\includegraphics[width=0.37\textwidth]{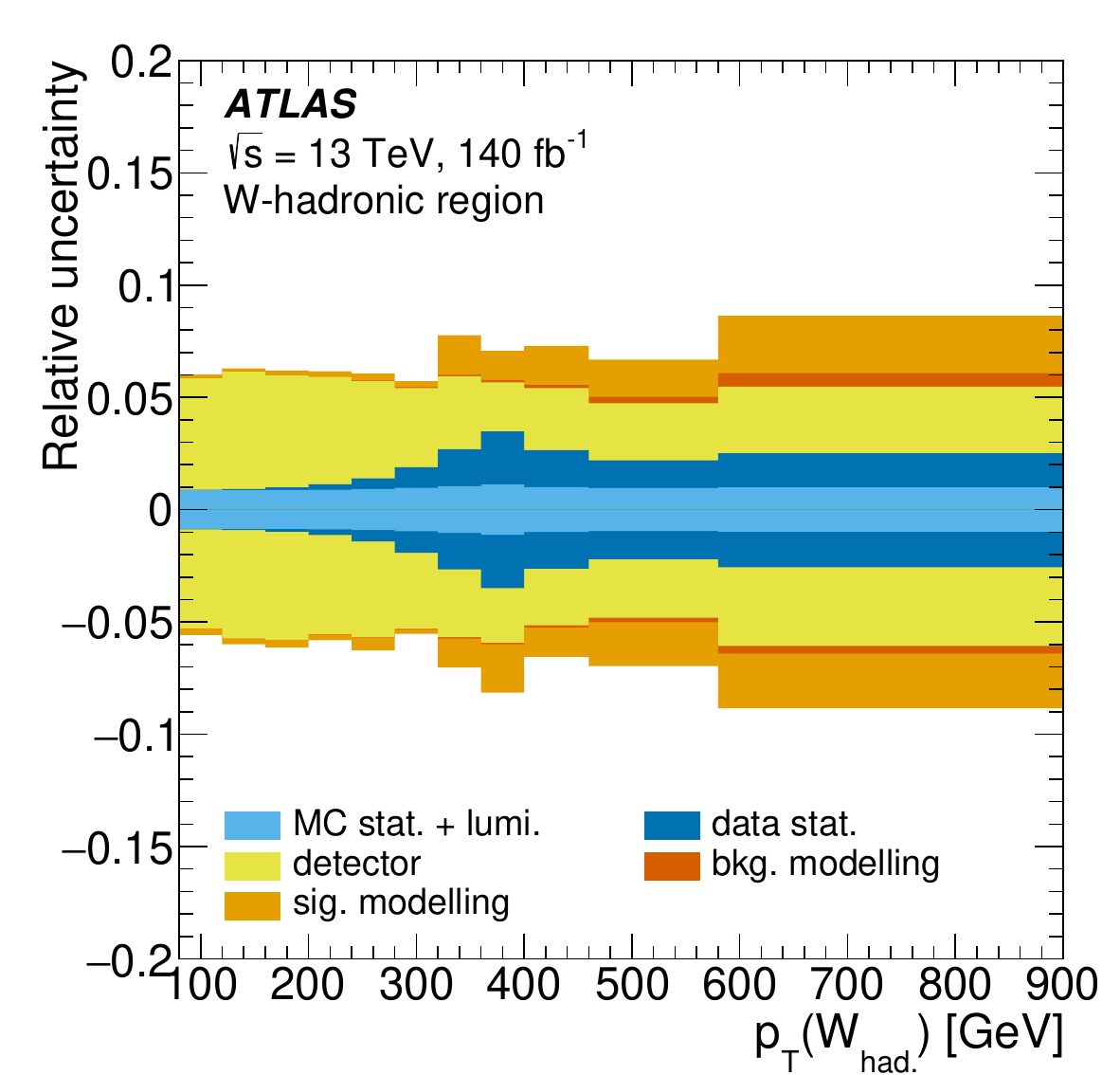}%
\label{fig:WR:ptwhad:uncertainties}
}
\subfloat[]{
\includegraphics[width=0.37\textwidth]{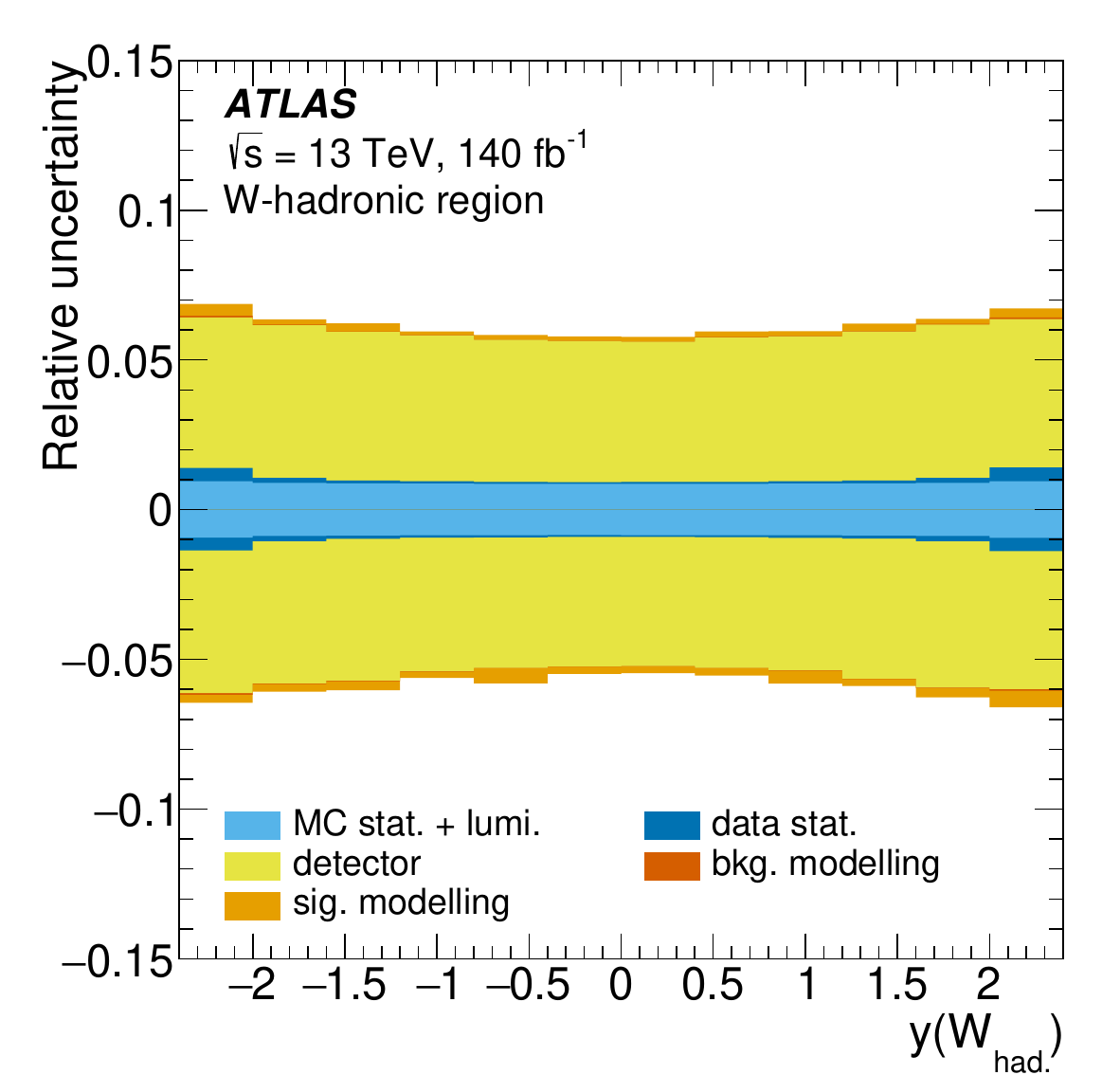}%
\label{fig:WR:ywhad:uncertainties}
}
\caption{Breakdown of the systematic uncertainties in the unfolded differential
cross-section measurement for
\protect\subref{fig:IR:mtblmin:uncertainties} \mtblmin for the inclusive region,
\protect\subref{fig:IR:mtlep:uncertainties} the transverse mass of the lepton, \mtlep, for the inclusive region,
\protect\subref{fig:SLR:mtblmin:uncertainties} \mtblmin for the search-like region,
\protect\subref{fig:SLR:mtlep:uncertainties} the transverse mass of the lepton, \mtlep, for the search-like region.
\protect\subref{fig:WR:ptwhad:uncertainties} the hadronic \Wboson candidate transverse momentum for the \WRlong region, and
\protect\subref{fig:WR:ywhad:uncertainties} the rapidity \yW for the \WRlong region.
\captionUncertainties}%
\label{fig:uncertainties:block2}
\end{figure}

\begin{figure}
\centering
\subfloat[]{
\includegraphics[width=0.37\textwidth]{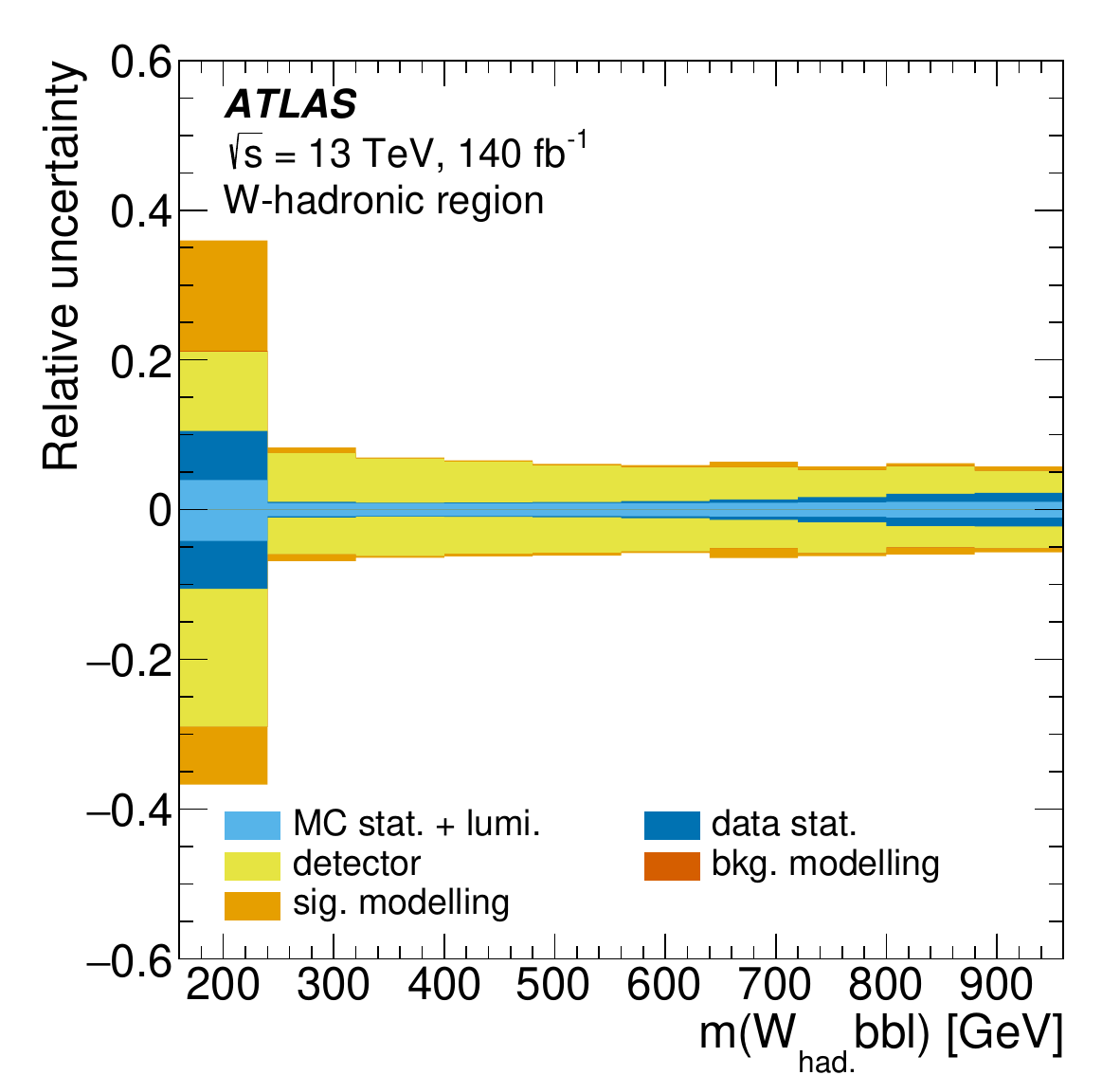}%
\label{fig:WR:mwhadbbl:uncertainties}
}
\subfloat[]{
\includegraphics[width=0.37\textwidth]{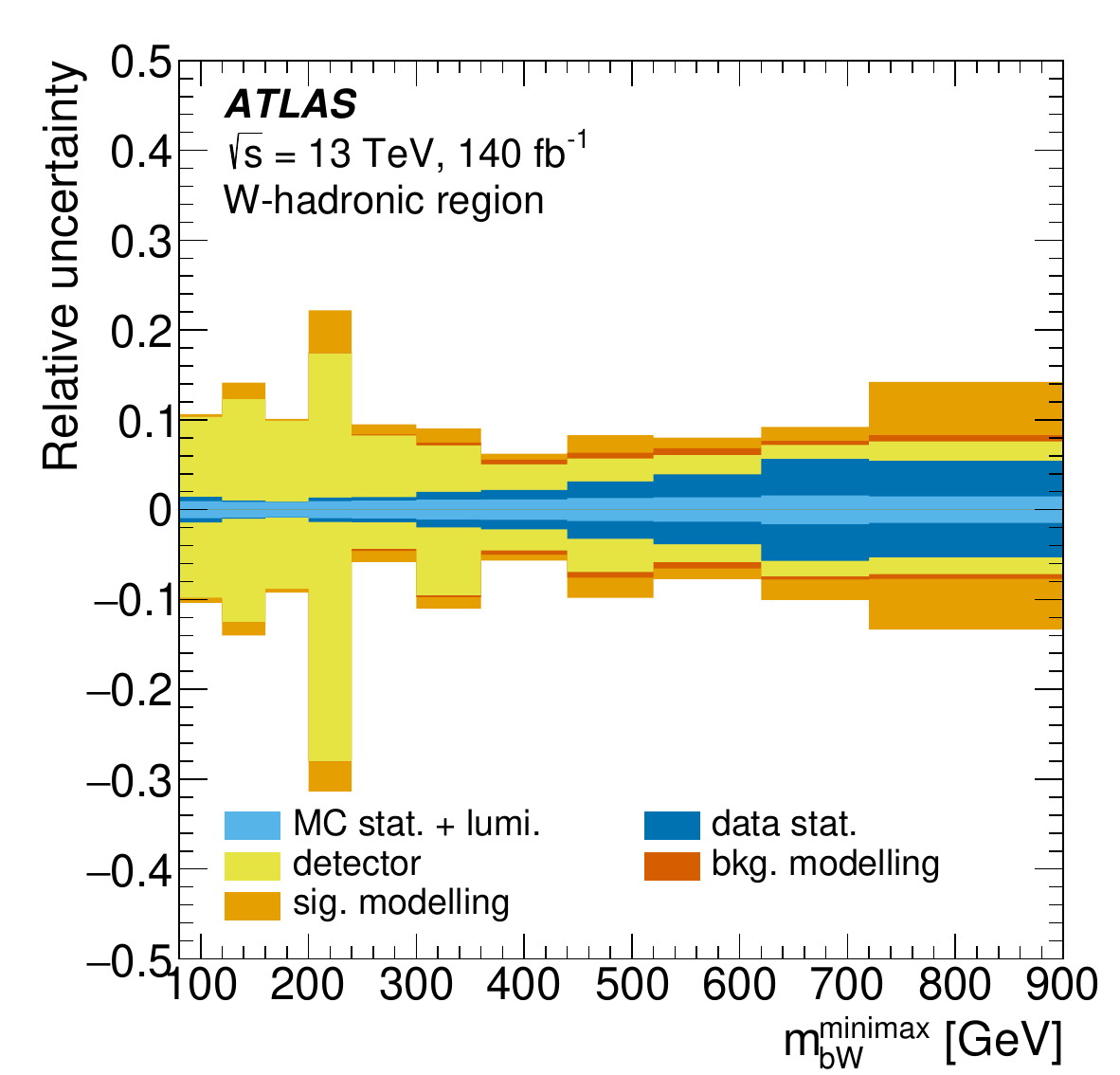}%
\label{fig:WR:minimax:uncertainties}
}
\caption{Breakdown of the systematic uncertainties in the unfolded differential
cross-section measurement for \protect\subref{fig:WR:mwhadbbl:uncertainties} the invariant mass of the \( Wbb\ell \) system
and \protect\subref{fig:WR:minimax:uncertainties} \mbWminimax for the \WRlong region.
\captionUncertainties}%
\label{fig:uncertainties:block3}
\end{figure}

A detailed breakdown of systematic uncertainties is presented in
\crefrange{fig:uncertainties:block1}{fig:uncertainties:block3}
for all observables discussed in this paper. Depending on the region
and observable the dominant systematic uncertainties are found to be detector
systematic uncertainties related to jet tagging and calibration, and modelling of
the signal processes. Observables that exhibit end-points, e.g. \mblmin, \mtblmin, \mblselected and \mtlep, are
particularly sensitive to variations of the jet energy scale and resolution,
which create sharp increases in the uncertainties around the top-quark
mass in \cref{fig:IR:mblmin:uncertainties,fig:WR:mblselected:uncertainties,fig:IR:mtblmin:uncertainties,fig:IR:mtlep:uncertainties}.
Uncertainties related to the modelling of
the \ttbar signal processes are also significant around those end-points, where different modelling of the parton
shower or the initial and final state radiation can lead to different shaping
of the distributions.

\FloatBarrier
\section{Differential cross-section results}%
\label{sec:results}

\begin{figure}[htbp]
\centering
\subfloat[]{
\includegraphics[width=0.35\textwidth]{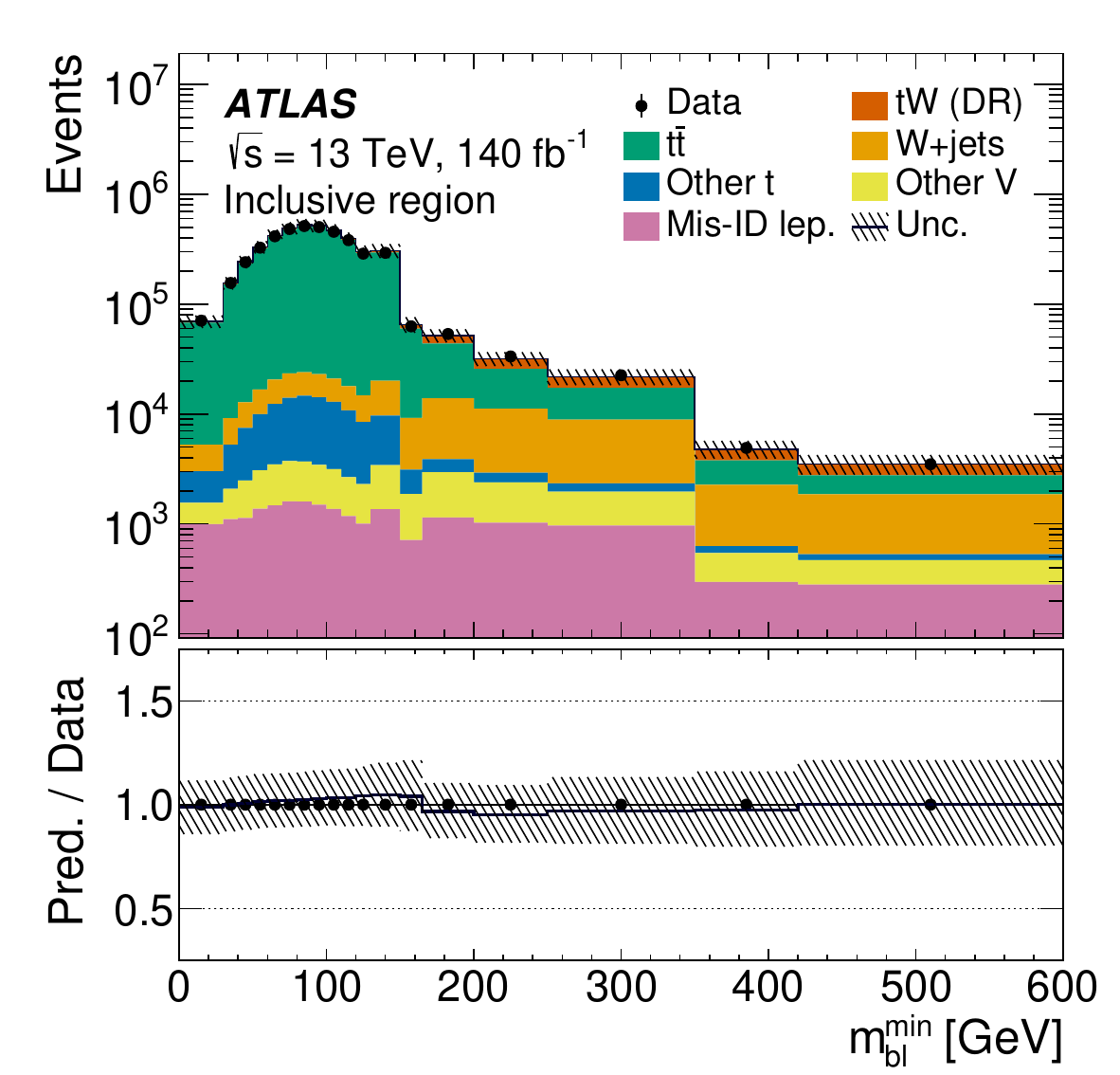}%
\label{fig:IR:mblmin:reco}
}
\subfloat[]{
\includegraphics[width=0.35\textwidth]{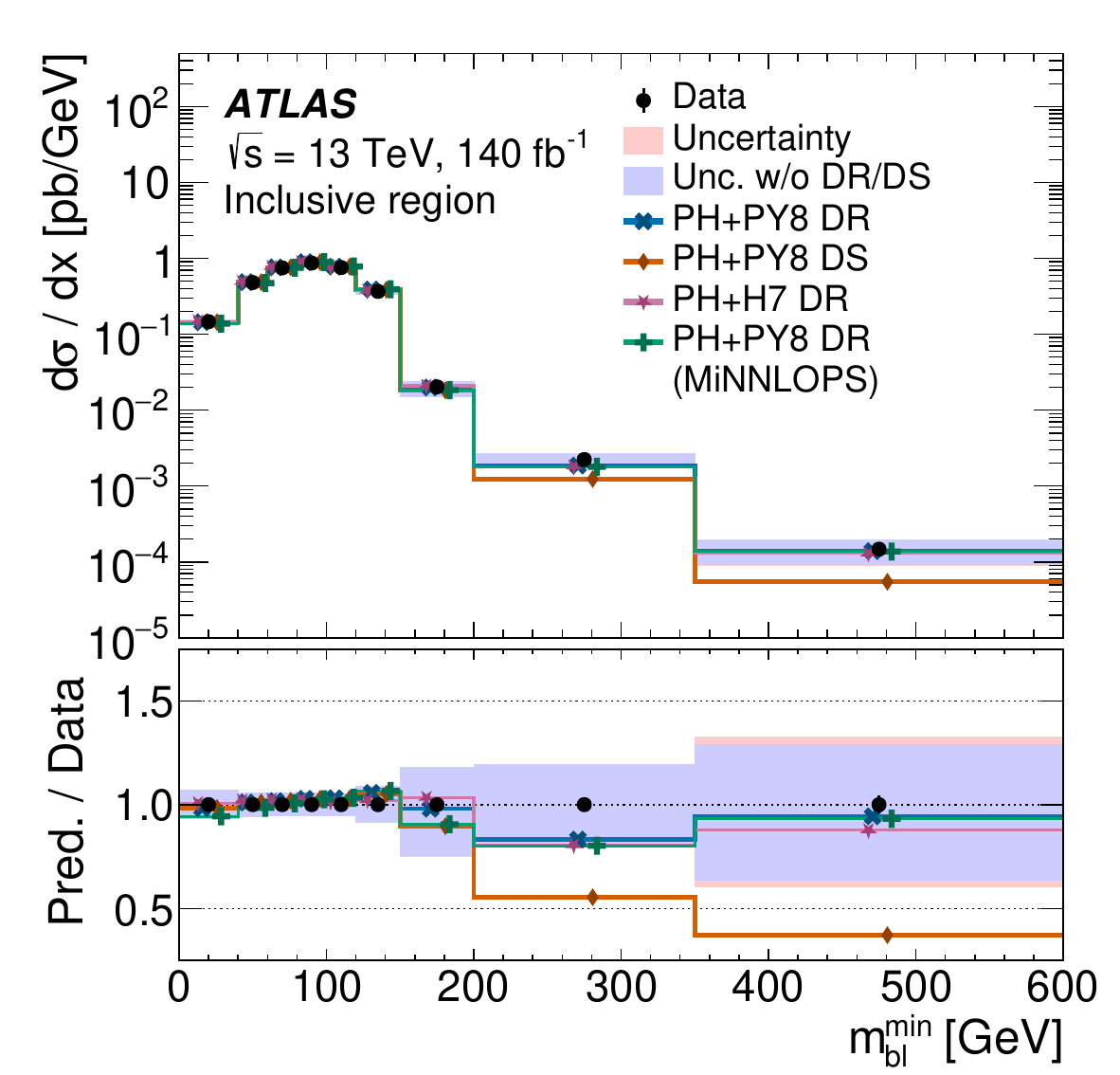}%
\label{fig:IR:mblmin:unfolded}
}
\\
\subfloat[]{
\includegraphics[width=0.35\textwidth]{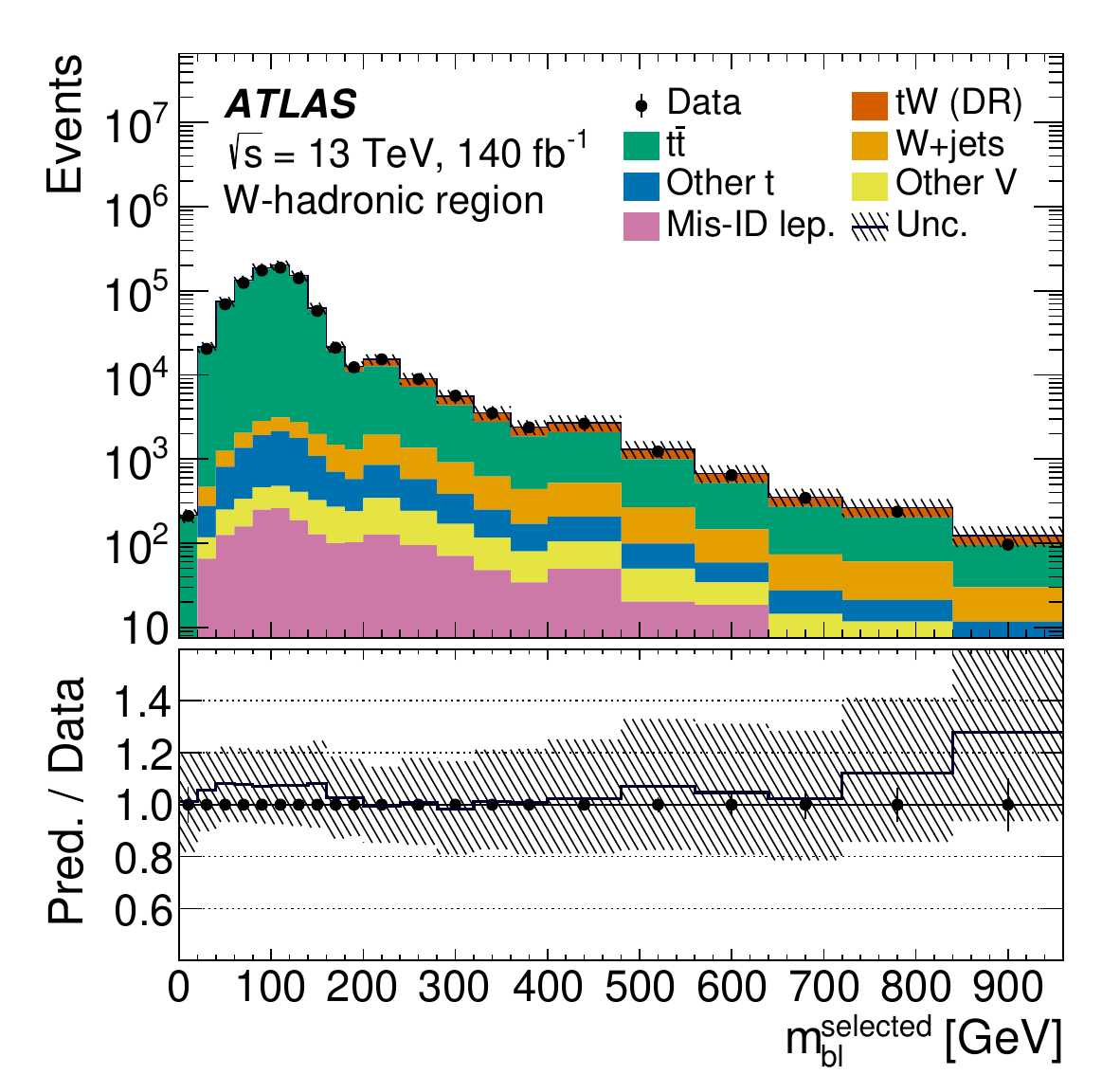}%
\label{fig:WR:mblselected:reco}
}
\subfloat[]{
\includegraphics[width=0.35\textwidth]{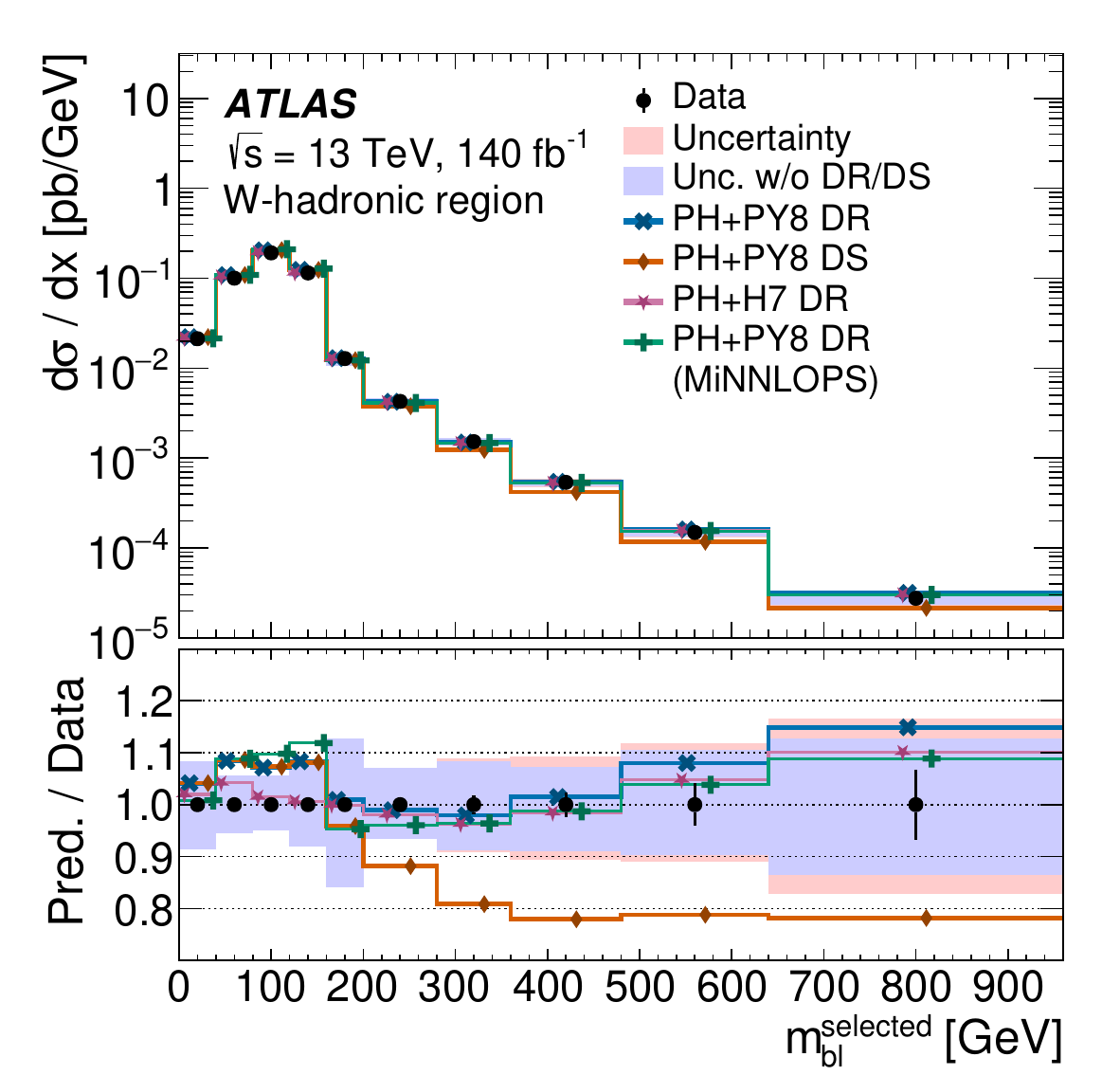}%
\label{fig:WR:mblselected:unfolded}
}
\\
\subfloat[]{
\includegraphics[width=0.35\textwidth]{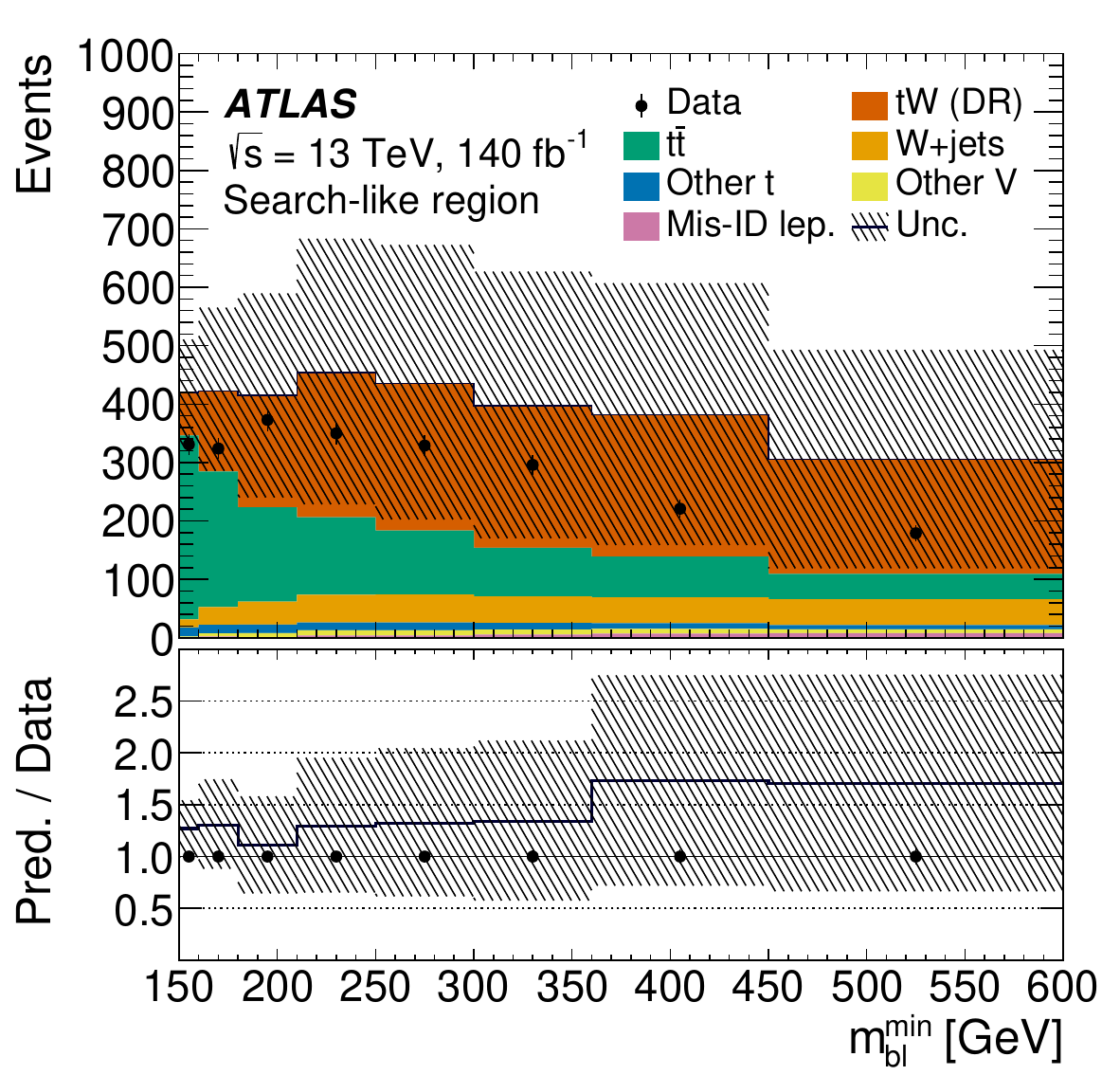}%
\label{fig:SLR:mblmin:reco}
}
\subfloat[]{
\includegraphics[width=0.35\textwidth]{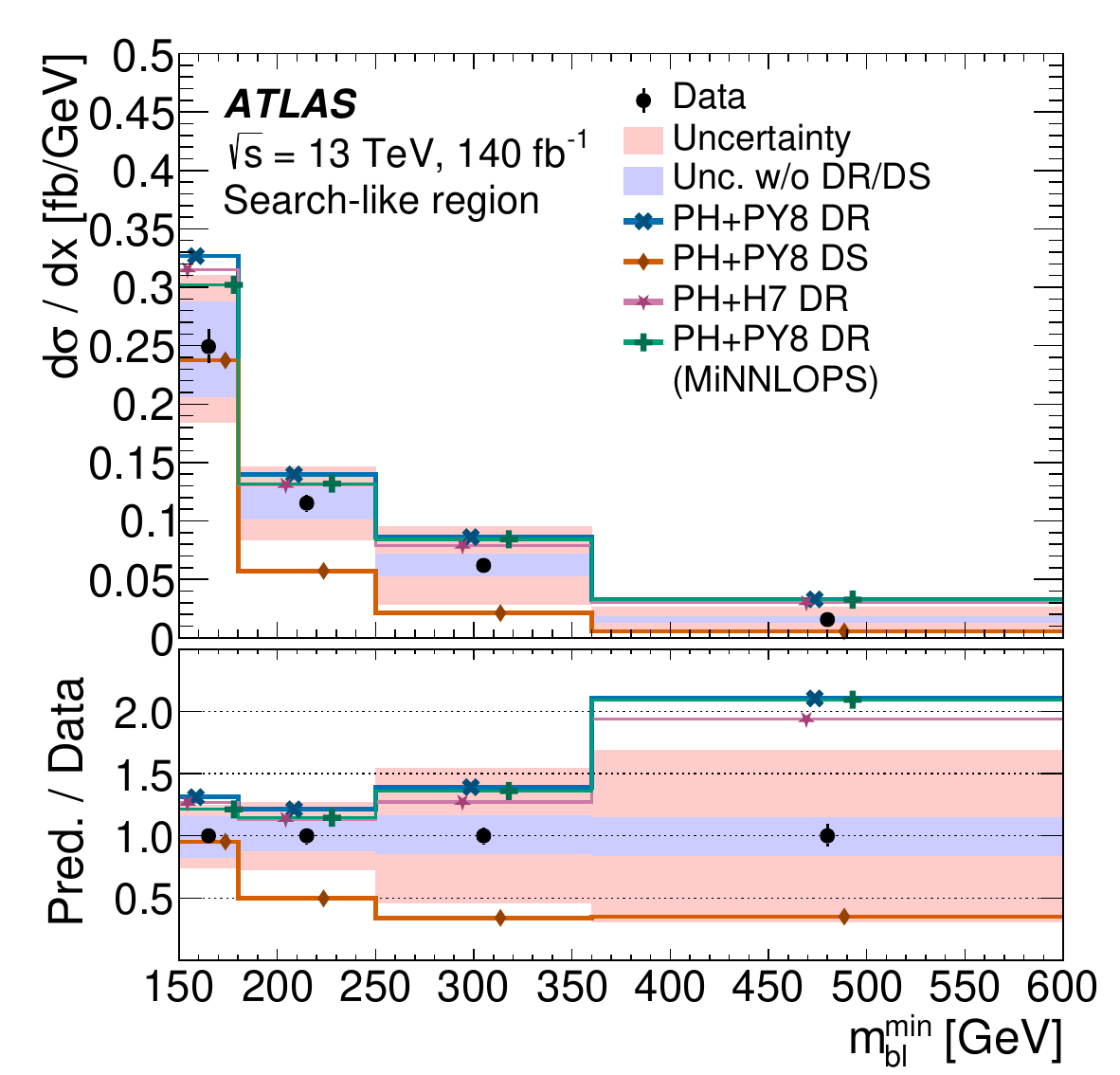}%
\label{fig:SLR:mblmin:unfolded}
}
\caption{The (\protect\subref{fig:IR:mblmin:reco}, \protect\subref{fig:WR:mblselected:reco}, \protect\subref{fig:SLR:mblmin:reco}) detector-level distributions
and (\protect\subref{fig:IR:mblmin:unfolded}, \protect\subref{fig:WR:mblselected:unfolded}, \protect\subref{fig:SLR:mblmin:unfolded}) unfolded differential cross-sections
of the invariant mass between the lepton and the \bjets.
The distributions are shown (\protect\subref{fig:IR:mblmin:reco} and \protect\subref{fig:IR:mblmin:unfolded}) in the inclusive region,
(\protect\subref{fig:WR:mblselected:reco} and \protect\subref{fig:WR:mblselected:unfolded}) in the \WRlong
and (\protect\subref{fig:SLR:mblmin:reco} and \protect\subref{fig:SLR:mblmin:unfolded}) in the search-like region.
\captionRatioRecoUnfolded
\captionSysBandExt
\captionUnfoldedComparison}%
\label{fig:results:mblmin}
\end{figure}

\begin{figure}[htbp]
\centering
\subfloat[]{
\includegraphics[width=0.35\textwidth]{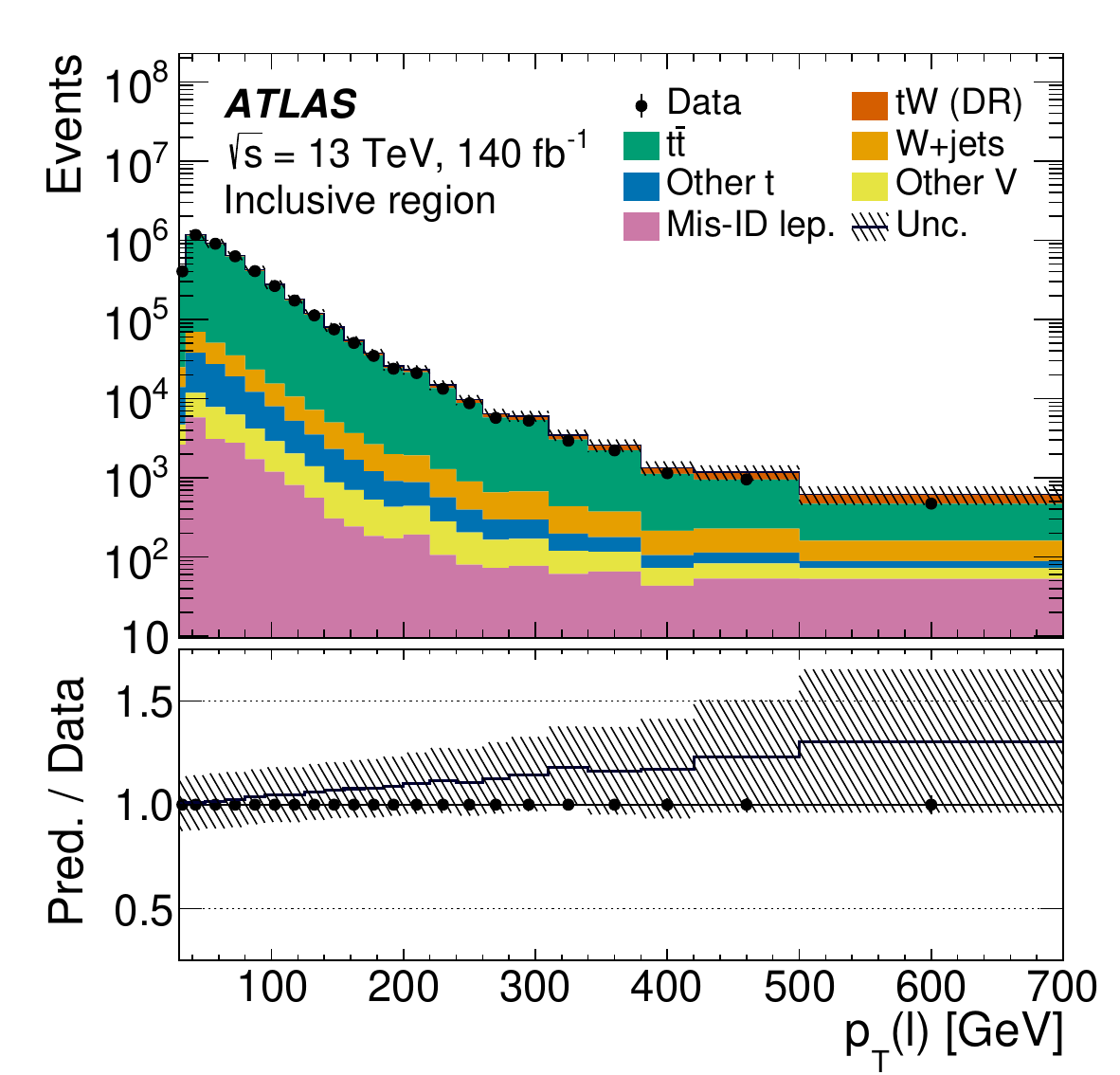}%
\label{fig:IR:ptl1:reco}
}
\subfloat[]{
\includegraphics[width=0.35\textwidth]{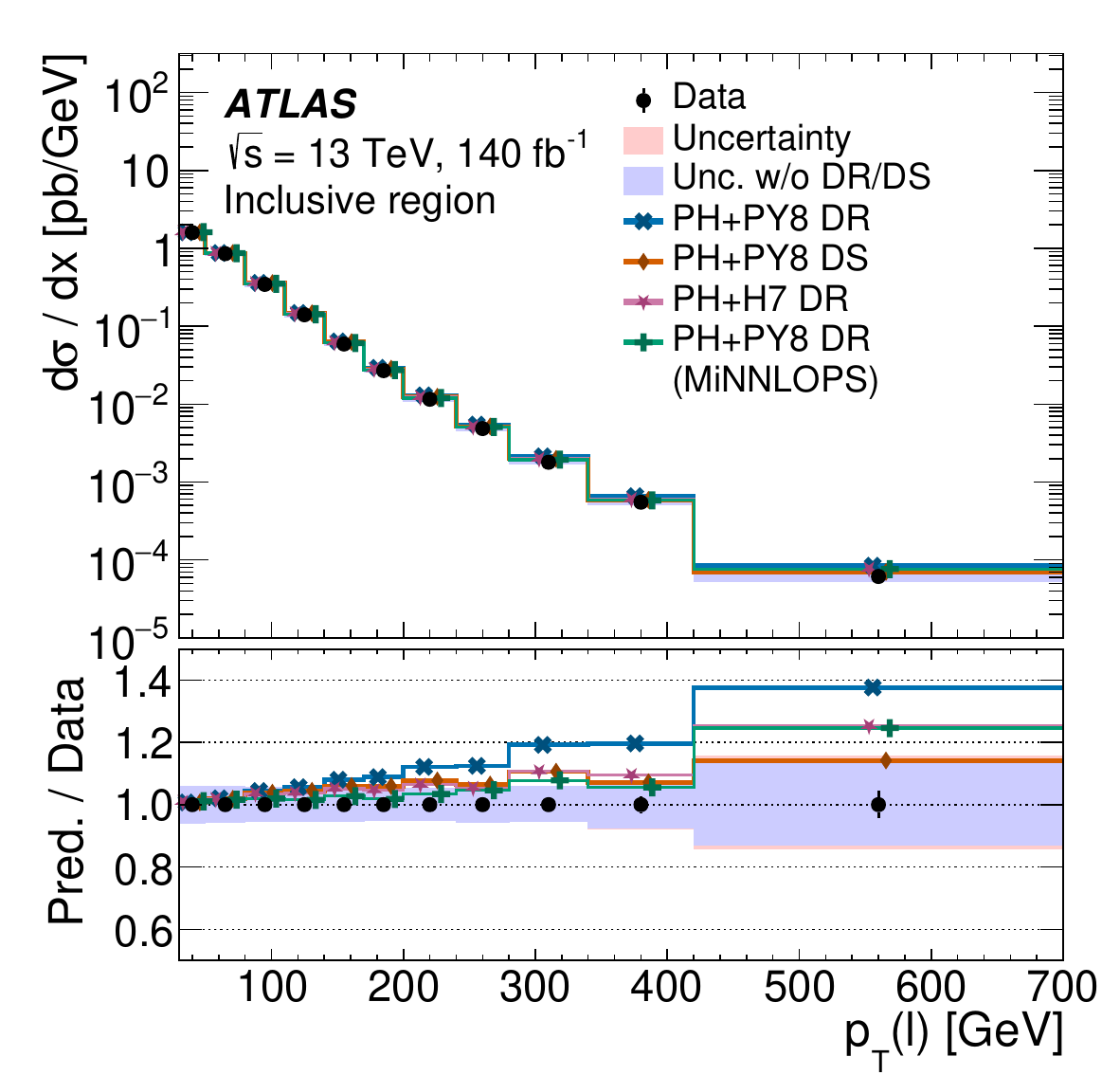}%
\label{fig:IR:ptl1:unfolded}
}
\\
\subfloat[]{
\includegraphics[width=0.35\textwidth]{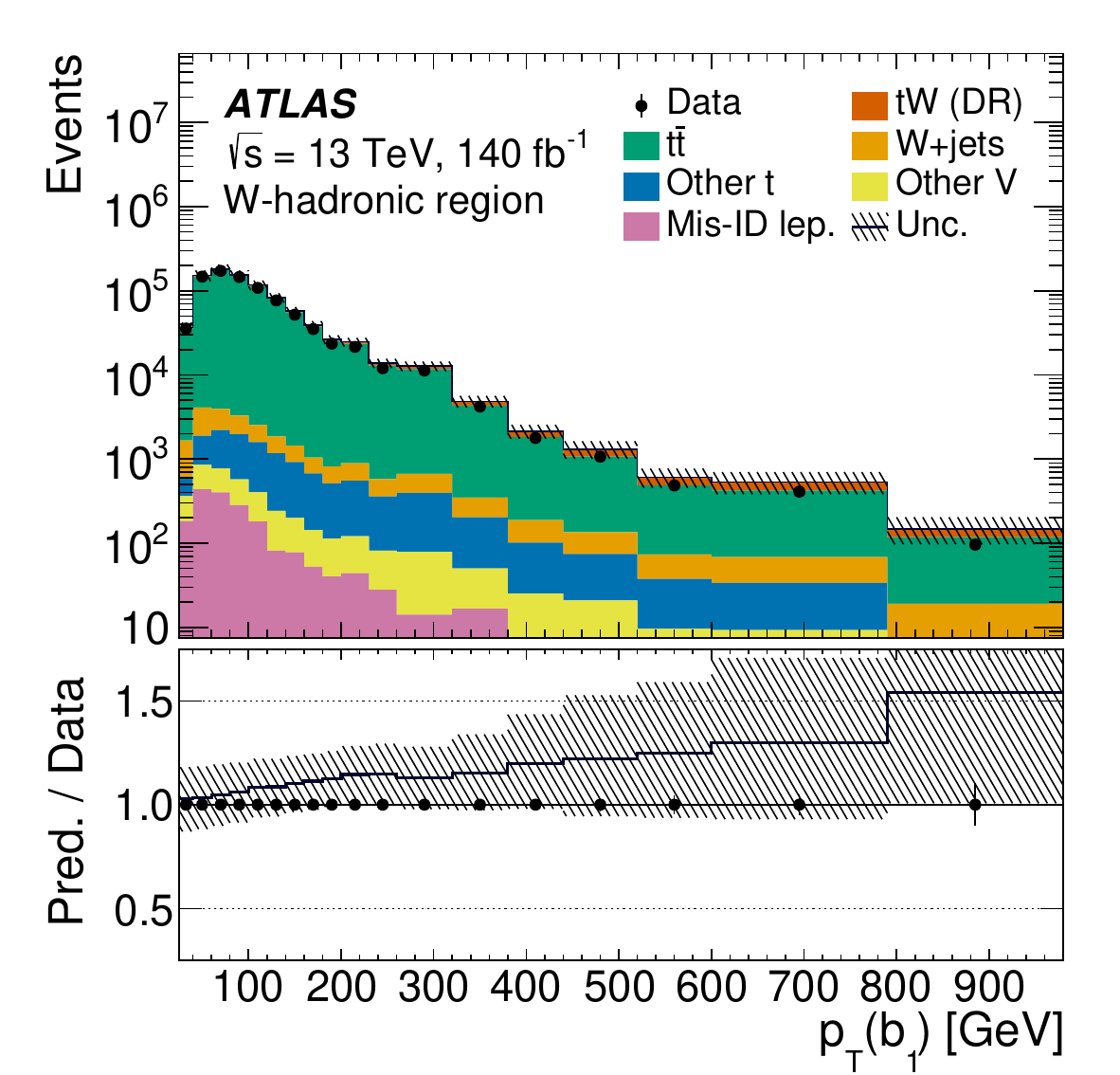}%
\label{fig:WR:ptb1:reco}
}
\subfloat[]{
\includegraphics[width=0.35\textwidth]{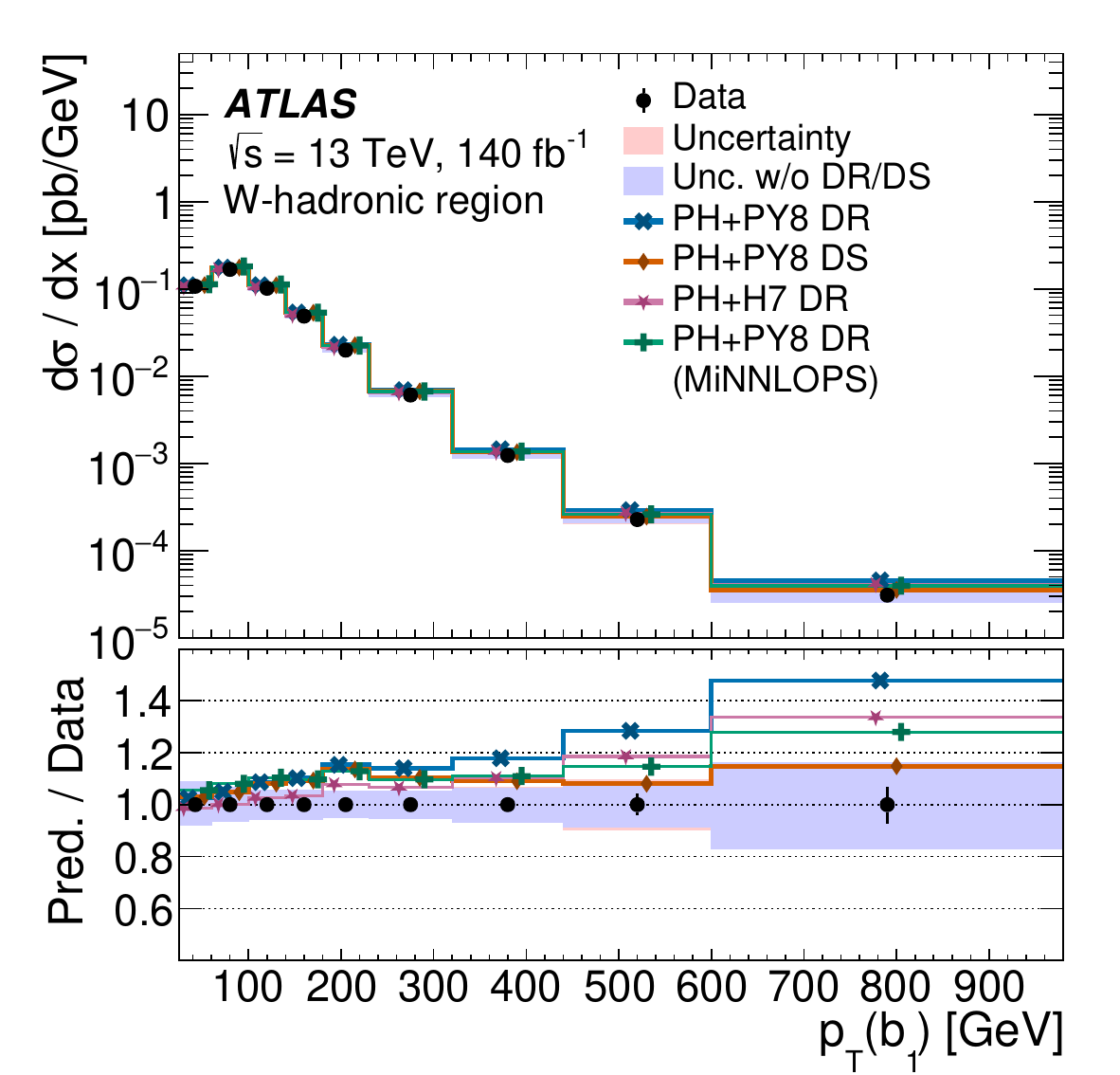}%
\label{fig:WR:ptb1:unfolded}
}
\\
\subfloat[]{
\includegraphics[width=0.35\textwidth]{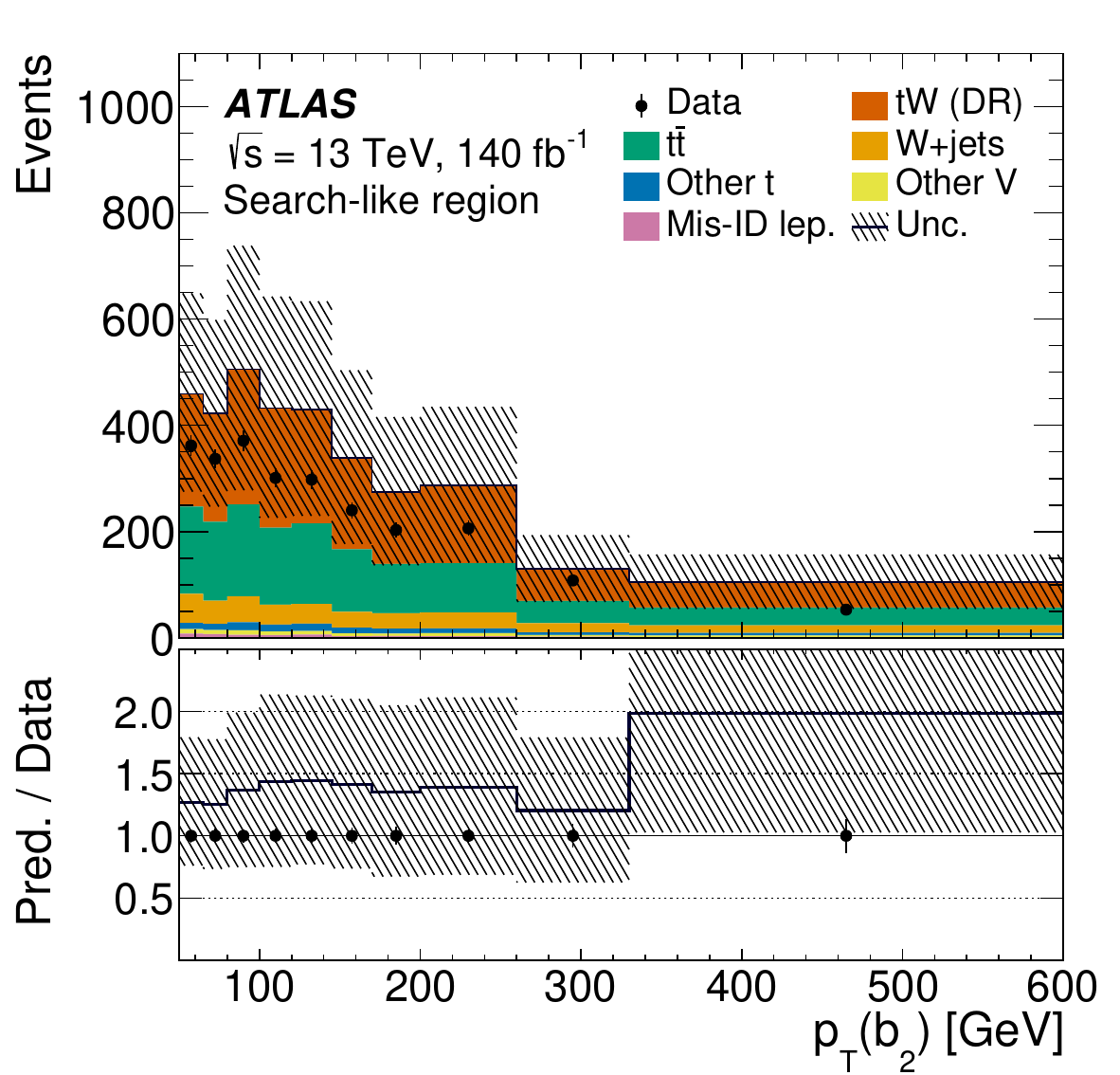}%
\label{fig:SLR:ptb2:reco}
}
\subfloat[]{
\includegraphics[width=0.35\textwidth]{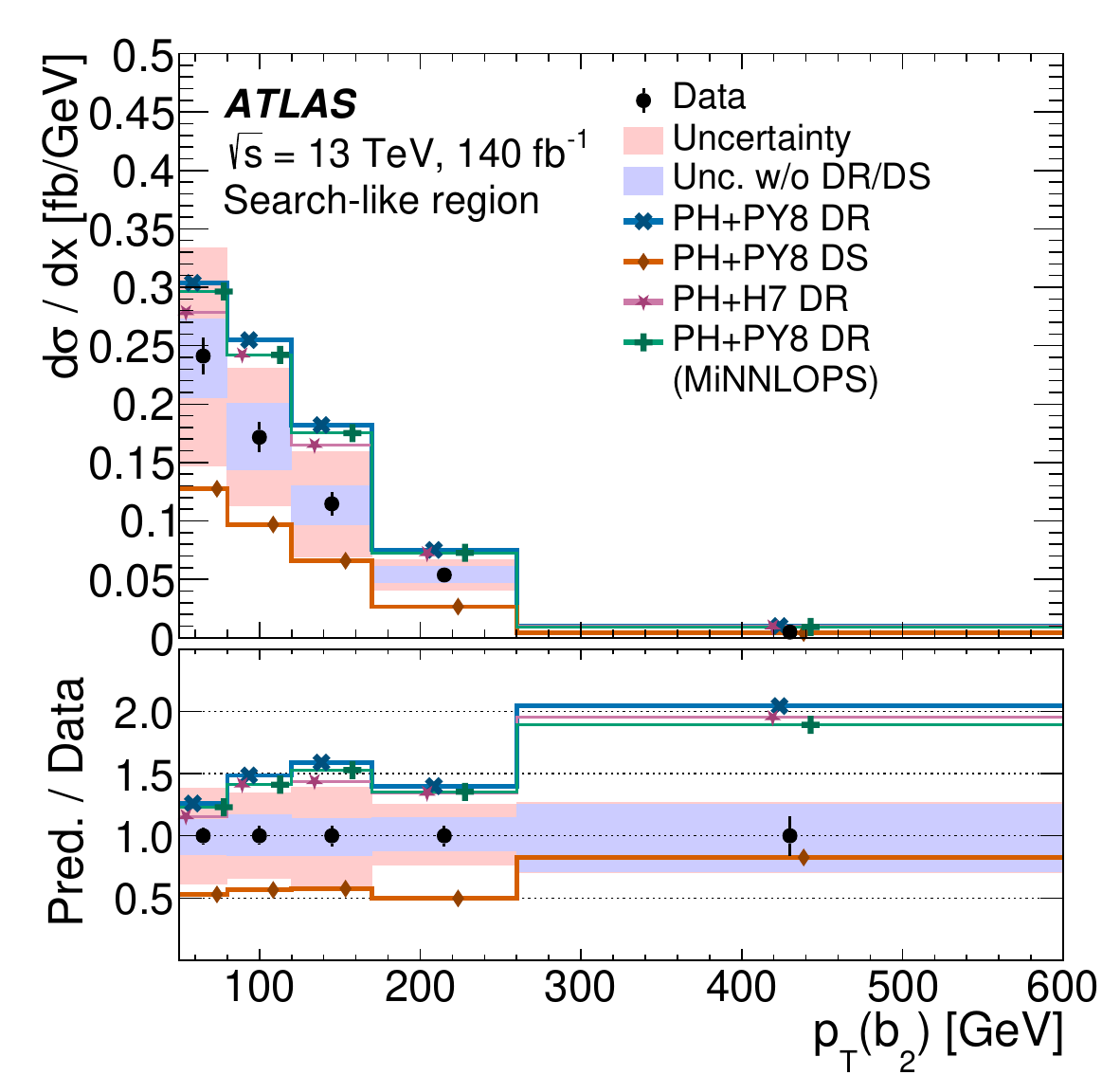}%
\label{fig:SLR:ptb2:unfolded}
}
\caption{The (\protect\subref{fig:IR:ptl1:reco}, \protect\subref{fig:WR:ptb1:reco}, \protect\subref{fig:SLR:ptb2:reco}) detector-level distributions
and (\protect\subref{fig:IR:ptl1:unfolded}, \protect\subref{fig:WR:ptb1:unfolded}, \protect\subref{fig:SLR:ptb2:unfolded}) unfolded differential cross-sections
of (\protect\subref{fig:IR:ptl1:reco} and \protect\subref{fig:IR:ptl1:unfolded}) the lepton \pt for the inclusive region,
(\protect\subref{fig:WR:ptb1:reco} and \protect\subref{fig:WR:ptb1:unfolded}) the leading \bjet \pt for the \WRlong region
and (\protect\subref{fig:SLR:ptb2:reco} and \protect\subref{fig:SLR:ptb2:unfolded}) the subleading \bjet \pt for the search-like region.
\captionRatioRecoUnfolded
\captionSysBandExt
\captionUnfoldedComparison}%
\label{fig:results:pt}
\end{figure}

\begin{figure}[htbp]
\centering
\subfloat[]{
\includegraphics[width=0.37\textwidth]{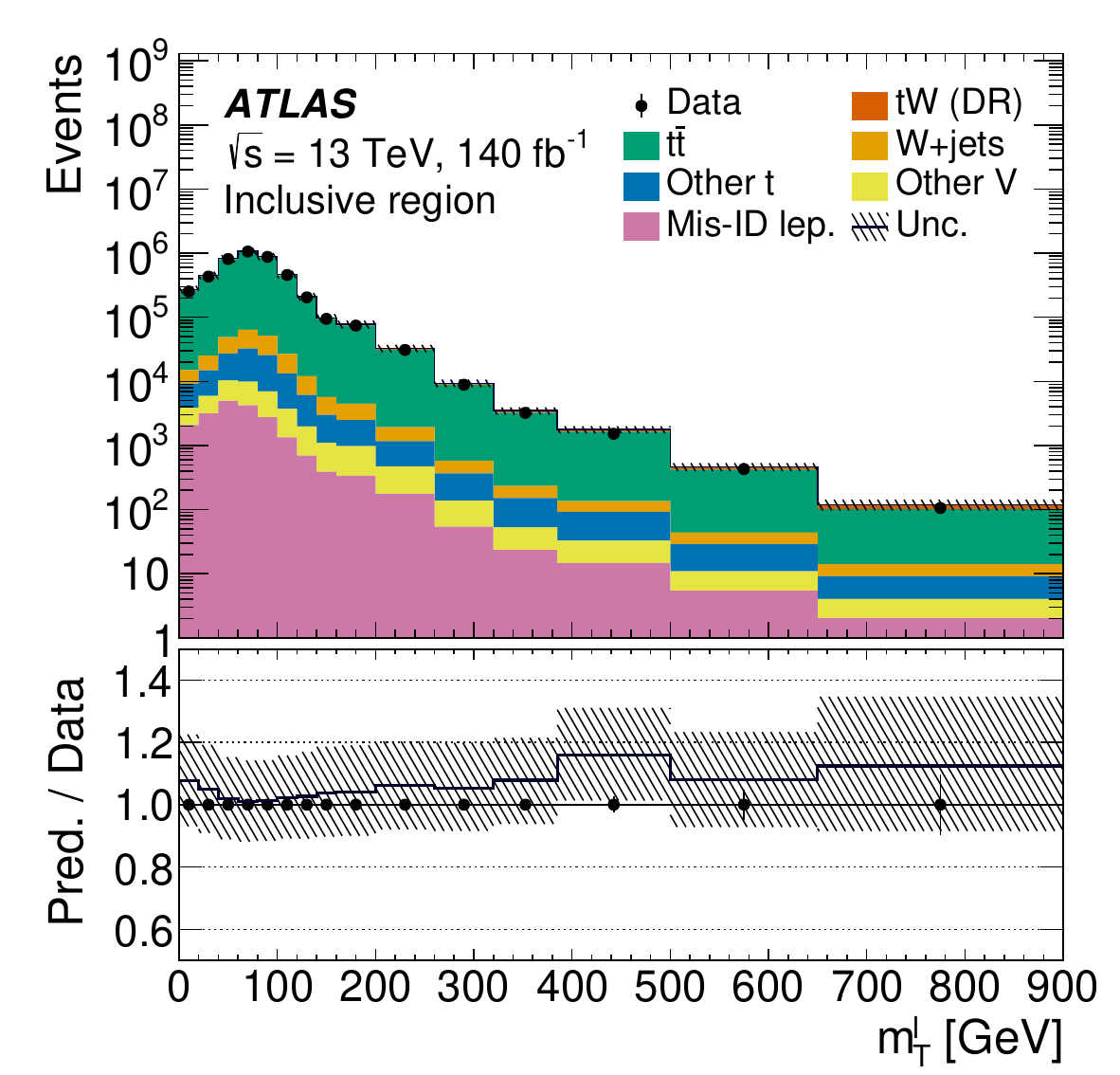}%
\label{fig:IR:mtlep:reco}
}
\subfloat[]{
\includegraphics[width=0.37\textwidth]{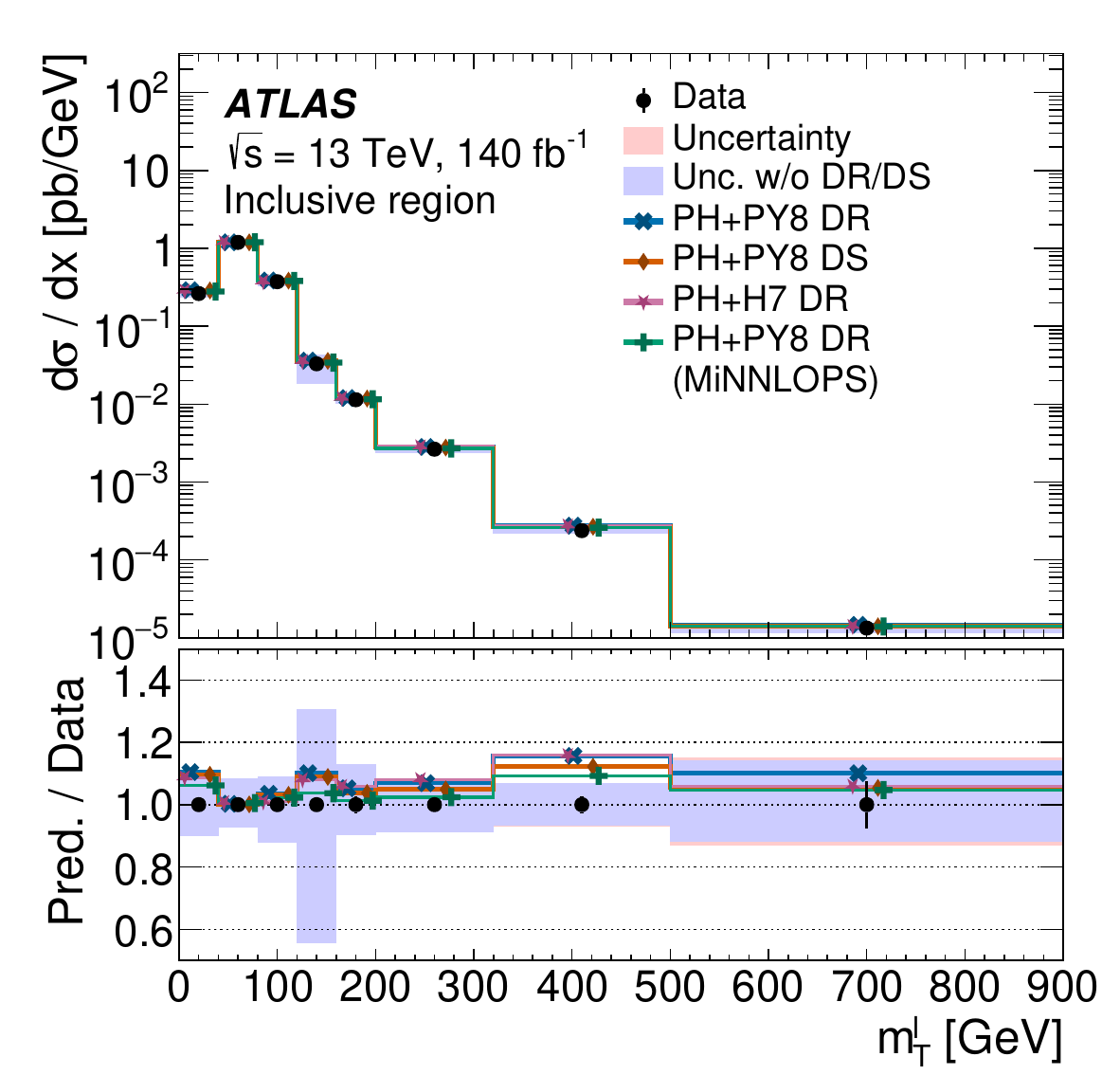}%
\label{fig:IR:mtlep:unfolded}
}
\\
\subfloat[]{
\includegraphics[width=0.37\textwidth]{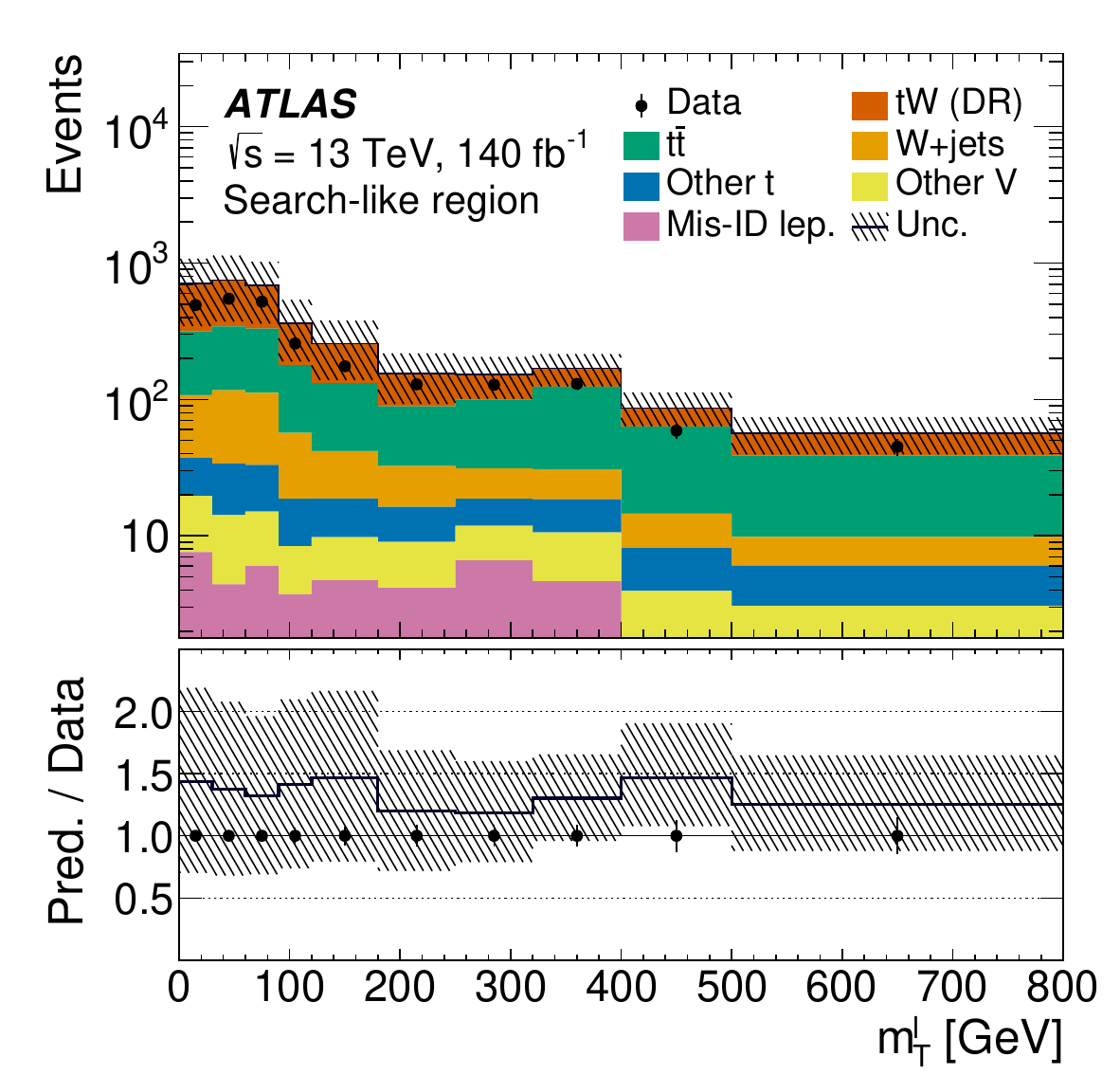}%
\label{fig:SLR:mtlep:reco}
}
\subfloat[]{
\includegraphics[width=0.37\textwidth]{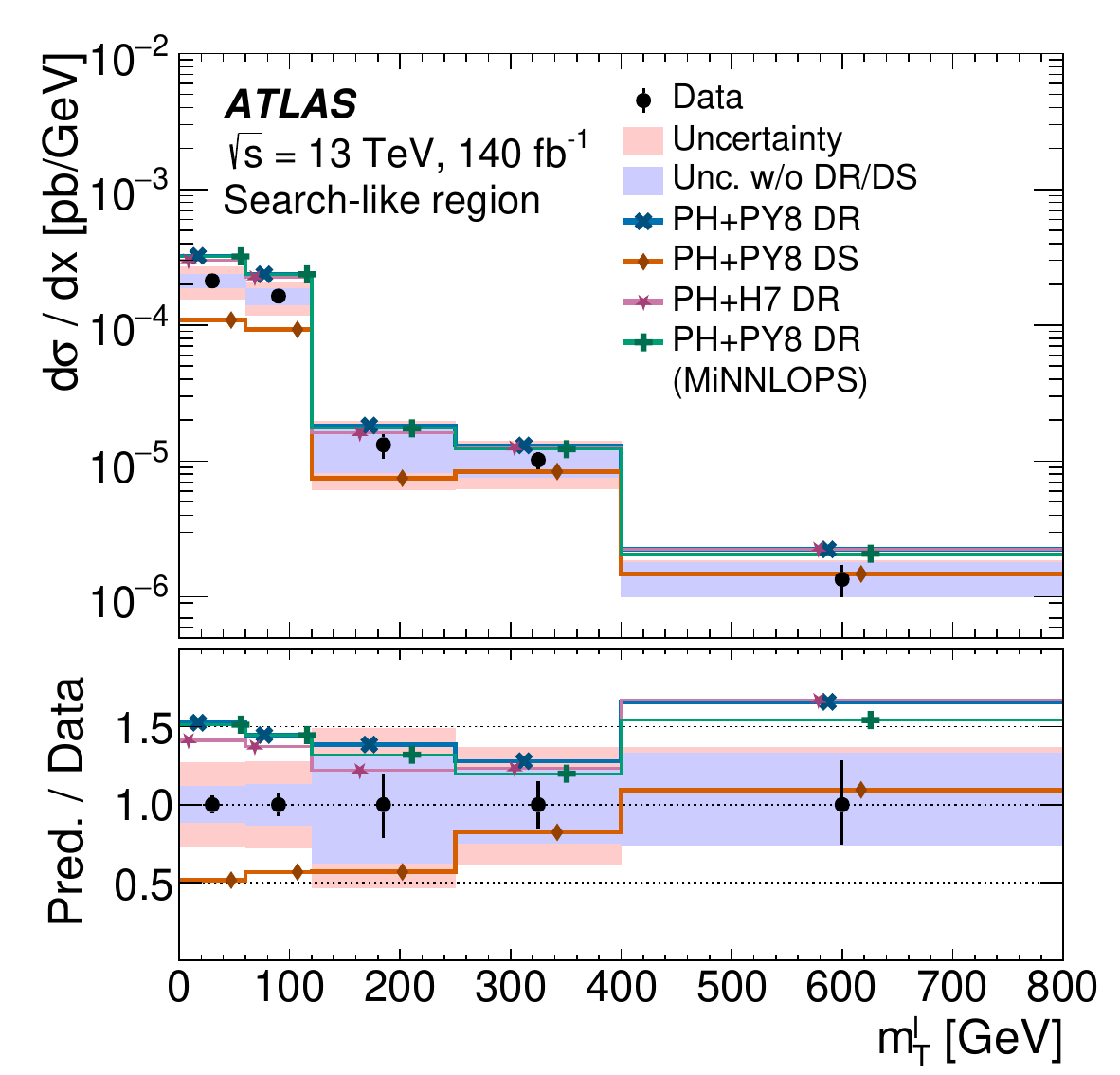}%
\label{fig:SLR:mtlep:unfolded}
}
\caption{The (\protect\subref{fig:IR:mtlep:reco}, \protect\subref{fig:SLR:mtlep:reco}) detector-level distributions and
(\protect\subref{fig:IR:mtlep:unfolded}, \protect\subref{fig:SLR:mtlep:unfolded}) unfolded differential cross-sections
of the lepton transverse mass, \mtlep.
The distributions are shown (\protect\subref{fig:IR:mtlep:reco} and \protect\subref{fig:IR:mtlep:unfolded}) in the inclusive region
and (\protect\subref{fig:SLR:mtlep:reco} and \protect\subref{fig:SLR:mtlep:unfolded}) in the search-like region.
\captionRatioRecoUnfolded
\captionSysBandExt
\captionUnfoldedComparison}%
\label{fig:results:mtlep}
\end{figure}

\begin{figure}[htbp]
\centering
\subfloat[]{
\includegraphics[width=0.37\textwidth]{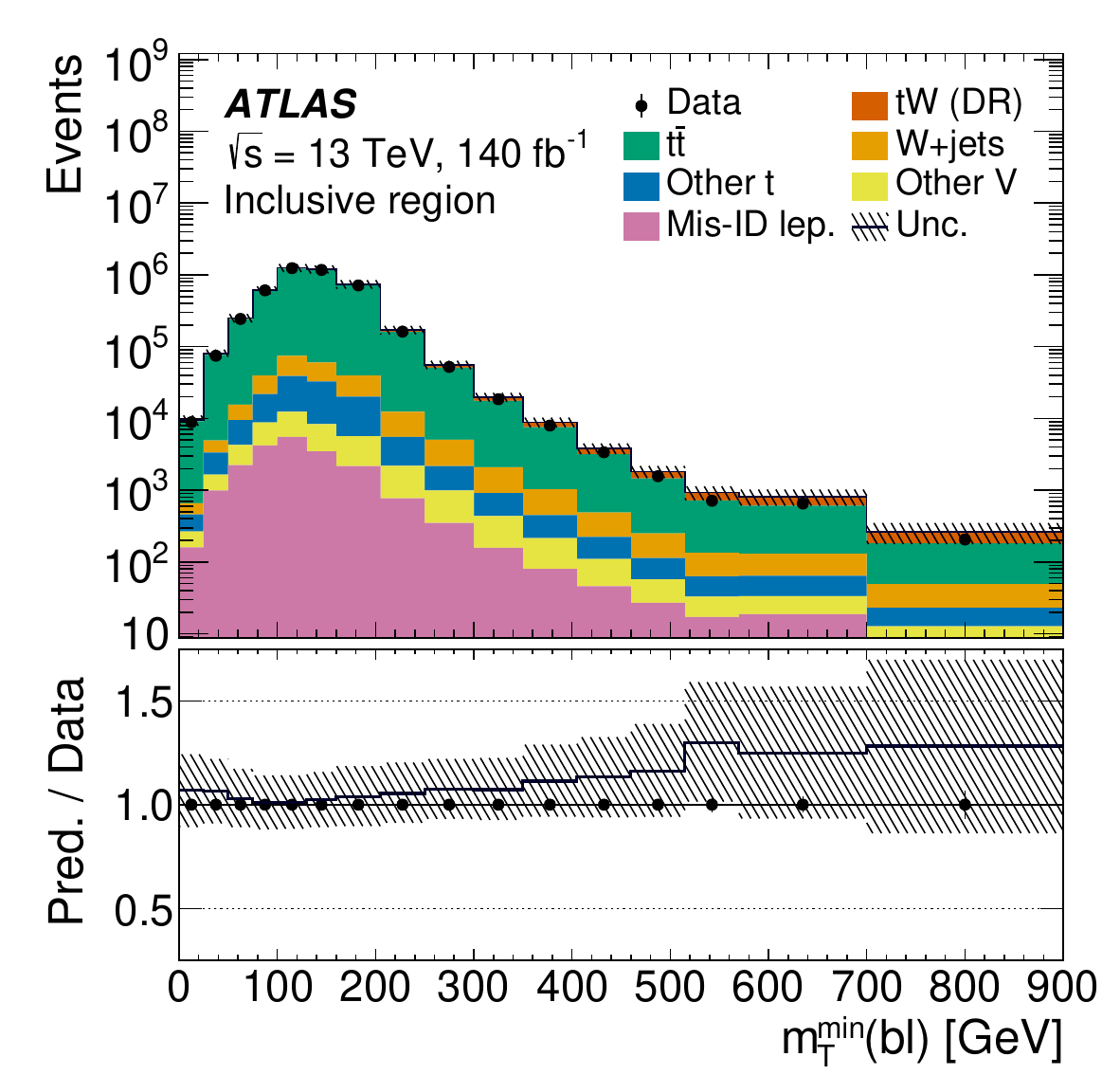}%
\label{fig:IR:mtblmin:reco}
}
\subfloat[]{
\includegraphics[width=0.37\textwidth]{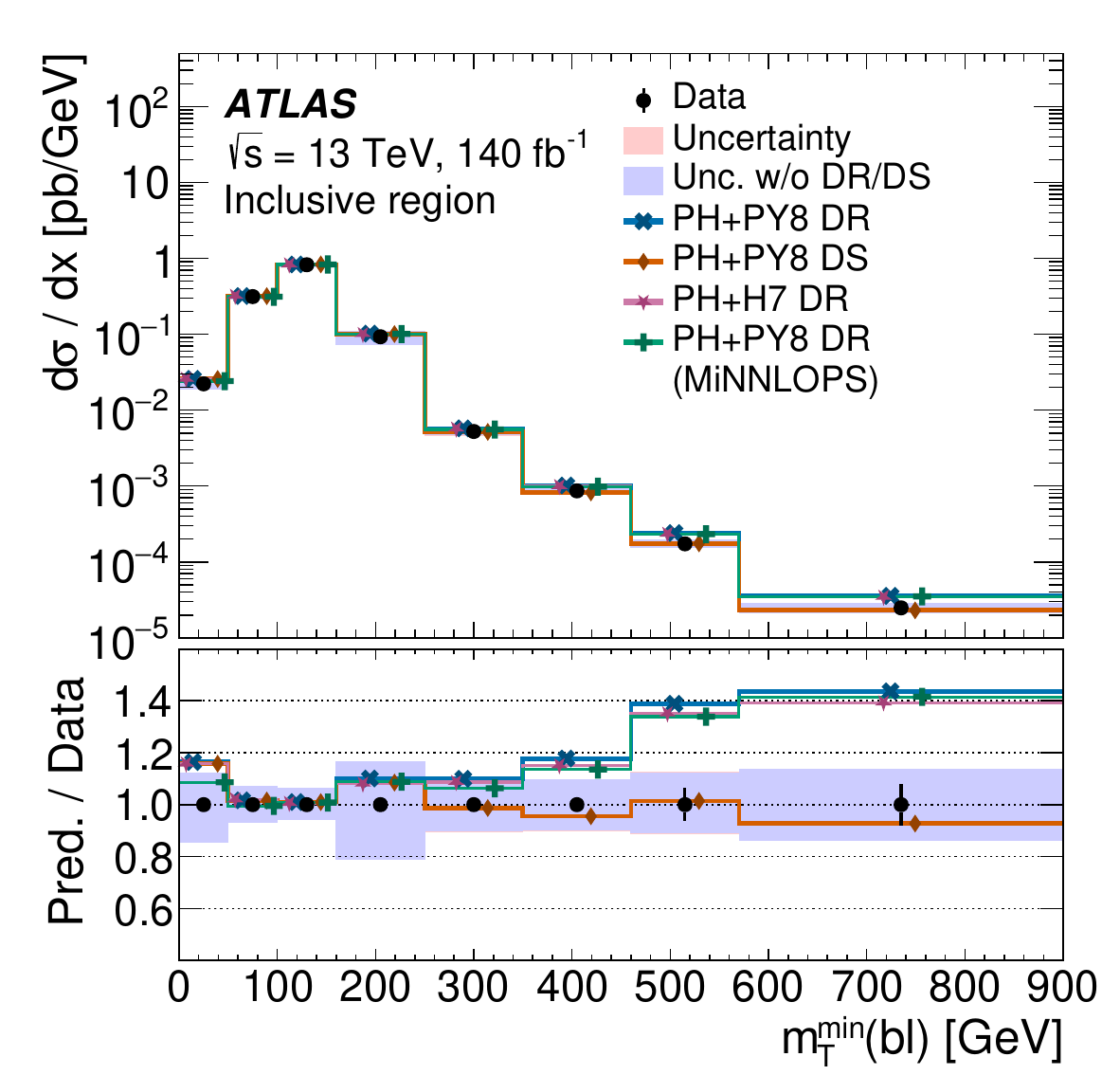}%
\label{fig:IR:mtblmin:unfolded}
}
\\
\subfloat[]{
\includegraphics[width=0.37\textwidth]{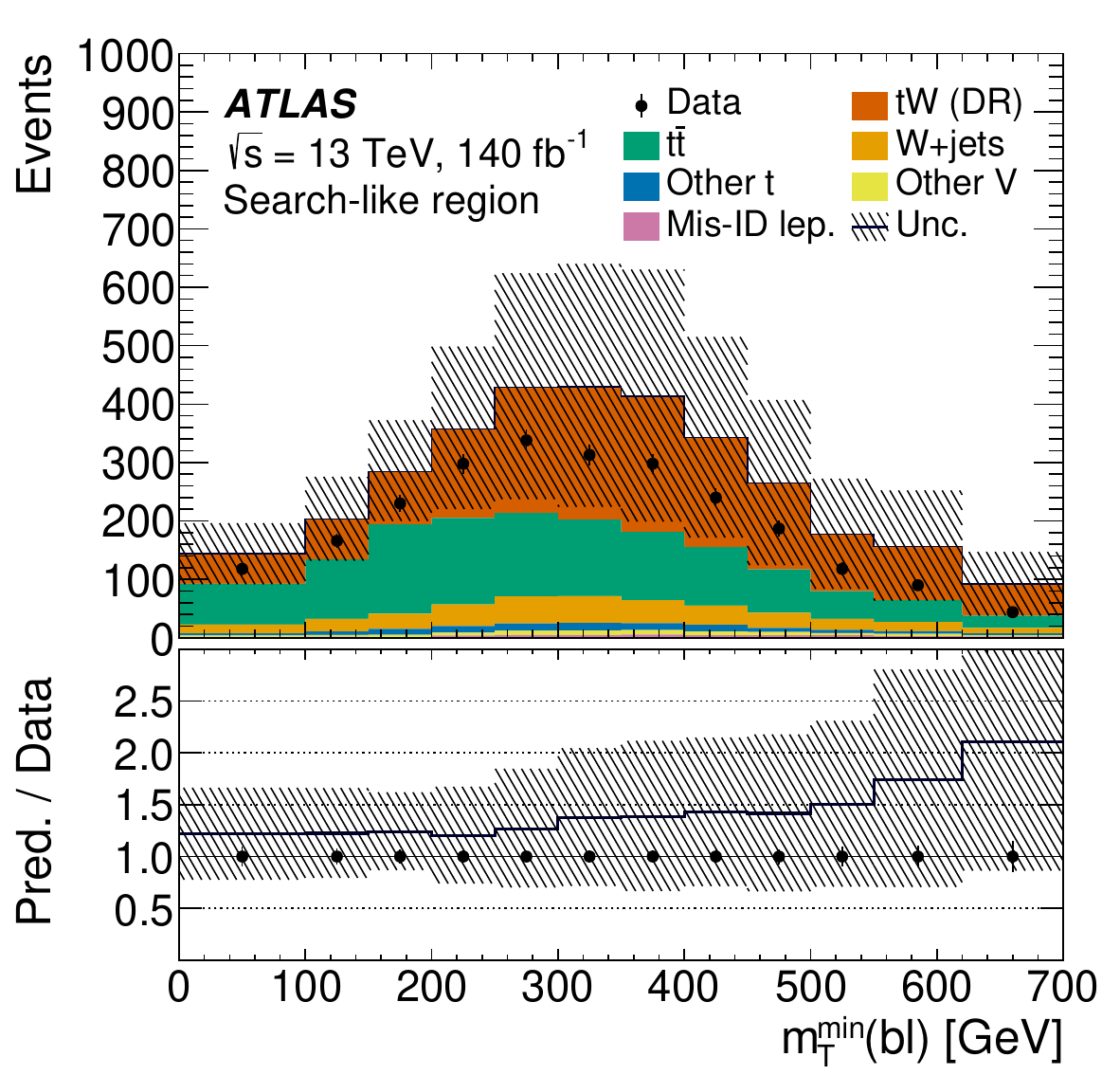}%
\label{fig:SLR:mtblmin:reco}
}
\subfloat[]{
\includegraphics[width=0.37\textwidth]{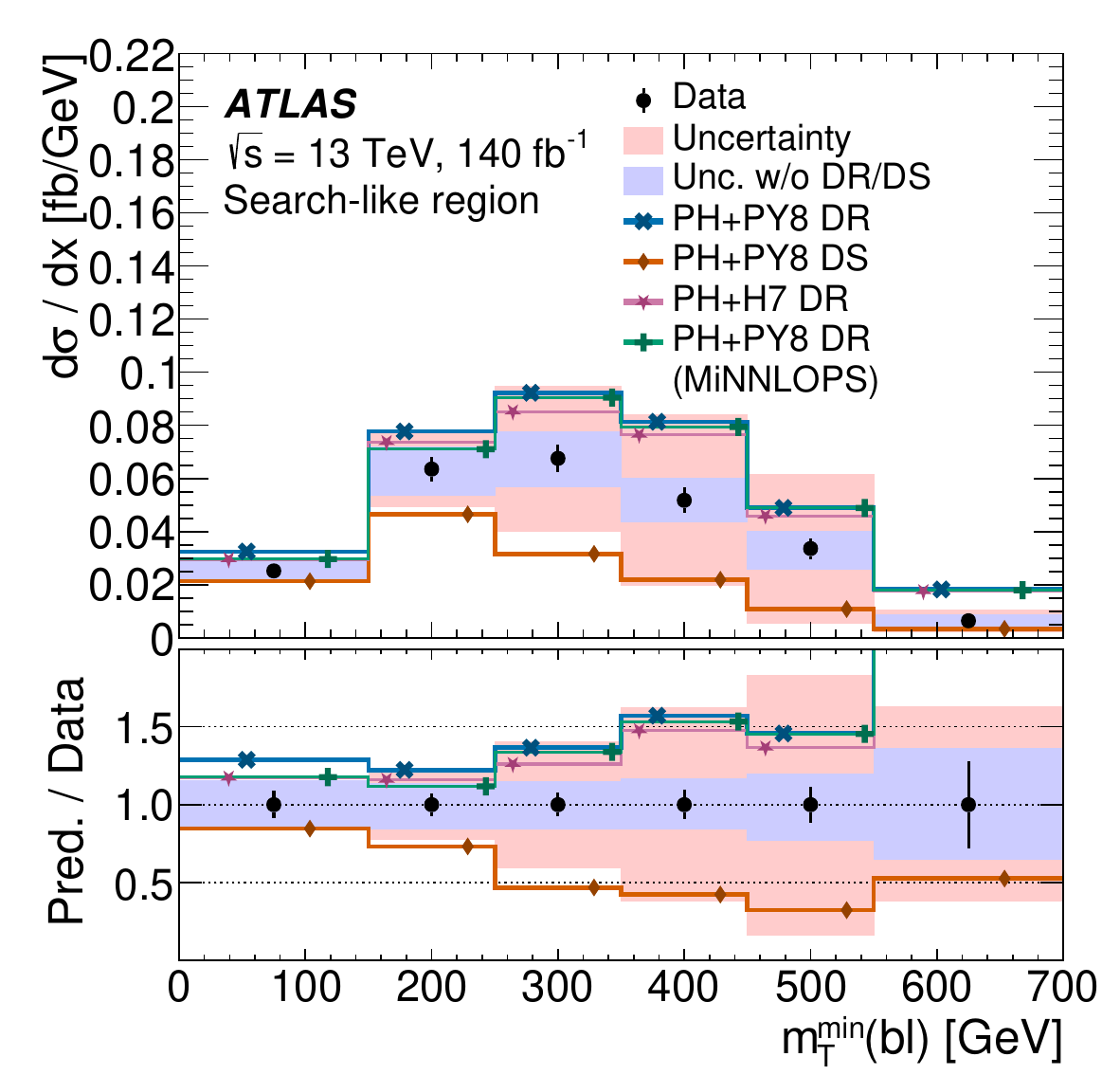}%
\label{fig:SLR:mtblmin:unfolded}
}
\caption{The (\protect\subref{fig:IR:mtblmin:reco}, \protect\subref{fig:SLR:mtblmin:reco}) detector-level distributions and
(\protect\subref{fig:IR:mtblmin:unfolded}, \protect\subref{fig:SLR:mtblmin:unfolded}) unfolded differential cross-sections
of the transverse component of the minimal transverse mass between the \met, the lepton and the \bjets in the event, \mtblmin.
The distributions are shown (\protect\subref{fig:IR:mtblmin:reco} and \protect\subref{fig:IR:mtblmin:unfolded}) in the inclusive region
and (\protect\subref{fig:SLR:mtblmin:reco} and \protect\subref{fig:SLR:mtblmin:unfolded}) in the search-like region.
\captionRatioRecoUnfolded
\captionSysBandExt
\captionUnfoldedComparison}%
\label{fig:results:mtblmin}
\end{figure}

The differential cross-sections are measured in the three measurement regions
(\IR, \WR and \SLR) for a set of kinematic variables. The variables are chosen
to probe the phase space where the interference effects between \ttbar and \tWb
processes are most prominent.
In the \WR region, the observables are chosen to represent all relevant kinematic
quantities of the \WbWb process.
The variables are defined at particle level
and are unfolded from the detector level using the procedure described
in the previous sections.
Selected unfolded differential cross-section results as a function of \mblmin and characteristic kinematic observables
for the three measurement regions are
presented in \cref{fig:results:mblmin,fig:results:pt}, together with their
detector-level distributions.
Cross-sections in the \IR and \SLR as a function of the  minimum transverse invariant mass between the \met, the lepton and the \bjets, \mtblmin,  and as a function of the transverse mass between the lepton and the \met, \mt,
are presented in \cref{fig:results:mtlep,fig:results:mtblmin}.

The observable presenting the largest overall disagreement between
the unfolded data and the simulation is \mblmin. This observable is
a proxy for the invariant mass of the leptonically decaying top quark and
is particularly sensitive to effects related to processes with virtual top
contributions in the \sampleB production. Detector-level and
particle-level distributions are presented in \cref{fig:results:mblmin}
for \IR, \WR and \SLR selections. The ratio panel of each figure
compares the reconstructed or unfolded data with the prediction from
simulation. The full set of systematic uncertainties affecting the unfolded data, arising from detector
and modelling sources, is presented as a shaded band.
In the detector-level results, only the nominal simulation prediction is shown,
which includes \tWb processes simulated with the DR approach.
Various unfolded \samplePlain differential cross-section predictions are shown in the figures.
The details of each prediction are given in \cref{subsec:simulation}.
They are: \PHPYDR (nominal prediction which includes the \tWb process with DR treatment),
\PHPYDS (prediction with the \tWb process simulated with DS treatment), MiNNLOPS (using NNLO QCD description for
the \ttbar process), and \PHHDR (using \HERWIG as shower model).

The unfolded results in \cref{fig:results:mblmin} show that none of the available predictions are well
suited to simultaneously describe all the kinematic properties of the phase space under investigation.
The various predictions exhibit a varying level of agreement in different observables and regions of the phase space.
The disagreement is particularly enhanced in those parts of the kinematic phase space that are not dominated exclusively
by doubly resonant top production, such as the
tails of the \mblmin observable for the \IR and \WR selection.
Additionally, a discrepancy is observed across the entire phase space and with a more pronounced deviation in the \SLR selection, because of the
interference-dominated phase space selected by this region. This discrepancy is about one standard deviation
if all systematic uncertainties are considered. The uncertainties in these selections are dominated by the modelling
of the singly resonant top quark production (DS/DR uncertainty%
).
Recent progress in event generators~\cite{Jezo:2023rht} could not yet be included in this work but will contribute to future studies of the modelling in this region.
The \PHPYDR and \PHPYDS predictions enclose the unfolded data distributions, with discrepancies at the level of approximately one standard deviation.
The unfolded results were found not to depend strongly on the choice of the nominal MC for unfolding.
The largest impact of this choice is on the magnitude of the out-of-acceptance corrections used in the unfolding, affecting the absolute cross-section.
As a result, when using the \PHPYDS predictions as the nominal MC for unfolding, the unfolded data moves away from the \PHPYDS prediction and closer to the \PHPYDR prediction.

The modelling of the lepton and the \ensuremath{b\text{-jets'}} momenta in the event crucially
characterises the phase space kinematic properties, as they are directly linked to the modelling of
the boost of the \tWb or \ttbar systems.
In the three representative distributions shown for the three measurement regions in \cref{fig:results:pt}
it can be observed that no prediction is within one sigma of the unfolded data in the \pt spectrum of the lepton
and the \bjets for highly energetic objects (\( \pt \gtrsim \qty{200}{\GeV} \)).
In the \IR  the discrepancy can be mitigated by \ttbar higher order
corrections (MiNNLOPS prediction) when doubly resonant top-quark production dominates
in the phase space. This is however not true everywhere, as for example the MiNNLOPS prediction exhibits slightly
worse modelling of the data than the nominal prediction for events with leading \bjet
momenta \( < \qty{200}{\GeV} \) in the \WR selection (\cref{fig:results:pt}\protect\subref{fig:WR:ptb1:unfolded}) and
the MiNNLOPS prediction is very similar to the nominal \PHPYDR prediction in the \SLR selection (\cref{fig:results:pt}\protect\subref{fig:SLR:ptb2:unfolded}).
Observables related to the invariant mass of the leptonic
\Wboson, like \mt, which are instead independent of the boost of the \tWb or \ttbar system, are
well modelled in the inclusive (\IR) phase space (see Figure~\ref{fig:results:mtlep}).
The \mtblmin observable is presented in \cref{fig:results:mtblmin}
for \IR and \SLR selections, as a proxy for the transverse component of
the invariant mass of the leptonically decaying top quark. Such an observable exploits the \met information
from the neutrino and is less sensitive to contaminations from the \Wjets
background in the  interference-enhanced phase space of the regions and therefore overall affected
by smaller systematic uncertainties. Despite the overlap between events in the
tail of the \mblmin and \mtblmin distributions in the inclusive
region (\cref{fig:results:mblmin}\protect\subref{fig:IR:mblmin:unfolded}
and \cref{fig:results:mtblmin}\protect\subref{fig:IR:mtblmin:unfolded}), the two distributions
exhibit opposite trends in terms of agreement between unfolded data and predictions.
The nominal \PHPYDR prediction models best the unfolded data for the \mblmin observable.
By contrast, the unfolded data for the \mtblmin observable is best described by the \PHPYDS prediction.
In the \SLR instead, the \PHPYDR and \PHPYDS predictions enclose the unfolded data distributions in similar
ways across the \mblmin and \mtblmin observables, with differences of about one standard deviation.

\begin{figure}[htbp]
\centering
\subfloat[]{
\includegraphics[width=0.37\textwidth]{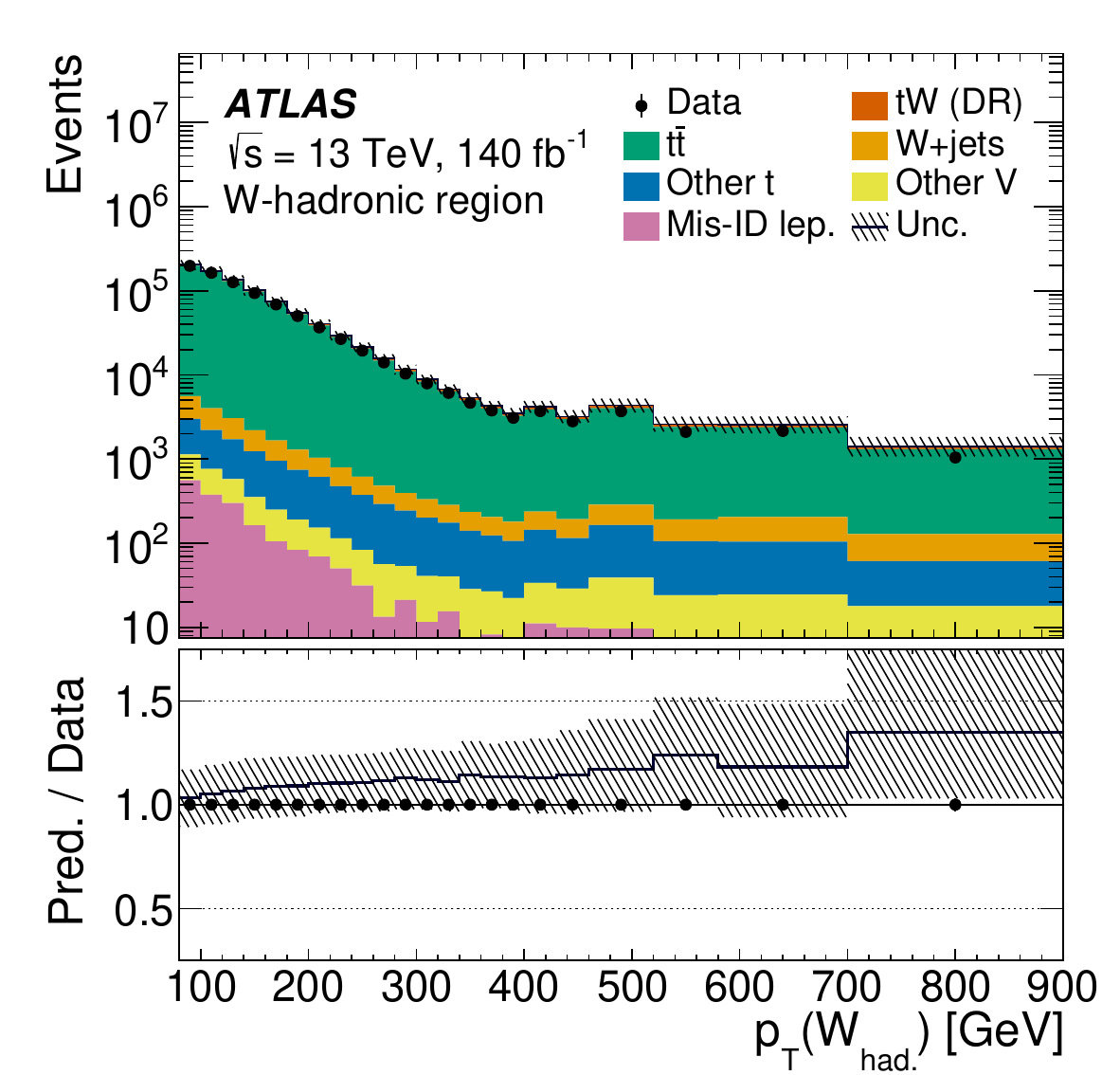}%
\label{fig:WR:ptwhad:reco}
}
\subfloat[]{
\includegraphics[width=0.37\textwidth]{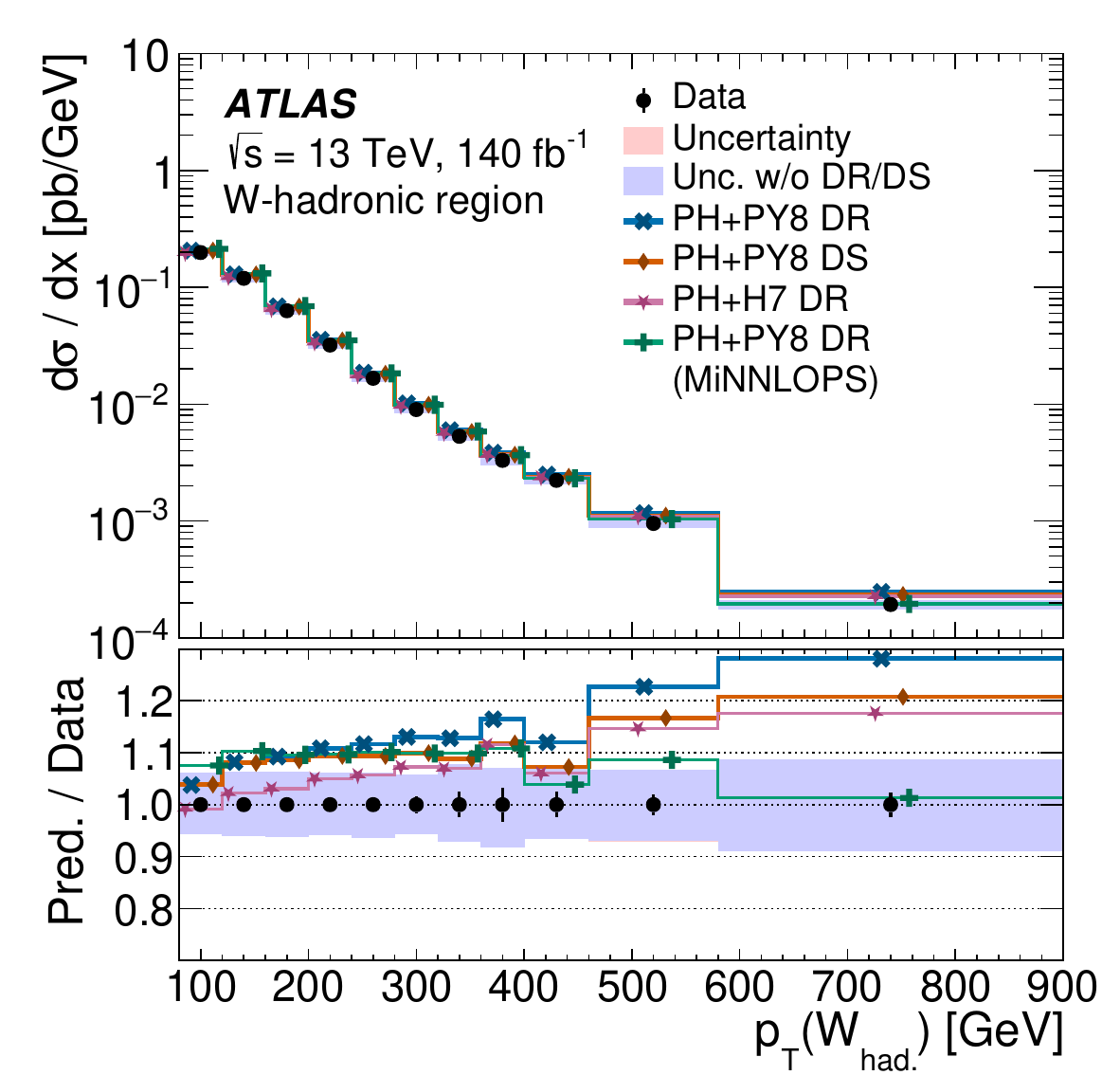}%
\label{fig:WR:ptwhad:unfolded}
}
\\
\subfloat[]{
\includegraphics[width=0.37\textwidth]{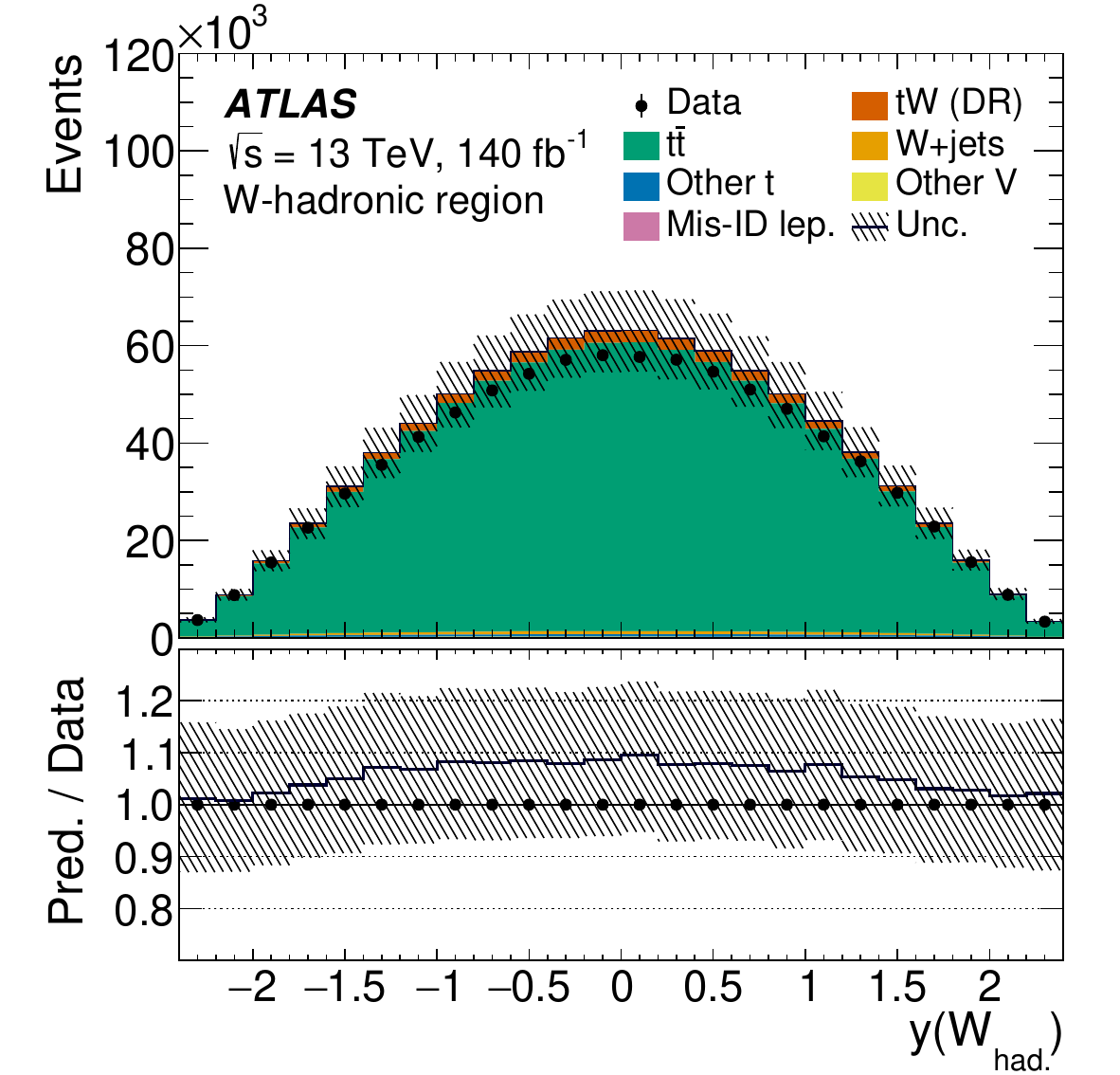}%
\label{fig:WR:ywhad:reco}
}
\subfloat[]{
\includegraphics[width=0.37\textwidth]{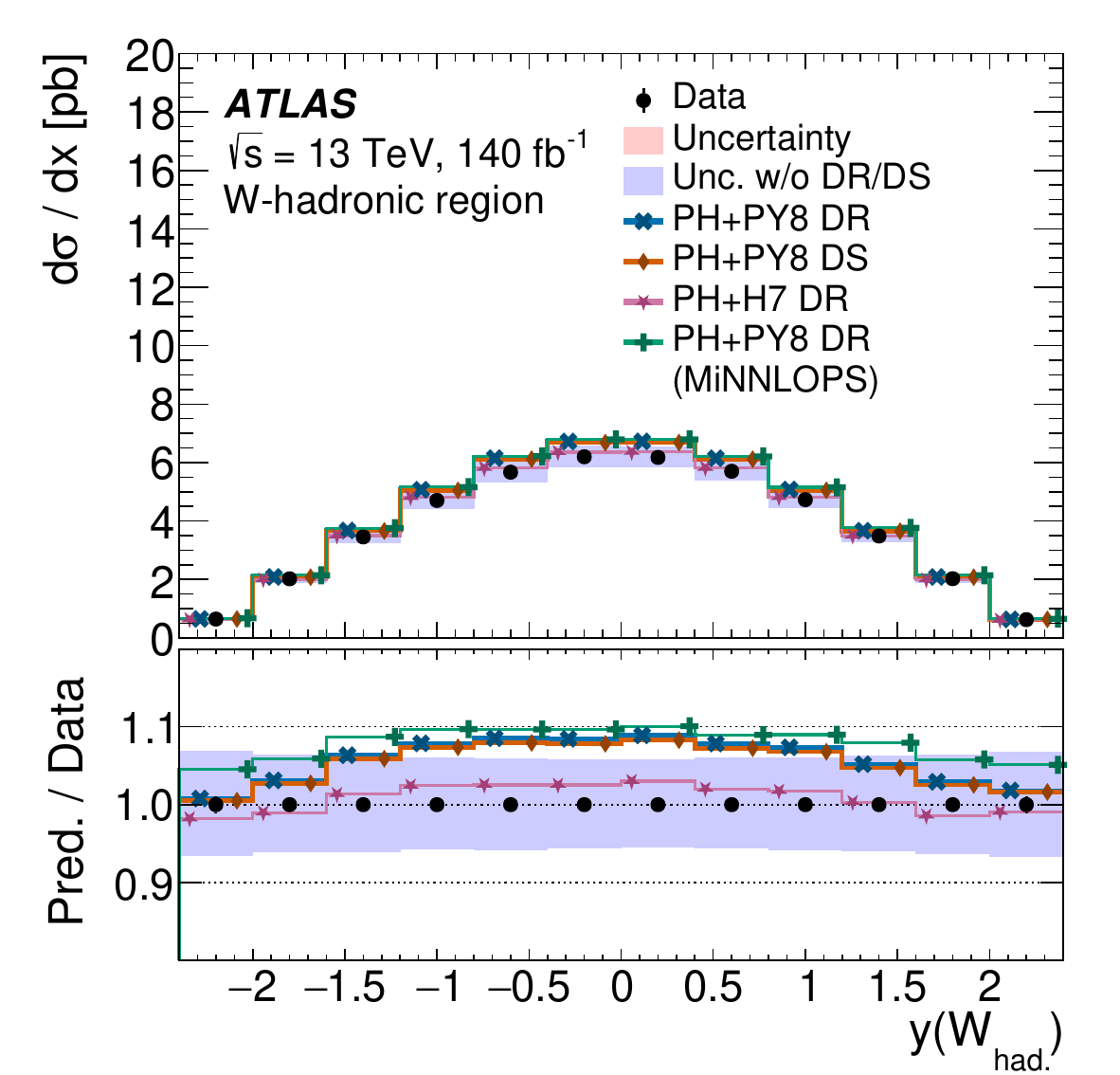}%
\label{fig:WR:ywhad:unfolded}
}
\caption{The (\protect\subref{fig:WR:ptwhad:reco}, \protect\subref{fig:WR:ywhad:reco}) detector-level distributions and
(\protect\subref{fig:WR:ptwhad:unfolded}, \protect\subref{fig:WR:ywhad:unfolded}) unfolded differential cross-sections
in the \WRlong region.
The distributions are shown for (\protect\subref{fig:WR:ptwhad:reco} and \protect\subref{fig:WR:ptwhad:unfolded}) the hadronic \Wboson candidate momentum \ptW
and (\protect\subref{fig:WR:ywhad:reco} and \protect\subref{fig:WR:ywhad:unfolded}) rapidity \yW.
\captionRatioRecoUnfolded
\captionSysBandExt
\captionUnfoldedComparison}%
\label{fig:results:WR}
\end{figure}

\begin{figure}[htbp]
\centering
\subfloat[]{
\includegraphics[width=0.37\textwidth]{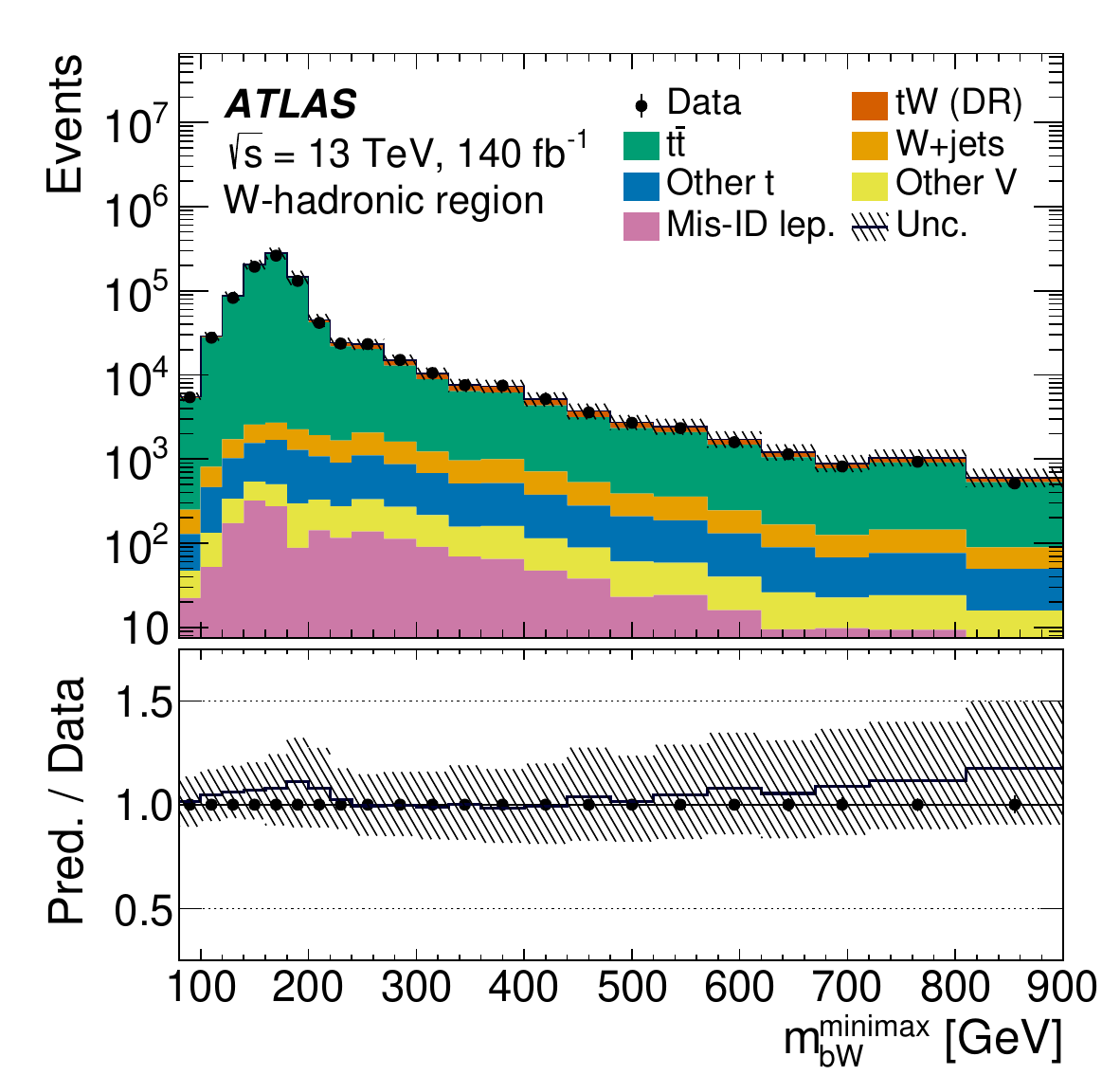}%
\label{fig:WR:minimax:reco}
}
\subfloat[]{
\includegraphics[width=0.37\textwidth]{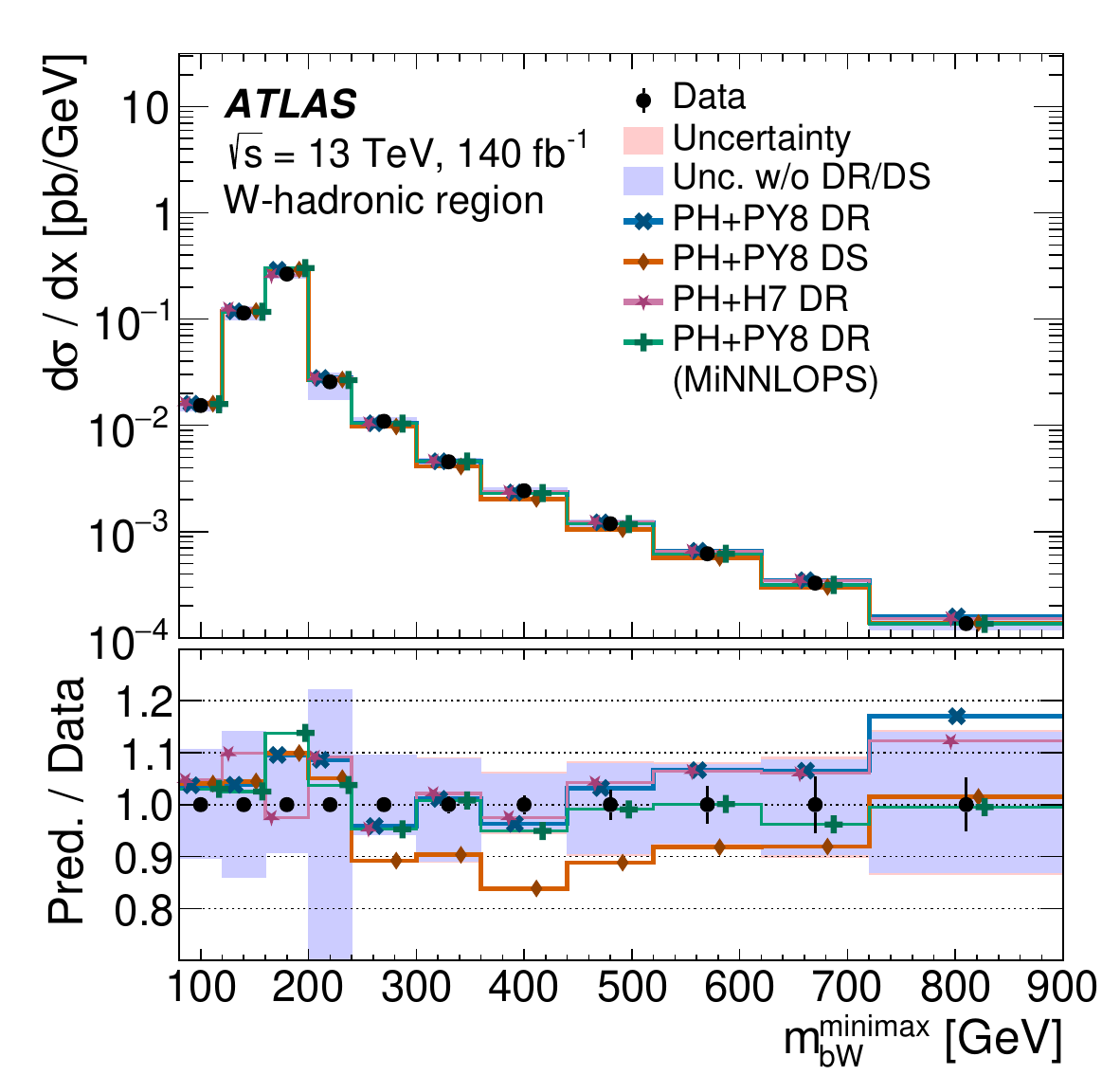}%
\label{fig:WR:minimax:unfolded}
}
\\
\subfloat[]{
\includegraphics[width=0.37\textwidth]{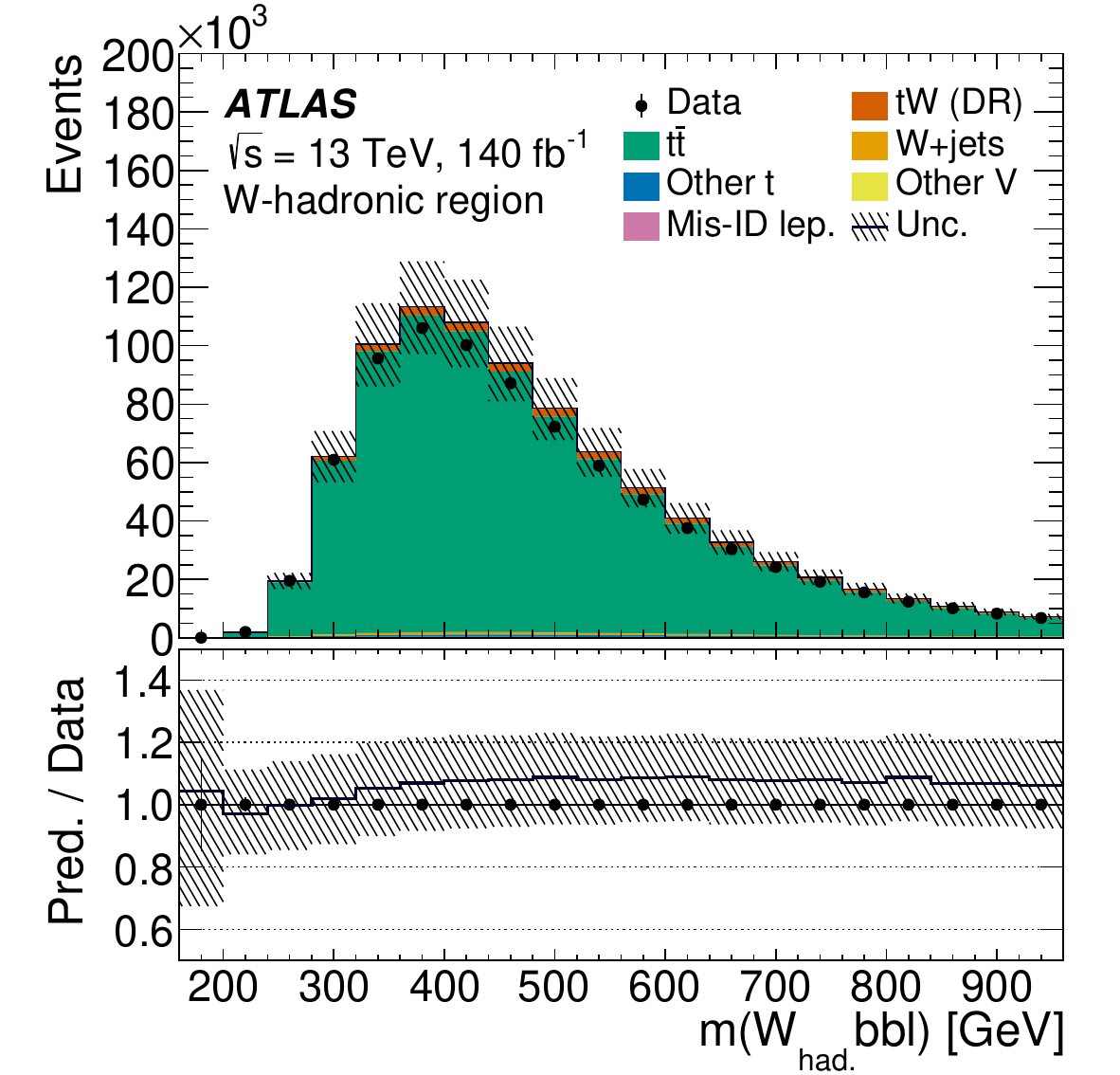}%
\label{fig:WR:mwhadbbl:reco}
}
\subfloat[]{
\includegraphics[width=0.37\textwidth]{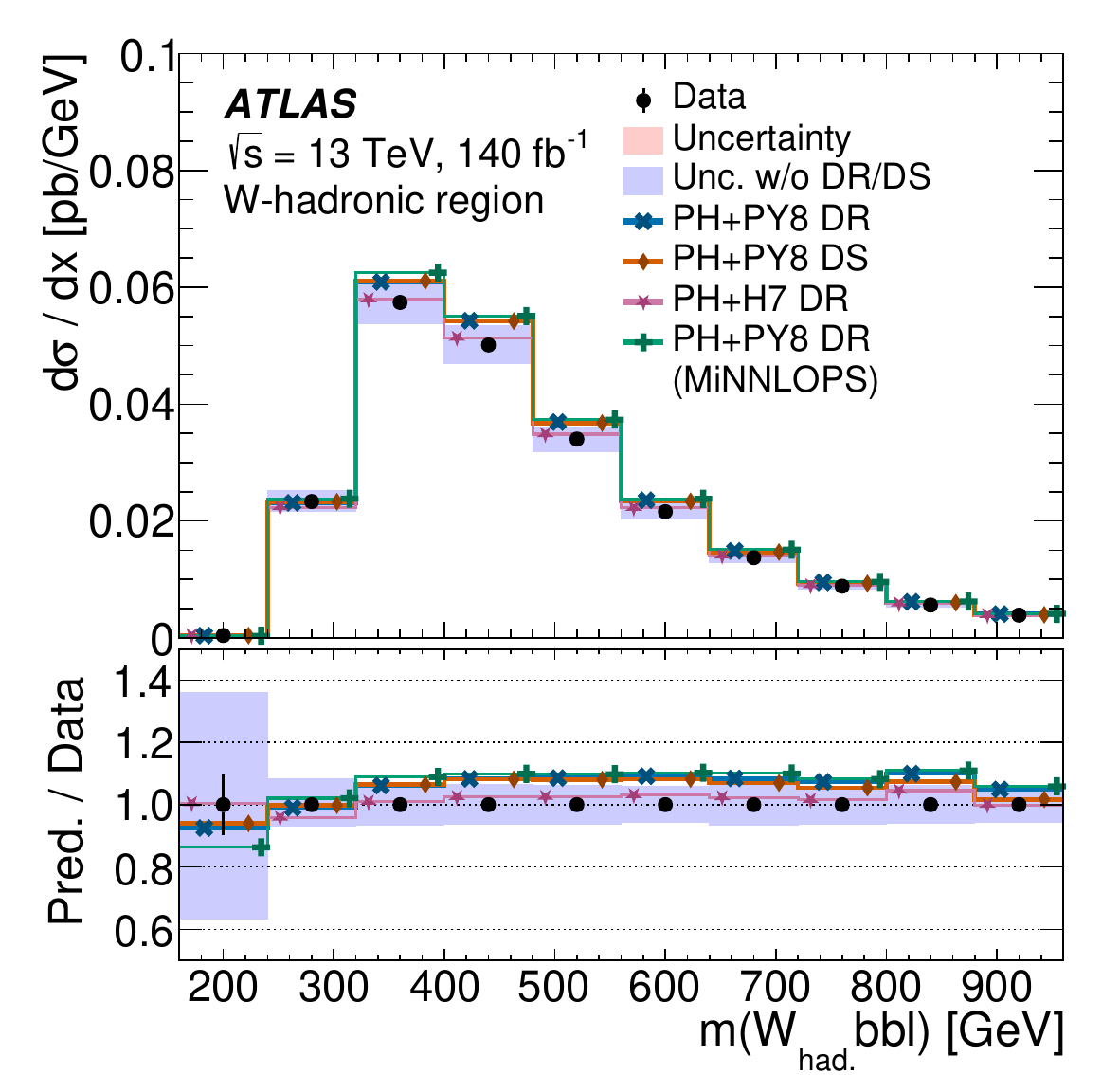}%
\label{fig:WR:mwhadbbl:unfolded}
}

\caption{The (\protect\subref{fig:WR:minimax:reco}, \protect\subref{fig:WR:mwhadbbl:reco}) detector-level distributions and
(\protect\subref{fig:WR:minimax:unfolded}, \protect\subref{fig:WR:mwhadbbl:unfolded}) unfolded differential cross-sections in the \WRlong region.
The distributions are shown for (\protect\subref{fig:WR:minimax:reco} and \protect\subref{fig:WR:minimax:unfolded}) \mbWminimax
and (\protect\subref{fig:WR:mwhadbbl:reco} and \protect\subref{fig:WR:mwhadbbl:unfolded}) the invariant mass of the \( Wbb\ell \) system.
\captionRatioRecoUnfolded
\captionSysBandExt
\captionUnfoldedComparison}%
\label{fig:results:WR2}
\end{figure}

\cref{fig:results:WR} presents the transverse momentum and rapidity of the hadronic W candidate in the
\WR\ selection. None of the prediction shows a good modelling of all the unfolded data, with the exception of
the alternative \PHHDR model. The discrepancy between the predictions and the unfolded data is particularly
enhanced for large \Wboson boson transverse momenta and central rapidity. The MiNNLOPS prediction exhibits
particularly large disagreements, especially in the overall normalisation in this selection.
The \mbWminimax and \mWbbl  observables are presented in \cref{fig:results:WR2}. The former allows to
probe the transition region of  the relative abundance of singly versus doubly resonant top-quark
processes (at the kinematic endpoint \(\mbWminimax < \sqrt{m_t^2 - m_W^2}\)). The latter allows to study
the modelling of the entire \WbWb system. Also in this case, the alternative \PHHDR model seems to provide the
best description of the unfolded data, while \PHPYDS exhibits a worse modelling, especially in the
region that is most sensitive to the interference effects between \ttbar and \tWb processes.

The compatibility of the unfolded distributions and the predictions
is quantitatively assessed by calculating \(\chi^2\) values, as presented in \crefrange{tab:results:IR:chi2}{tab:results:SLR:chi2}.
All uncertainties (statistical and systematic, both symmetrised) associated with the unfolded distributions,
described in \cref{subsec:systematics} are included, taking into account their correlations,
with no additional uncertainties included on the predictions under test.

The corresponding \(p\)-values are calculated inclusively for each distribution and reflect
the various level of disagreement observed between the unfolded data and the simulation
in the various observables and regions studied in this paper.

\begin{table}
\centering
\caption{\(\chi^2\) and \(p\)-values for each variable (rows) for each
prediction (columns) for the absolute differential cross-section in the \IR.
All systematic uncertainties described in \cref{subsec:systematics} are included,
with no additional uncertainties included on the predictions under test.}%
\label{tab:results:IR:chi2}
\resizebox{\textwidth}{!}{%

\bgroup
\hfuzz=6.2pt
\begin{tabular}{%
c|
S[table-format=3.2, round-mode=places, round-precision=2]@{\: / \:} S[table-format=2.0] S[table-format=2.2, round-mode=places, round-precision=2, round-minimum=0.01] |
S[table-format=3.2, round-mode=places, round-precision=2]@{\: / \:} S[table-format=2.0] S[table-format=2.2, round-mode=places, round-precision=2, round-minimum=0.01] |
S[table-format=3.2, round-mode=places, round-precision=2]@{\: / \:} S[table-format=2.0] S[table-format=2.2, round-mode=places, round-precision=2, round-minimum=0.01] |
S[table-format=3.2, round-mode=places, round-precision=2]@{\: / \:} S[table-format=2.0] S[table-format=2.2, round-mode=places, round-precision=2, round-minimum=0.01]
}
\toprule

\textbf{Inclusive region}            &
\multicolumn{3}{c|}{\PHPYDR}         &
\multicolumn{3}{c|}{\PHPYDS}         &
\multicolumn{3}{c|}{\PHHDR}          &
\multicolumn{3}{c}{\PHPYDR (MiNNLOPS)}                                                                                                                    \\

\midrule

Observable                           &
\multicolumn{2}{c}{\(\chi^2\) / NDF} & {\(p\)-value} &
\multicolumn{2}{c}{\(\chi^2\) / NDF} & {\(p\)-value} &
\multicolumn{2}{c}{\(\chi^2\) / NDF} & {\(p\)-value} &
\multicolumn{2}{c}{\(\chi^2\) / NDF} & {\(p\)-value}                                                                                                      \\

\midrule

\mblmin                              & 7.26598       & 9  & 0.609448    & 10.1545 & 9  & 0.338124  & 5.99588 & 9  & 0.74033   & 23.7228 & 9  & 0.00476182 \\
\( \pt(\ell) \)                      & 31.6705       & 11 & 0.000861307 & 19.7018 & 11 & 0.0496012 & 13.1069 & 11 & 0.286398  & 7.54228 & 11 & 0.753624   \\
\mtlep                               & 11.196        & 8  & 0.190837    & 7.90864 & 8  & 0.442445  & 12.1294 & 8  & 0.145523  & 3.68198 & 8  & 0.884617   \\
\mtblmin                             & 21.168        & 8  & 0.00671427  & 7.49451 & 8  & 0.484335  & 16.5413 & 8  & 0.0352563 & 17.3389 & 8  & 0.0267673  \\

\bottomrule
\end{tabular}
\egroup


}
\end{table}

\begin{table}
\centering
\caption{\(\chi^2\) and \(p\)-values for each variable (rows) for each
prediction (columns) for the absolute differential cross-section in the \WR.
All systematic uncertainties described in \cref{subsec:systematics} are included,
with no additional uncertainties included on the predictions under test.}%
\label{tab:results:WR:chi2}
\resizebox{\textwidth}{!}{%

\bgroup
\hfuzz=6.2pt
\begin{tabular}{%
c|
S[table-format=3.2, round-mode=places, round-precision=2]@{\: / \:} S[table-format=2.0] S[table-format=2.2, round-mode=places, round-precision=2, round-minimum=0.01] |
S[table-format=3.2, round-mode=places, round-precision=2]@{\: / \:} S[table-format=2.0] S[table-format=2.2, round-mode=places, round-precision=2, round-minimum=0.01] |
S[table-format=3.2, round-mode=places, round-precision=2]@{\: / \:} S[table-format=2.0] S[table-format=2.2, round-mode=places, round-precision=2, round-minimum=0.01] |
S[table-format=3.2, round-mode=places, round-precision=2]@{\: / \:} S[table-format=2.0] S[table-format=2.2, round-mode=places, round-precision=2, round-minimum=0.01]
}
\toprule

\textbf{\WRlong region}              &
\multicolumn{3}{c|}{\PHPYDR}         &
\multicolumn{3}{c|}{\PHPYDS}         &
\multicolumn{3}{c|}{\PHHDR}          &
\multicolumn{3}{c}{\PHPYDR (MiNNLOPS)}                                                                                                                   \\

\midrule

Observable                           &
\multicolumn{2}{c}{\(\chi^2\) / NDF} & {\(p\)-value} &
\multicolumn{2}{c}{\(\chi^2\) / NDF} & {\(p\)-value} &
\multicolumn{2}{c}{\(\chi^2\) / NDF} & {\(p\)-value} &
\multicolumn{2}{c}{\(\chi^2\) / NDF} & {\(p\)-value}                                                                                                     \\

\midrule

\mblselected                         & 15.6475       & 10 & 0.110178  & 26.8389 & 10 & 0.00276155 & 6.95079 & 10 & 0.730083 & 30.5309 & 10 & 0.000701024 \\
\( \pt(b_1)\)                        & 25.4971       & 9  & 0.0024678 & 20.5649 & 9  & 0.0147288  & 10.1871 & 9  & 0.335548 & 13.6197 & 9  & 0.136509    \\
\ptW                                 & 20.7901       & 11 & 0.035612  & 14.3157 & 11 & 0.216012   & 10.9906 & 11 & 0.444048 & 7.28061 & 11 & 0.775925    \\
\yW                                  & 14.8128       & 12 & 0.251835  & 14.1345 & 12 & 0.292204   & 7.95907 & 12 & 0.78832  & 11.9939 & 12 & 0.446168    \\
\mbWminimax                          & 22.3745       & 11 & 0.021627  & 44.2412 & 11 & 0.001      & 12.7152 & 11 & 0.312349 & 34.3254 & 11 & 0.001       \\
\mWbbl                               & 5.36793       & 10 & 0.865285  & 7.83251 & 10 & 0.645193   & 2.07505 & 10 & 0.995732 & 4.53726 & 10 & 0.919874    \\

\bottomrule
\end{tabular}
\egroup


}
\end{table}
\begin{table}
\centering
\caption{\(\chi^2\) and \(p\)-values for each variable (rows) for each
prediction (columns) for the absolute differential cross-section in the \SLR.
All systematic uncertainties described in \cref{subsec:systematics} are included,
with no additional uncertainties included on the predictions under test.}%
\label{tab:results:SLR:chi2}
\resizebox{\textwidth}{!}{%

\bgroup
\hfuzz=6.2pt
\begin{tabular}{%
c|
S[table-format=3.2, round-mode=places, round-precision=2]@{\: / \:} S[table-format=2.0] S[table-format=2.2, round-mode=places, round-precision=2, round-minimum=0.01] |
S[table-format=3.2, round-mode=places, round-precision=2]@{\: / \:} S[table-format=2.0] S[table-format=2.2, round-mode=places, round-precision=2, round-minimum=0.01] |
S[table-format=3.2, round-mode=places, round-precision=2]@{\: / \:} S[table-format=2.0] S[table-format=2.2, round-mode=places, round-precision=2, round-minimum=0.01] |
S[table-format=3.2, round-mode=places, round-precision=2]@{\: / \:} S[table-format=2.0] S[table-format=2.2, round-mode=places, round-precision=2, round-minimum=0.01]
}
\toprule

\textbf{Search-like region}          &
\multicolumn{3}{c|}{\PHPYDR}         &
\multicolumn{3}{c|}{\PHPYDS}         &
\multicolumn{3}{c|}{\PHHDR}          &
\multicolumn{3}{c}{\PHPYDR (MiNNLOPS)}                                                                                                                    \\

\midrule

Observable                           &
\multicolumn{2}{c}{\(\chi^2\) / NDF} & {\(p\)-value} &
\multicolumn{2}{c}{\(\chi^2\) / NDF} & {\(p\)-value} &
\multicolumn{2}{c}{\(\chi^2\) / NDF} & {\(p\)-value} &
\multicolumn{2}{c}{\(\chi^2\) / NDF} & {\(p\)-value}                                                                                                 \\

\midrule

\mblmin                              & 9.5685        & 4 & 0.0483586  & 7.57456 & 4 & 0.108466 & 9.20547  & 4 & 0.0561639 & 10.7282  & 4 & 0.0297947 \\
\(\pt(b_2)\)                         & 17.1077       & 5 & 0.00430009 & 4.12235 & 5 & 0.531938 & 14.7153  & 5 & 0.0116506 & 13.0025  & 5 & 0.0233554 \\
\mtlep                               & 0.413232      & 4 & 0.981379   & 4.23575 & 4 & 0.37504  & 0.302684 & 4 & 0.98964   & 0.583059 & 4 & 0.964928  \\
\mtblmin                             & 11.5386       & 6 & 0.0730889  & 2.76441 & 6 & 0.837782 & 10.0129  & 6 & 0.124109  & 10.4483  & 6 & 0.106997  \\

\bottomrule
\end{tabular}
\egroup


}
\end{table}


\FloatBarrier

\section{Conclusions}%
\label{sec:conclusions}

This paper presented a differential cross-section measurement of the
coherent production of singly and doubly resonant top quarks in \WbWb events
using \qty{140}{\ifb} of \pp collisions at
a center-of-mass energy of \qty{13}{\TeV}, collected with the ATLAS experiment
in the years from 2015 to 2018 during Run 2 of the LHC.\@
Events requiring one lepton, three or four jets, of which at least two \bjets, were considered.
Three measurement regions were defined: the inclusive region, the \WRlong region,
and the search-like region.
The analysis selections, fiducial volumes and observables were optimised to isolate a
phase space that probes the kinematic properties of the \sampleB
processes and their interference, actively suppressing the pure
doubly resonant top-quark production.

Differential cross-sections obtained at particle-level from unfolded data
are derived for various observables.
The results are compared with predictions from several simulations,
including the full set of uncertainties affecting the unfolded data and no additional systematic uncertainties in the predictions.
The results indicate that none of the considered
predictions can simultaneously accurately describe all observables in the fiducial phase-space.
For most of the phase space, observables connected to the transverse momentum of
the \tW/\ttbar system are described better by the
MiNNLOPS prediction, which describes \ttbar processes at NNLO QCD accuracy.
These observations are compatible with the findings of other
recent analyses targeting dileptonic final states~\cite{TOPQ-2020-14,TOPQ-2024-12}.

The measurement is limited by the uncertainties in the signal modelling,
in particular those related to the treatment of the \sampleB interference,
uncertainties related to the calibration of jets and \bjets.
When comparing with the measurement performed in the dilepton channel~\cite{TOPQ-2020-14}, a
larger number of observables sensitive to interference effects has been studied.
The overall findings are consistent, with discrepancies between data and predictions at the
level of one standard deviation when the comparison between the DS and DR removal schemes
is included as modelling uncertainty, increasing to two to three standard deviations when excluding it.
Furthermore, statistical uncertainties in the single-lepton measurement are significant
only in the tails of the distributions.
This measurement represents the first differential cross-section measurement
of the coherent production of singly and doubly resonant top-quarks in final states with exactly one lepton.
The results of this study provide valuable insights into the modelling
of the \WbWb process in the fiducial phase-space and can be used to assess the modelling accuracy
of future Monte Carlo simulation developments.
%
%
%


%
\section*{Acknowledgements}
%

%
%

%
%

We thank CERN for the very successful operation of the LHC and its injectors, as well as the support staff at
CERN and at our institutions worldwide without whom ATLAS could not be operated efficiently.

The crucial computing support from all WLCG partners is acknowledged gratefully, in particular from CERN, the ATLAS Tier-1 facilities at TRIUMF/SFU (Canada), NDGF (Denmark, Norway, Sweden), CC-IN2P3 (France), KIT/GridKA (Germany), INFN-CNAF (Italy), NL-T1 (Netherlands), PIC (Spain), RAL (UK) and BNL (USA), the Tier-2 facilities worldwide and large non-WLCG resource providers. Major contributors of computing resources are listed in Ref.~\cite{ATL-SOFT-PUB-2026-001}.

We gratefully acknowledge the support of ANPCyT, Argentina; YerPhI, Armenia; ARC, Australia; BMWFW and FWF, Austria; ANAS, Azerbaijan; CNPq and FAPESP, Brazil; NSERC, NRC and CFI, Canada; CERN; ANID, Chile; CAS, MOST and NSFC, China; Minciencias, Colombia; MEYS CR, Czech Republic; DNRF and DNSRC, Denmark; IN2P3-CNRS and CEA-DRF/IRFU, France; SRNSFG, Georgia; BMFTR, HGF and MPG, Germany; GSRI, Greece; RGC and Hong Kong SAR, China; ICHEP and Academy of Sciences and Humanities, Israel; INFN, Italy; MEXT and JSPS, Japan; CNRST, Morocco; NWO, Netherlands; RCN, Norway; MNiSW, Poland; FCT, Portugal; MNE/IFA, Romania; MSTDI, Serbia; MSSR, Slovakia; ARIS and MVZI, Slovenia; DSI/NRF, South Africa; MICIU/AEI, Spain; SRC and Wallenberg Foundation, Sweden; SERI, SNSF and Cantons of Bern and Geneva, Switzerland; NSTC, Taipei; TENMAK, T\"urkiye; STFC/UKRI, United Kingdom; DOE and NSF, United States of America.

Individual groups and members have received support from BCKDF, CANARIE, CRC and DRAC, Canada; CERN-CZ, FORTE and PRIMUS, Czech Republic; COST, ERC, ERDF, Horizon 2020, ICSC-NextGenerationEU and Marie Sk{\l}odowska-Curie Actions, European Union; Investissements d'Avenir Labex, Investissements d'Avenir Idex and ANR, France; DFG and AvH Foundation, Germany; Herakleitos, Thales and Aristeia programmes co-financed by EU-ESF and the Greek NSRF, Greece; BSF-NSF and MINERVA, Israel; NCN and NAWA, Poland; La Caixa Banking Foundation, CERCA and AGAUR programs from Generalitat de Catalunya and PROMETEO and GenT Programmes Generalitat Valenciana, Spain; G\"{o}ran Gustafssons Stiftelse, Sweden; The Royal Society and Leverhulme Trust, United Kingdom; Eric and Wendy Schmidt Fund for Strategic Innovation, United States of America.

In addition, individual members wish to acknowledge support from CERN: European Organization for Nuclear Research (CERN DOCT); Chile: Agencia Nacional de Investigaci\'on y Desarrollo (ANID FONDECYT reg. 1230987, FONDECYT 1230812, FONDECYT 1240864, Fondecyt 3240661, Fondecyt Regular 1240721); China: Chinese Ministry of Science and Technology (MOST-2023YFA1605700, MOST-2023YFA1609300), National Natural Science Foundation of China (NSFC - 12175119, NSFC 12275265); Czech Republic: Czech Science Foundation (GACR - 24-11373S), Ministry of Education Youth and Sports (ERC-CZ-LL2327, FORTE CZ.02.01.01/00/22\_008/0004632), PRIMUS Research Programme (PRIMUS/21/SCI/017); EU: H2020 European Research Council (ERC - 101002463); European Union: European Research Council (BARD No. 101116429, ERC - 948254, ERC 101089007), European Regional Development Fund (HE COFUND GA No.101081355, ERDF), European Union, Future Artificial Intelligence Research (FAIR-NextGenerationEU PE00000013), Italian Center for High Performance Computing, Big Data and Quantum Computing (ICSC, NextGenerationEU), Marie Sklodowska-Curie Actions (ERC Horizon-2020, GA 956086, GAP-101168829); France: Agence Nationale de la Recherche (ANR-21-CE31-0013, ANR-21-CE31-0022, ANR-22-EDIR-0002, ANR-24-CE31-0504-01); Germany: Deutsche Forschungsgemeinschaft (DFG - 469666862, DFG - CR 312/5-2); China: Research Grants Council (GRF); Italy: Istituto Nazionale di Fisica Nucleare (ICSC, NextGenerationEU), Ministero dell'Università e della Ricerca (NextGenEU 153D23001490006 M4C2.1.1, NextGenEU I53D23000820006 M4C2.1.1, NextGenEU I53D23001490006 M4C2.1.1, SOE2024\_0000023); Japan: Japan Society for the Promotion of Science (JSPS KAKENHI  JP25H0063, JSPS KAKENHI JP22H01227, JSPS KAKENHI JP22H04944, JSPS KAKENHI JP22KK0227, JSPS KAKENHI JP24K23939, JSPS KAKENHI JP24KK0251, JSPS KAKENHI JP25H00650, JSPS KAKENHI JP25H01291, JSPS KAKENHI JP25K01011, JSPS KAKENHI JP25K01023); Norway: Research Council of Norway (RCN-314472); Poland: Ministry of Science and Higher Education (IDUB AGH, POB8, D4 no 9722), Polish National Science Centre (NCN 2021/42/E/ST2/00350, NCN OPUS 2023/51/B/ST2/02507, NCN OPUS nr 2022/47/B/ST2/03059, NCN UMO-2019/34/E/ST2/00393, UMO-2022/47/O/ST2/00148, UMO-2023/49/B/ST2/04085, UMO-2023/51/B/ST2/00920, UMO-2024/53/N/ST2/00869); Portugal: Foundation for Science and Technology (FCT); Spain: Agència de Gestió d'Ajuts Universitaris i de Recerca. (AGAUR - 2023 BP 00141), Generalitat Valenciana (ASFAE/2022/008), Ministry of Science and Innovation (MCIN \& NextGenEU PCI2022-135018-2, MICIN \& FEDER PID2021-125273NB, RYC2019-028510-I, RYC2020-030254-I, RYC2021-031273-I, RYC2022-038164-I), Ministerio de Ciencia, Innovación y Universidades/Agencia Estatal de Investigaci\'on (EU NextGenerationEU (PRTR-C17.I1), PID2022-142604OB-C22); Sweden: Carl Trygger Foundation (Carl Trygger Foundation CTS 22:2312), Swedish Research Council (Swedish Research Council 2023-04654, VR 2021-03651, VR 2022-03845, VR 2022-04683, VR 2023-03403, VR 2024-05451), Knut and Alice Wallenberg Foundation (KAW 2018.0458, KAW 2022.0358, KAW 2023.0366); Switzerland: Swiss National Science Foundation (SNSF - PCEFP2\_194658); United Kingdom: Royal Society (NIF-R1-231091); United States of America: U.S. Department of Energy (ECA DE-AC02-76SF00515), Neubauer Family Foundation.

%
%


%
%
%
\printbibliography  %
\clearpage
 
\begin{flushleft}
\hypersetup{urlcolor=black}
{\Large The ATLAS Collaboration}

\bigskip

\AtlasOrcid[0000-0002-6665-4934]{G.~Aad}$^\textrm{\scriptsize 103}$,
\AtlasOrcid[0000-0001-7616-1554]{E.~Aakvaag}$^\textrm{\scriptsize 17}$,
\AtlasOrcid[0000-0002-5888-2734]{B.~Abbott}$^\textrm{\scriptsize 121}$,
\AtlasOrcid[0000-0002-0287-5869]{S.~Abdelhameed}$^\textrm{\scriptsize 84b}$,
\AtlasOrcid[0000-0002-1002-1652]{K.~Abeling}$^\textrm{\scriptsize 55}$,
\AtlasOrcid[0000-0001-5763-2760]{N.J.~Abicht}$^\textrm{\scriptsize 49}$,
\AtlasOrcid[0000-0002-8496-9294]{S.H.~Abidi}$^\textrm{\scriptsize 30}$,
\AtlasOrcid[0009-0003-6578-220X]{M.~Aboelela}$^\textrm{\scriptsize 45}$,
\AtlasOrcid[0000-0002-9987-2292]{A.~Aboulhorma}$^\textrm{\scriptsize 36e}$,
\AtlasOrcid[0000-0001-5329-6640]{H.~Abramowicz}$^\textrm{\scriptsize 155}$
% ...
\bigskip
\\

% ...
$^{17}$Department for Physics and Technology, University of Bergen, Bergen; Norway.\\
% ...
$^{30}$Physics Department, Brookhaven National Laboratory, Upton NY; United States of America.\\
% ...
$^{36}$$^{(a)}$Facult\'e des Sciences Ain Chock, Universit\'e Hassan II de Casablanca;$^{(b)}$Facult\'{e} des Sciences, Universit\'{e} Ibn-Tofail, K\'{e}nitra;$^{(c)}$Facult\'e des Sciences Semlalia, Universit\'e Cadi Ayyad, LPHEA-Marrakech;$^{(d)}$LPMR, Facult\'e des Sciences, Universit\'e Mohamed Premier, Oujda;$^{(e)}$Facult\'e des sciences, Universit\'e Mohammed V, Rabat;$^{(f)}$Institute of Applied Physics, Mohammed VI Polytechnic University, Ben Guerir; Morocco.\\
% ...
$^{45}$Physics Department, Southern Methodist University, Dallas TX; United States of America.\\
% ...
$^{49}$Fakult\"{a}t Physik, Technische Universit{\"a}t Dortmund, Dortmund; Germany.\\
% ...
$^{55}$II. Physikalisches Institut, Georg-August-Universit\"{a}t G\"ottingen, G\"ottingen; Germany.\\
% ...
$^{84}$$^{(a)}$Khalifa University of Science and Technology, Abu Dhabi;$^{(b)}$New York University Abu Dhabi, Abu Dhabi;$^{(c)}$United Arab Emirates University, Al Ain;$^{(d)}$University of Sharjah, Sharjah; United Arab Emirates.\\
% ...
$^{103}$CPPM, Aix-Marseille Universit\'e, CNRS/IN2P3, Marseille; France.\\
% ...
$^{121}$Homer L. Dodge Department of Physics and Astronomy, University of Oklahoma, Norman OK; United States of America.\\
% ...
$^{155}$Raymond and Beverly Sackler School of Physics and Astronomy, Tel Aviv University, Tel Aviv; Israel.\\
% ...
$^{a}$ Also at Affiliated with an institute formerly covered by a cooperation agreement with CERN.\\
$^{b}$ Also at An-Najah National University, Nablus; Palestine.\\
$^{c}$ Also at Borough of Manhattan Community College, City University of New York, New York NY; United States of America.\\
$^{d}$ Also at Center for Interdisciplinary Research and Innovation (CIRI-AUTH), Thessaloniki; Greece.\\
$^{e}$ Also at Centre of Physics of the Universities of Minho and Porto (CF-UM-UP); Portugal.\\
$^{f}$ Also at CERN, Geneva; Switzerland.\\
% ...

\end{flushleft}


%

%
%

%

%
%
%
%

\end{document}